\documentclass[twocolumn,amssymb, nobibnotes, aps, prd, superscriptaddress]{revtex4-1}

\usepackage[utf8]{inputenc}
\usepackage{graphicx}
\usepackage{mathtools}
\usepackage{amsmath}
\usepackage{amsfonts}
\usepackage{amssymb}
\usepackage{mathrsfs}  
\usepackage{float}
\usepackage{xcolor}
\usepackage[export]{adjustbox}
\definecolor{amethyst}{rgb}{0.6, 0.4, 0.8}
\usepackage{pgfplots} 
\usepackage{tikz-3dplot}
\pgfplotsset{compat=1.6,ylabsh/.style={every axis y label/.style={at={(0,0.5)}, xshift=#1, rotate=90}}}
\usetikzlibrary{positioning}
\usetikzlibrary{shapes}
\usetikzlibrary{shapes.geometric}
\usetikzlibrary{decorations.text}
\usetikzlibrary{decorations.pathreplacing,shapes.misc}
\usetikzlibrary{matrix}
\usetikzlibrary{pgfplots.statistics, pgfplots.colorbrewer} 
\usetikzlibrary{
  arrows.meta,                        
  backgrounds,                        
  ext.paths.ortho,                    
  ext.positioning-plus,                
  ext.node-families.shapes.geometric, 
  calc}         
\pgfplotsset{cycle list/BuPu-5}
\usepackage{pgfplotstable}
\usepackage{caption}
\newcommand*{\addheight}[2][.5ex]{
	\raisebox{0pt}[\dimexpr\height+(#1)\relax]{#2}
}
\usepackage{array}
\newcommand{\PreserveBackslash}[1]{\let\temp=\\#1\let\\=\temp}
\newcolumntype{C}[1]{>{\PreserveBackslash\centering}p{#1}}

\begin{document}

\title{A two-dimensional vertex model for curvy cell-cell interfaces at the subcellular scale}

	\author{Kyungeun Kim}
	\affiliation{Physics Department, Syracuse University}
	\author{J. M. Schwarz}
	\affiliation{Physics Department, Syracuse University}
	\affiliation{Indian Creek Farm, Ithaca, NY 14850}
	\author{Martine Ben Amar}
	\affiliation{Laboratoire de Physique de l'Ecole normale sup\'erieure, ENS, Universit\'e PSL, CNRS, Sorbonne Universit\'e, Universit\'e Paris cit\'e, F-75005 Paris, France}
	\affiliation{Institut Universitaire de Canc\'erologie, Facult\'e de m\'edecine, Sorbonne Universit\'e, 91 Bd de l'H\^opital,  75013 Paris, France}

\begin{abstract}
Cross-sections of cell shapes in a tissue monolayer typically resemble a tiling of convex polygons. Yet, examples exist where the polygons are not convex with curved cell-cell interfaces, as seen in the adaxial epidermis. To date, two-dimensional vertex models predicting the structure and mechanics of cell monolayers have been mostly limited to convex polygons. To overcome this limitation, we introduce a framework to study curvy cell-cell interfaces at the subcellular scale within vertex models by using a parameterized curve between vertices that is expanded in a Fourier series and whose coefficients represent additional degrees of freedom. This extension to non-convex polygons allows for cells with same shape index, or dimensionless perimeter, to be, for example, either elongated or globular with lobes. In the presence of applied, anisotropic stresses, we find that local, subcellular curvature, or buckling, can be energetically more favorable than larger scale deformations involving groups of cells. Inspired by recent experiments, we also find that local, subcellular curvature at cell-cell interfaces emerges in a group of cells in response to the swelling of additional cells surrounding the group. Our framework, therefore, can account for a wider array of multi-cellular responses to constraints in the tissue environment.

\end{abstract}
\maketitle

\section{Introduction}
  
 \hspace{0.5cm}Isolated cells take on a multitude of convex and non-convex shapes as determined by their underlying cytoskeletal morphology and their environment. Cells packed together to form a tissue can take on different shapes (from isolated cells), given the additional interactions between them. Interestingly, cells in a number of tissues take on convex polyhedral shapes, such as truncated octahedrons~\cite{MENTON1976,ALLEN1976,Yokouchi2016}. The observation of these convex polyhedra in cultured lung epithelial tissue and other tissues motivated a comparison with soap bubble arrangements, which then led to the birth of vertex models~\cite{honda1980much,Honda1982}. In vertex models, tissue cells are treated as deformable polyhedrons forming a space-filling packing with some energetic cost to the cellular deformation~\cite{fletcher2014vertex,honda2022vertex}.  Indeed, vertex modeling has provided key insights into understanding how cells interact with each other in tissues~\cite{Farhadifar2007,Staple2010,Okuda2012,guillot2013mechanics,Bi2015,Bi2016,kim2018universal,krajnc2018fluidization,sussman2020interplay,fiore2020mechanics,Sahu2020,tong2022linear,perez2023tension,Zhang2023,Staddon2023}. For instance, a predicted rigidity transition in disordered tissues~\cite{Bi2015} was discovered during a period of fast tail growth in Zebrafish embryos~\cite{Mongera2018,Lenne2022}. Cells in the mesodermal progenitor zone (MPZ) act as a liquid with cells exchanging neighbors. However, when cells become part of the presomitic mesoderm (PSM), they act more as solids, with a decrease in cell-to-cell contact variations. {\it In vitro} tissues have also been shown to exhibit glassy behavior and superelasticity, with these properties also emerging in vertex models~\cite{Angelini2011,latorre2018active,sussman2018anomalous}.

To date, vertex modeling of tissues has focused on polygons or polyhedrons with straight edges. And yet, there are numerous examples of both plant and animal tissues containing cells with curved edges \cite{Malinverno2017,Belteton2021,rigato2022}. See Fig. 1. Curved edges are possible presumably due to the underlying reorganization of the actomyosin cortex. Moreover, they allow cells to readily increase their surface area/perimeter while maintaining their volume/area. For instance, a study focusing on the rigidity transition in a monolayer of MCF10A cells found that in the control case, the cell-cell interfaces were curved at the sub-cellular scale~\cite{Malinverno2017}.  However, when they elevated a regulator of endocytosis, RAB5A, the cell-cell interface became less curved and the cells could more easily pass each other to form a fluid. Intriguingly, RAB5A not only regulates endocytosis by mediating the fusion rate of endosomes with endocytic vesicles, it also triggers lamellipodia formation in fibroblasts and so is involved with actin cytoskeletal reorganization~\cite{spaargaren1999rab5}. As lamellipodia are typically broad protrusions involving a branched dendritic network~\cite{svitkina1999arp2,gopinathan2007branching}, the curved interfaces straighten out. 

 More recently, experiments by Rigato, {\it et al.}  \cite{rigato2022} discovered curved cell-cell interfaces in a cell monolayer of the larval Drosophila epidermis.  This epidermis is made up of larval epithelial cells and histoblasts. The histoblasts typically cluster in groups of 5-17 cells. These groups are surrounded by larval epithelial cells. The histoblasts begin as straight-edged polyhedrons, however, over time their apical surface actively shrinks, causing their adherens junctions to fold and form a characteristic wavy pattern on the apical surface.  As larval epithelial cells expand, histoblasts are forced to adjust their shape to accommodate their own growth within the shrinking space available, inducing a type of folding of the actomyosin cortex with each cell. This dynamic change in histoblast morphology underscores the importance of cell confinement in the larval epidermis, which is presumably at the heart of this remarkable cell shape transformation.
 
 Based on the prevalence of cells in essentially close-packed tissues with curved cell-cell interfaces and the recent study demonstrating a transition from straight to curved edges, we modify the classic version of the two-dimensional vertex model \cite{honda2004three,shraiman2005mechanical,lin2017dynamic} to create a version suitable for tissues with curved cell-cell interfaces at the sub-cellular scale. We do so by replacing a straight line between two vertices with a parameterized curve described by several Fourier coefficients. This approach will allow us to go beyond simple arcs between vertices~\cite{Ishimoto2014} and does not involve increasing the number of vertices with straight edges (shared or not shared between cells) to allow for cell shapes that change convexity~\cite{Perrone2016,boromand2018jamming,vetter2023polyhoop}. We will explore under what conditions straight cell-cell interfaces are energetically favorable and under what conditions curved cell-cell interfaces are energetically favorable. Our exploration will focus on a simple ordered tiling as well as the experimental situation described in the work of Rigato {\it et al.} It will yield novel mechanical insights into the multitude of ways cells regulate their shape at the sub-cellular (or intra-cellular scale) leading to new emergent phenomena at the tissue scale. 
 
 \begin{figure*}
 	\captionsetup{singlelinecheck = false, justification=raggedright}
\begin{center}    
    \includegraphics[height=4cm,valign=b]{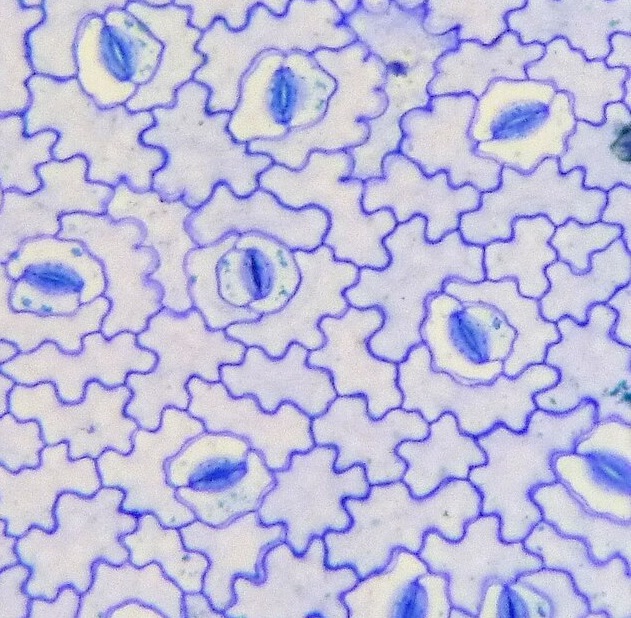}
    \hspace{1.0cm}
    \includegraphics[height=4.0cm,valign=b]{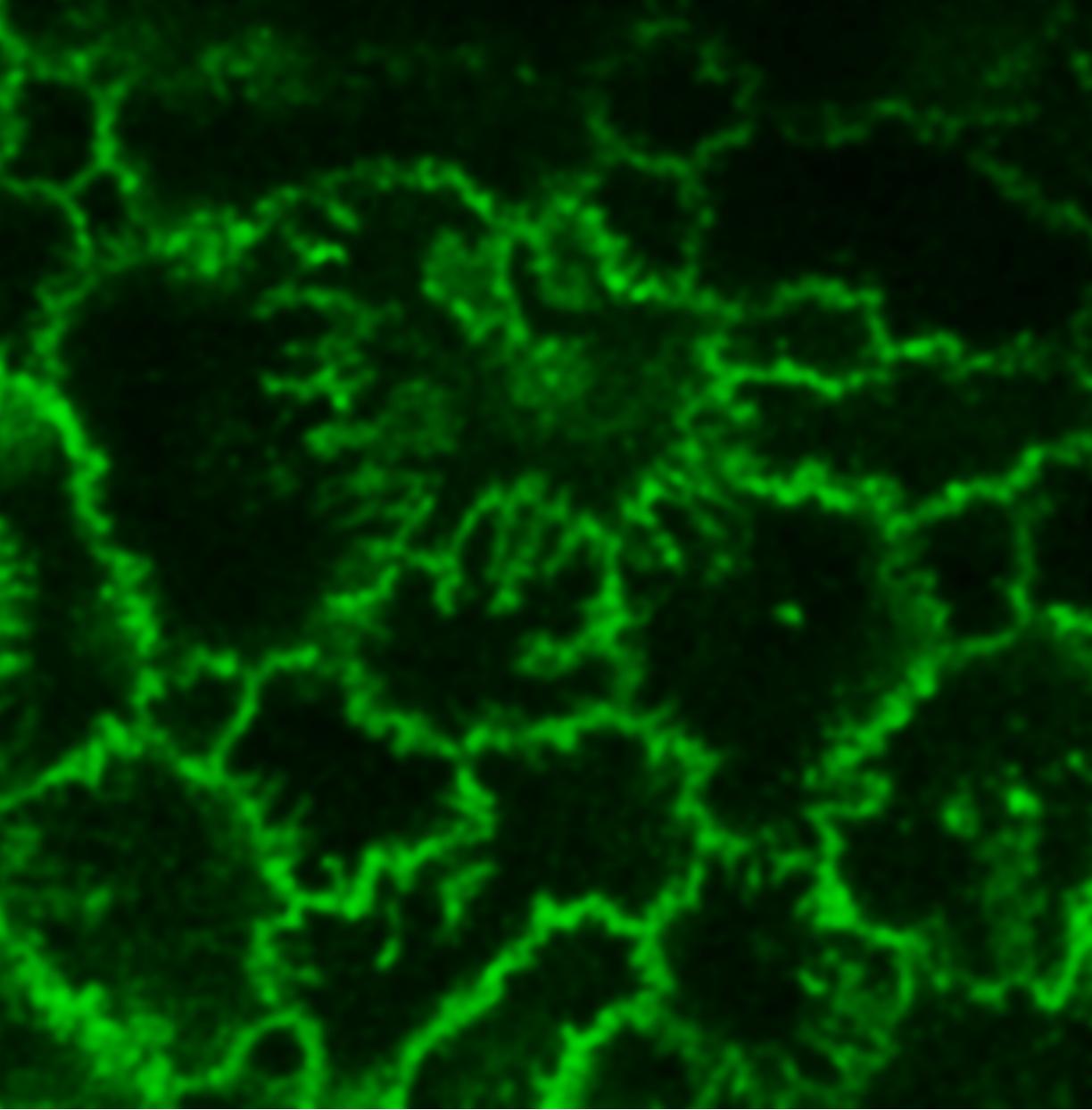}
    \hspace{1.0cm}
    \includegraphics[height=3.5cm,valign=b]{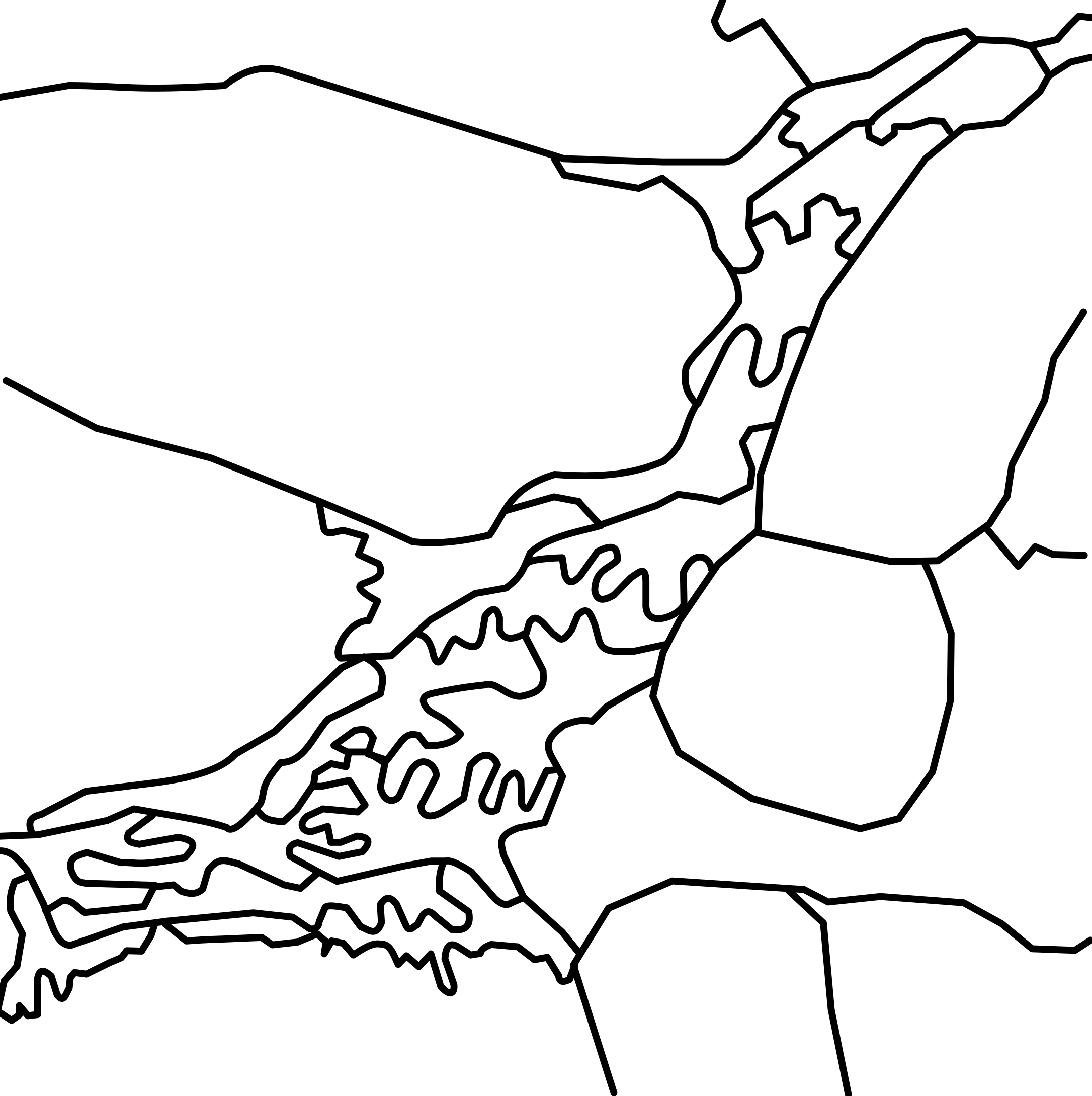}
   
			\caption{{\it Curvy cell-cell interfaces in plant and animal cells.} Left: Cross-section of a monolayer of plant cells. This is a cropped image originally taken by Karl Az and is licensed under CC BY 4.0 \cite{estomatos}. Middle: Cross-section of a monolayer of MCF10A cells with cell-cell interfaces fluorescently labelled in green. (Unpublished image with permission from G. Scita). Right: Drawing of a cross section of a monolayer of cells in the larval Drosophila dermis during development based on images from  Ref.~\cite{rigato2022}. }
\label{fig:1}
\end{center}
\end{figure*}
	\section{Model and Methodology}
	\hspace{0.5cm} While the mechanics of solid-like cellular tissue can be represented by continuous elastic fields at large enough wave lengths\cite{ben2019physics}, here, we consider a cellular-based approach to study the structure of cells as their cell-cell interfaces deviate from straight geometries to curved geometries in the presence of applied stresses. Let us begin with the mechanical energy $E_\alpha$ for the $\alpha$th cell within a cellular packing~\cite{Farhadifar2007,Staple2010,Bi2015} with cross-sectional area $A_\alpha$ and cross-sectional perimeter $P_\alpha$:  
\begin{equation}
\label{single_cell_energy}
\quad E_\alpha = K_{A} (A_\alpha-A_{0})^2 + K_P P_\alpha^2 + \gamma P_\alpha.
\end{equation}
The first term captures the cell's bulk elasticity ~\cite{hufnagel2007mechanism,zehnder2015cell} with $K_{A}$ denoting area stiffness and $A_0$ the target area. The second term in the cell mechanical energy 
is quadratic in the cell cross-sectional perimeter $P_\alpha$ and
 models the active contractility of the intracellular actomyosin cortex with elastic constant $K_P$~\cite{Farhadifar2007}. Finally, the last term represents an interfacial tension $\gamma$ set by a competition between the cortical tension and the energy of cell-cell adhesion~\cite{bonnet2012mechanical,manning_2010} between two cells in contact. Note that for a single cell, $\gamma$ denotes the surface tension of a cell. For simplicity, we assume that all four parameters---$K_A$, $A_0$, $K_P$, and $\gamma$---are the same for each cell.
 
 The addition of a constant to the mechanical energy, which does not affect the forces, allows one to complete the square to obtain:
 \begin{equation}
E_\alpha = K_{A} (A_\alpha-A_{0})^2 + K_P {(P_\alpha - P_{0})}^2,\label{eq-e_tot}
\end{equation}
with $P_{0} = -\gamma / (2 K_P)$ as the effective target perimeter.  This energy can be non-dimensionalized such that 
\begin{equation}
\label{scaled_e_tot}
\mathcal{\epsilon} =\frac{1}{K_A A_0^2} E_\alpha = (a_\alpha-1)^2 + \frac{K_P}{(K_A A_0)}(p_\alpha - p_0)^2
\end{equation}
where $a_\alpha = A_\alpha/A_0$ and $p_\alpha = P_\alpha /\sqrt{A_0}$  are the rescaled shape functions for area and perimeter. Moreover, $p_0 = P_0/\sqrt{A_0}$ is the \textit{target shape index}, a quantity that plays a crucial role in determining the mechanics of tissues in two dimensions. To be specific, 
a regular hexagon corresponds to $p_0^{hex}=2 \sqrt{2} \sqrt[4]{3}\approx3.72$ and a regular pentagon to $p_0^{pent}=2 \sqrt{5} {(5-2 \sqrt{5})}^{1/4}\approx3.81$. Wavy cell shapes are also expected to have target shape indices larger than the circle as they also deviate from the circle in a manner different from a convex polygon. 
	\begin{figure*}
 	\captionsetup{singlelinecheck = false, justification=raggedright}
 \centering
\begin{tabular}{cc}
    \addheight{\includegraphics[width=75mm]{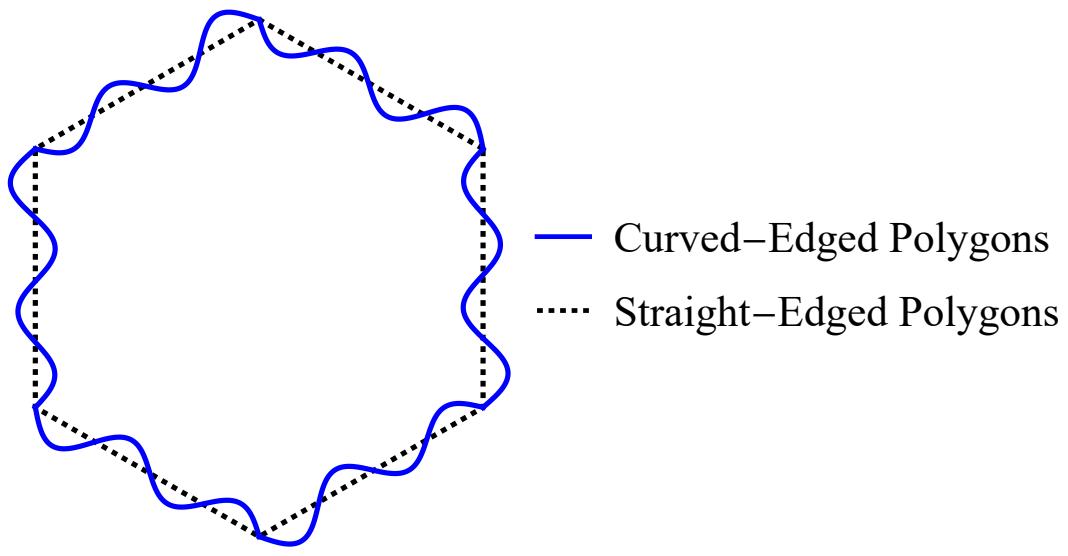}
		} & \addheight{\includegraphics[width=50mm]{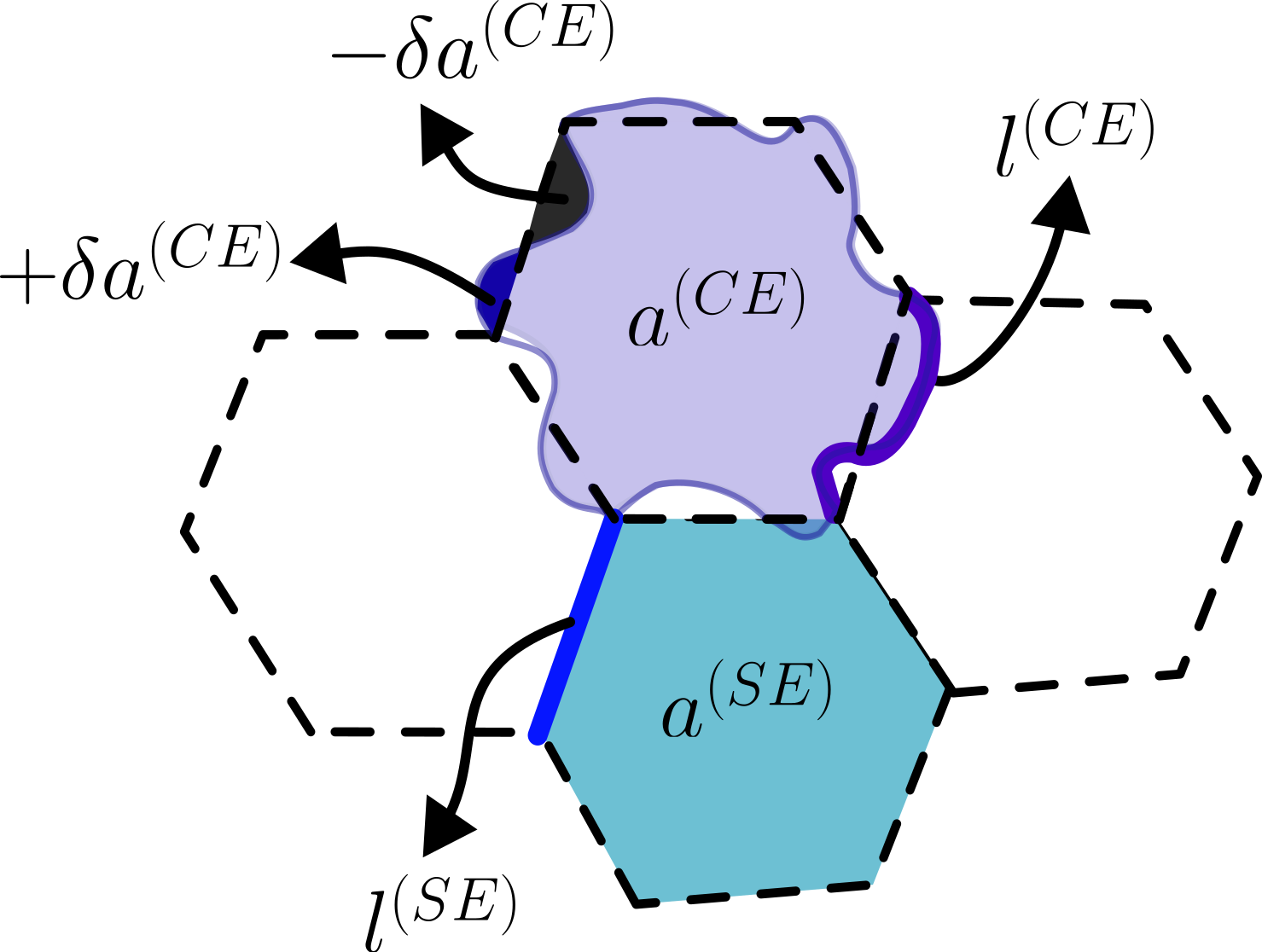}
		} \\
    (a) Straight-edged/curved-edged polygons of the cell. & (b) Example for area and length components of CE/SE. 
\end{tabular}
\caption{{\it Representation of straight-edged/curved-edged polygons.} (a) A polygon can be constructed with curved edges and with straight edges. (b) Area and length components can be obtained separately for the curved and straight edges. Note that $\pm\delta a^{(CE)}$ represents relative area differences as compared to straight-edged polygons. Signs indicate whether area was subtracted from, or added, to area of the straight-edged polygon.}
			\label{fig:vm-cell}
	\end{figure*}

A deformable cell is now described as a combination of a straight-edge polygon (SE) and a curved-edge polygon (CE), as shown in Fig. \ref{fig:vm-cell}. As the curved-edges extend the cellular geometry by incorporating a continuous field, which introduces, in principle, infinite degrees of freedom at each edge, in practice, the additional degrees of freedom to capture curvature between two vertices will be constrained in several ways. First, we chose the mathematical expression $c^{(1)}\sin\theta+c^{(2)}\sin2\theta+c^{(3)}\sin3\theta+c^{(4)}\sin4\theta$ to parameterize each curved edge. In addition to constraining the endpoints to match those of the straight edge solutions in certain cases, the sinusoidal function more readily ensures that $\delta a^{(CE)}$ in Fig. \ref{fig:vm-cell} (b) is zero even if the edge length changes. While this choice is not necessary, it makes it easier to calculate the $c^{(r)}$'s. We will address additional constraints below. Also, we exclude overhangs in our formulation for now.

For the curved edge model, we connect vertex pairs using a parameterized curve within the range $t\in [0,1]$. The number of polygon vertices is equal to the number of edges for the single cell, so we define coefficients $c^{(j)}_{\alpha,i}$ to describe the $j$-th coefficient of the cell edge $i$ between vertices $i$ and $i+1$ and for cell $\alpha$ (equivalently, the $k$-th edge of the entire cell array). We now use an index $k$ to represent edges explicitly. The polygon's area (and perimeter) depends on indices $\alpha$ and $k$. Given our parameterization, the coordinate for a vertex $i$ associated with a curved edge $\mathbf{v}_k(t)$ and its associated curve $c_k(t)$ in terms of the Cartesian coordinate vector components $(x,y)$: 
\begin{align}
		c_k(t)&=c_{\alpha,i}(t)\nonumber\\
  &=c_{\alpha,i}^{(1)} \sin(\pi t) + c_{\alpha,i}^{(2)} \sin(2\pi t)\nonumber\\&+c_{\alpha,i}^{(3)} \sin(3\pi t) + c_{\alpha,i}^{(4)} \sin(4\pi t)\label{eq-sin}\\
		\textbf{v}_{k}(t)&=\textbf{v}_{\alpha,i}(t)=(f_{\alpha,i}(t),g_{\alpha,i}(t)),\\
		f_{\alpha,i}(t)&=(1 - t) x_{\alpha,i} + 
		t x_{\alpha,i+1} \nonumber\\
 & + \frac{c_{\alpha,i}(t)(x_{\alpha,i} + 
			x_{\alpha,i+1})}{\sqrt{
				(x_{\alpha,i} + x_{\alpha,i+1})^2 + (y_{\alpha,i} + y_{\alpha,i+1})^2}}
		,\label{eq08}\\
		g_{\alpha,i}(t)&=(1 - t) y_{\alpha,i} + 
		t y_{\alpha,i+1}\nonumber\\
 & + \frac{c_{\alpha,i}(t)(y_{\alpha,i} + 
			y_{\alpha,i+1})}{\sqrt{
				(x_{\alpha,i} + x_{\alpha,i+1})^2 + (y_{\alpha,i} + y_{\alpha,i+1})^2}},\nonumber\label{eq09}\\.
\end{align}
where the bold type indicates a vector quantity. This parameterization has certain singularities for several specifically angled edges, such as a horizontal line, in which case a different parameterization can be used.  Please see the Appendix for a few more details.
The straight-edge equivalent for $\textbf{v}_k^{(SE)}$ is:
\begin{align}
		\textbf{v}_{k}^{(SE)}(t)&={\textbf{v}}^{(SE)}_{\alpha,i}(t)=(f^{(SE)}_{\alpha,i}(t),g^{(SE)}_{\alpha,i}(t)),\\
		f^{(SE)}_{\alpha,i}(t)&=(1 - t) x_{\alpha,i} + 
		t x_{\alpha,i+1},\\
		g^{(SE)}_{\alpha,i}(t)&=(1 - t) y_{\alpha,i} + 
		t y_{\alpha,i+1}.
	\end{align}
With this information, in addition to computing their perimeter, we can also trivially compute their area by determining the polygon center, given by the average of the vertex coordinates, and then dividing up the polygon into triangles each with areas $A^{(SE)}_{\alpha,i}$ and whose endpoints consist of the polygon center and the two vertices $i$ and $i+1$, or 
\begin{equation}
	A^{(SE)}_{\alpha,i}=\frac{1}{2!}\begin{vmatrix}
			x_{\alpha,i}&y_{\alpha,i} & 1 \\ 
			x_{\alpha,i+1}&y_{\alpha,i+1} & 1\\
			x_{\alpha,c}&y_{\alpha,c}& 1
		\end{vmatrix},
 \end{equation}
where vertical lines indicate the determinant.  The total polygon area is the sum of such triangular areas. 

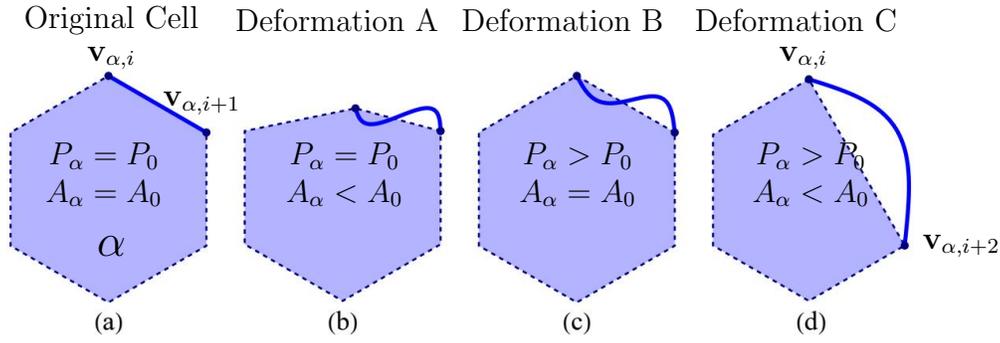
\begin{figure*}
	\captionsetup{singlelinecheck = false, justification=raggedright}
	\begin{center}
		\begin{tikzpicture}
			\node[anchor=south west,inner sep=0] (image) at (0,0) {\includegraphics[width=0.5\textheight]{"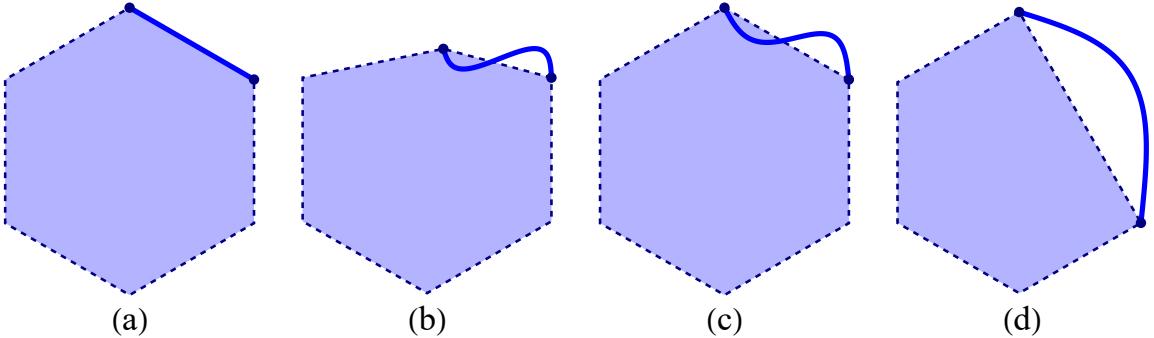"}};
			\node[align=center,black,thick,font={\large}] at (1.4,3.7) {$\textbf{v}_{\alpha,i}$};
			\node[align=center,black,thick,font={\large}] at (2.6,3.1) {$\textbf{v}_{\alpha,i+1}$};
			\node[align=center,black,thick,font={\large}] at (10.6,3.7) {$\textbf{v}_{\alpha,i}$};
			\node[align=center,black,thick,font={\large}] at (12.7,1.2) {$\textbf{v}_{\alpha,i+2}$};
			\node[align=center,black,thick,font={\large}] at (1.3,2.4) {$P_\alpha=P_0$};
			\node[align=center,black,thick,font={\large}] at (1.3,1.9) {$A_\alpha=A_0$};
			\node[align=center,black,thick,font={\large}] at (4.5,2.4) {$P_\alpha=P_0$};
			\node[align=center,black,thick,font={\large}] at (4.5,1.9) {$A_\alpha<A_0$};
			\node[align=center,black,thick,font={\large}] at (7.6,2.4) {$P_\alpha>P_0$};
			\node[align=center,black,thick,font={\large}] at (7.6,1.9) {$A_\alpha=A_0$};
			\node[align=center,black,thick,font={\large}] at (10.7,2.4) {$P_\alpha>P_0$};
			\node[align=center,black,thick,font={\large}] at (10.7,1.9) {$A_\alpha<A_0$};
			\node[align=center,black,thick,font={\LARGE}] at (1.4,1.2) {$\alpha$};
			\node[align=center,black,thick,font={\large}] at (1.4,4.2) {Original Cell};
			\node[align=center,black,thick,font={\large}] at (4.4,4.2) {Deformation A};
			\node[align=center,black,thick,font={\large}] at (7.4,4.2) {Deformation B};
			\node[align=center,black,thick,font={\large}] at (10.5,4.2) {Deformation C};
		\end{tikzpicture}
	\end{center}
	\caption{{\it Potential types of deformations in the curved edge model.} (a) Original cell shape. (b) Deformation type A (perimeter preserved): initial side lengths are conserved while Euclidean distance of pairs $(i,i+1)$ decreases. (c) Deformation type B (area conserved): side lengths are increased while the Euclidean distance of pairs $(i,i+1)$ is conserved. (d) One of the possible cases in deformation type C (vertex removed): the junction angle between a line $(\textbf{v}_i,\textbf{v}_{i+1})$ and $(\textbf{v}_{i+1},\textbf{v}_{i+2})$ becomes zero while the sum of initial side length of pairs $(i,i+1)$ and $(i+1,i+2)$ is conserved (a point $\textbf{v}_{i+1}$} is hidden inside of the straight line).
	\label{fig:vertex-model-type}
\end{figure*}

A similar approach to compute the cell area for the curved edge model can be done using the parameterization for each curved edge in a polygon. However, there are several possibilities now that the Euclidean distance between vertices and the edge length, or contour length, between vertices can be different.  In  Fig. \ref{fig:vertex-model-type}, deformation type A denotes cell side contraction where initial side lengths are maintained but the Euclidean distance of pairs $(i,i+1)$ is reduced. In other words, the area of the cell decreases while its perimeter remains the same. Deformation type B refers to the elongation of cell edges while maintaining the same Euclidean distance between pairs $(i,i+1)$, or the cell area remaining fixed while its perimeter increases. Deformation type C results in the cell perimeter increasing while the area decreases, as shown in Fig. \ref{fig:vertex-model-type}, as the junction angle at a vertex flattens. For the time being, we will not explore in detail the subcellular mechanisms driving these shape changes other than to state that defomration type B change, for example, could be induced by radially-oriented microtubules or localized branched actin structures~\cite{bouchet2017microtubules,gopinathan2007branching}. Moreover, the retraction of microtubules lead to an increase in acto-myosin cortex contraction, observed in individual migrating cells, which may lead to deformation type A~\cite{kopf2020microtubules}. In any event, subcellular curvature may arise from different biophysical mechanisms that involve both internal and external forces. 
	
To compute the area under each parameterized curve to find the total area of the cell, we find the areas deviating from the straight edged model, keeping track of signs, and add them to the straight edge polygon area. The sign of area segments is positive when a curve segment is located outside the straight edge polygon version and negative when the curve segment is located inside. See Fig. \ref{fig:vm-cell}(b). Specifically, the deviation in area $\delta a_{\alpha}$ for polygon $\alpha$, for a Euclidean distance between to $2$  vertices denoted by $d$ and index $j$ denoting breaking up each curved edge into subsections, is given by
	\begin{align}
		\delta a_{\alpha}^{(CE)}&=\sum_j dt\cdot
		\begin{cases}
			d({\textbf{v}}_{\alpha,i}(j\cdot dt),{\textbf{v}}^{(SE)}_{\alpha,i}(j\cdot dt)),& \text{(outside)}\nonumber\\
			-d({\textbf{v}}_{\alpha,i}(j\cdot dt),{\textbf{v}}^{(SE)}_{\alpha,i}(j\cdot dt))          & \text{(inside).}
		\end{cases}\label{eq13}\\
	\end{align}
    For lengths of shared edges between cells $\alpha$ and $\beta$ and corresponding $(i,i+1)$ and $(i',i'+1)$ vertex pairs, 
\begin{align}
		l_{k}^{(CE)}&=	\begin{cases}
			\sum_j 	d({\textbf{v}}_{\alpha,i}(j\cdot dt),{\textbf{v}}_{\alpha,i}({j+1}\cdot dt)),& \text{(for cell $\alpha$)}\nonumber\\
			\sum_j 	d({\textbf{v}}_{\beta,i'}(j\cdot dt),{\textbf{v}}_{\beta,i'}({j+1}\cdot dt))& \text{(for cell $\beta$)}.
		\end{cases}\\
	\end{align}
    The result for $v_{\alpha,i}=v_{\beta,i^{\prime}},v_{\alpha,i+1}=v_{\beta,i^{\prime}+1}$ is identical for either choice. Thus, we will simply note above as 
	\begin{align}
		l_{k}^{(CE)}=
		\sum_j 	d({\textbf{v}}_{k}(j\cdot dt),{\textbf{v}}_{k}({j+1}\cdot dt))\label{eq14}
	\end{align} 
	To numerically determine $dt$ we set substantive deviations of $c^{(1)}=c^{(2)}=c^{(3)}=c^{(4)}=0.05$ and adopted $dt=\frac{1}{40}$ to balance computational speed with precision up to two decimal places to compute cell areas and perimeters. Smaller $dt$s will yield higher precision results.  

\tikzset{
  basic box/.style={
    shape=rectangle, rounded corners, align=center, draw=#1, fill=#1!25},
  header node/.style={
    node family/width=header nodes,
    font=\strut\large,
    text depth=+.3ex, fill=white, draw},
  header/.style={
    inner ysep=+1.5em,
    append after command={
      \pgfextra{\let\TikZlastnode\tikzlastnode}
      node [header node] (header-\TikZlastnode) at (\TikZlastnode.north) {#1}
      node [span=(\TikZlastnode)(header-\TikZlastnode)]
           at (fit bounding box) (h-\TikZlastnode) {}
    }
  },
  fat black line/.style={ultra thick, black}
}
\begin{figure*}
		\captionsetup{singlelinecheck = false, justification=raggedright}
        	\begin{tikzpicture}[
          node distance=1cm and 1.2cm,
          thick,
          nodes={align=center},
          >={Latex[scale=.9]},
          ortho/install shortcuts]
    \node[basic box=cyan, header=Straight-Edged Model, anchor=north] at (0,0) (SE) {Energy $\mathcal{A}^{(SE)}$ for area of polygon $\alpha$, $a^{(SE)}_{\alpha}$\\
Energy $\mathcal{P}^{(SE)}$ for perimeter of polygon $\alpha$, $p^{(SE)}_{\alpha}$\\
    Energy $\mathcal{R}^{(SE)}$ for initial positions of polygon $\alpha$};
    \node[basic box=blue, header=Curved-Edged Model, anchor=north] at (0,-3.5)
  (CE) {Energy $\mathcal{A}^{(CE)}$ for area of polygon $\alpha$, $a^{(CE)}_{\alpha}$\\
    (or energy $\mathcal{A_\delta}^{(CE)}$ for area change from SE model, $\delta a^{(CE)}_{\alpha}$)\\
Energy $\mathcal{P}^{(CE)}$ for perimeter of polygon $\alpha$, $p^{(CE)}_{\alpha}$\\
    Energy $\mathcal{R}^{(CE)}$ for initial state of curved edges in $\alpha$};
    \node[ basic box=lightgray, header=Simultaneous Minimization, anchor=north] at (9,-2.5) (E1) {Set the total energy to be $E^{(SE)}+E^{(CE)}$\\
    and minimize.};
    \node[ basic box=lightgray, header=Independent Minimization, anchor=north] at (9,0.7) (E2) {Set the total energy to be $E^{(SE)}$ or $E^{(CE)}$\\
    and minimize.\\
    Straight-Edged Model:  $E^{(SE)}=\mathcal{A}^{(SE)}+\mathcal{P}^{(SE)}+\cdots$\\
    Curved-Edged Model:  $E^{(CE)}=\mathcal{A}^{(CE)}+\mathcal{P}^{(CE)}+\cdots$};
    \node[ basic box=lightgray, header=Sequential Minimization, anchor=north] at (9,-4.8) (E3) {Minimize $E^{(SE)}$, then minimize\\ $E^{(CE)}$ with the solution of $E^{(SE)}$\\ as the initial condition of $E^{(CE)}$,\\
    (vertices can be fixed after SE minimization or not).};
        \end{tikzpicture}
			\caption{{\it Computational methods for evaluating cell shapes from given energies.} Either SE model, CE model, or SE+CE model can be taken for analysis and based on the problem at hand, optimization can be done once or separately.}
			\label{fig:vm-cell2}
	\end{figure*}
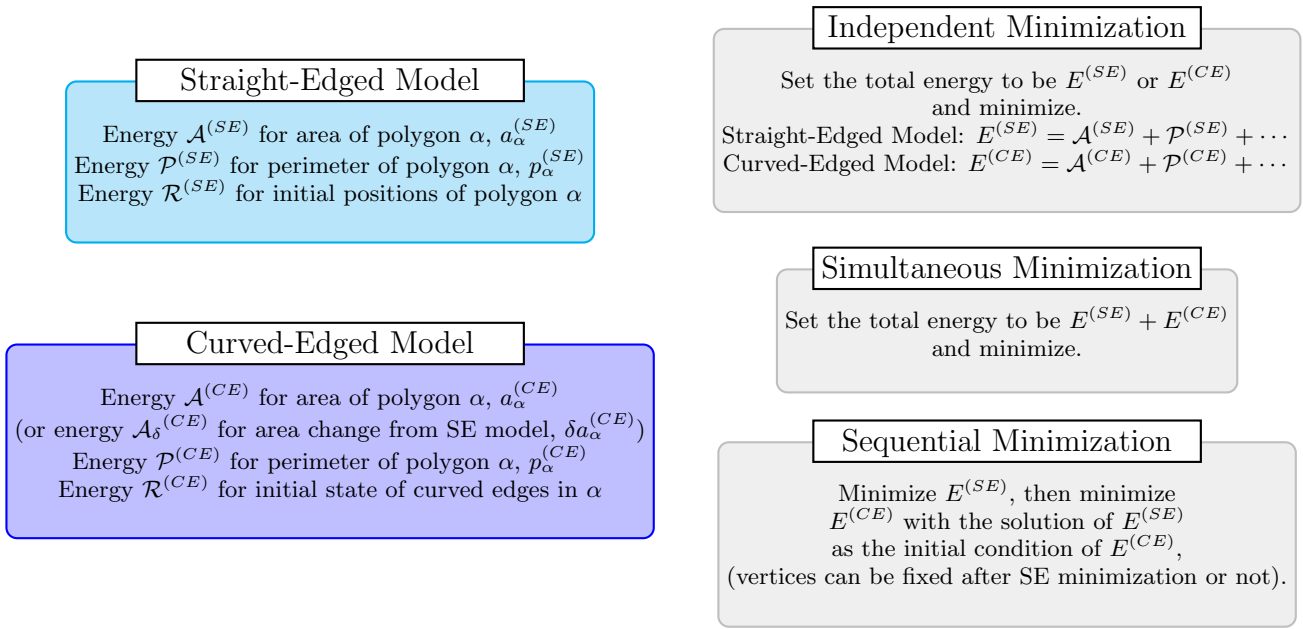
We also impose additional soft constraints on both the straight-edged (SE) and curved-edged (CE) models with the additional resistance contributions to the energy of each cell. For instance, we impose a vertex resistance energy function in which vertex positions are quadratically penalized to deviate from some target position as controlled by coefficient $K_R$ and summed over vertices $i$. Such energy contributions model the friction of the epithelium on a substrate, which can be either organic or inorganic, and represents the third dimension of space discarded in our 2D approach. In {\it in vitro} experiments the substrate can be rigid or soft, while in {\it in vivo} experiments the substrate is the extracellular matrix attached to the cell monolayer\cite{ackermann2022modeling,rigato2022}. The quadratic choice is inspired by the Winkler model \cite{winkler1867lehre}, which is very popular in solid mechanics because of its simplicity (the links are represented by a collection of springs), although more sophisticated extensions have been proposed.  Therefore, the SE energy function takes the form:
$E^{(SE)} = \mathcal{A}^{(SE)} + \mathcal{P}^{(SE)} + \mathcal{R}^{(SE)}$. 

For models with continuously curved edges, and as mentioned previously, there is a vertex resistance energy function enforcing fixed positions for two endpoints of a curved edge by quadratically penalizing deviations from target positions  and controlled by coefficient $K_{R1}$. An edge curvature resistance function ($\mathcal{R}_2$) exists as well by quadratically penalizing too large deviations from target coefficients in the sinusoidal harmonics defining the curves, and summed over edges $k$ and harmonics $j$ to assist with the minimization process. Note that $\mathcal{R}_2$ can be considered as a measure of the wall stiffness since large values limit their distortion. Formally, the CE model energy function takes the form: $E^{(CE)} = \mathcal{A}^{(CE)} + \mathcal{P}^{(CE)} + \mathcal{R}^{(CE)} \quad \mbox{with} \quad \mathcal{R}^{(CE)}=\mathcal{R}^{(CE)}_1 +\mathcal{R}^{(CE)}_2$.

While both models share similar energy components, the resistance term ($\mathcal{R}$) in the SE model addresses two potential scenarios. The first is vertex displacement resistance, which penalizes quadratic deviations of vertex positions due to friction of the substrate. For the curved-edged model, we represent the curved edges using sinusoidal harmonics with four coefficients ($c^{(j)}$). In our analysis, we set small initial values for these coefficients (for example, $c^{(j)}= 0.01$) using the $\mathcal{R}_2^{(CE)}$ function with small $K_{R2}$. The target Fourier coefficients are empirically chosen to constrain the amount of curvature along the cell-cell interface to more readily find local minima.  Indeed, there are physical constraints on the curvature due to the structure of the cytoskeleton and its ability to form filopodia and lamellipodia.  As we do not have a multiscale model containing more detailed information about cytoskeletal organization, hence the empirical choice. With fixed positions for straight-edged cells, the CE model involves four more unknowns (harmonic coefficients) per curved edge, or more degrees of freedom. To ensure solution convergence with these additional degrees of freedom, we utilize the $\mathcal{R}$ function as an additional constraint. We have also studied how strongly the energy minimized solutions depend on these additional energy contributions.  Please see the Appendix for more discussion on the relationship between degrees of freedom and constraints. If we assume that we have $2V$ number of unknowns for position vectors and $4E$ unknowns for curvy edge functions for $V$ number of vertices and $E$ number of edges, by adding the resistance function $\mathcal{R}_1=\frac{1}{2}K_{R1}\cdot\sum_{i}((x_{i}-x_{i}(0))^2+(y_{i}-y_{i}(0))^2)$ and $\mathcal{R}_2=K_{R2}\cdot\sum_i(c_i-c_t)^2$ we introduce an equal number of constraints to the system, which stabilizes the simulation.

Finally, in terms of the energy minimization procedure, as elaborated on in Fig. \ref{fig:vm-cell2}, we illustrate our scheme for evaluating the energy using polygons with straight edges and polygons with curved edges. We may first minimize the straight-edged model and then the curved-edge model.  Alternatively, we can also minimize the two models simultaneously. 

\section{Properties of the Curved-Edged Model}
One measure characterizing cell shape is the cell \emph{shape index} $p$, or its dimensionless perimeter. We now compare minimal energy solutions for both straight-edged solutions and curved-edge solutions for single cells and for ordered multi-cellular tilings, focusing on regular, hexagonal cells for simplicity.  
\begin{figure*}
	\captionsetup{singlelinecheck = false, justification=raggedright}
	\centering{
		\resizebox{80mm}{!}{\includegraphics{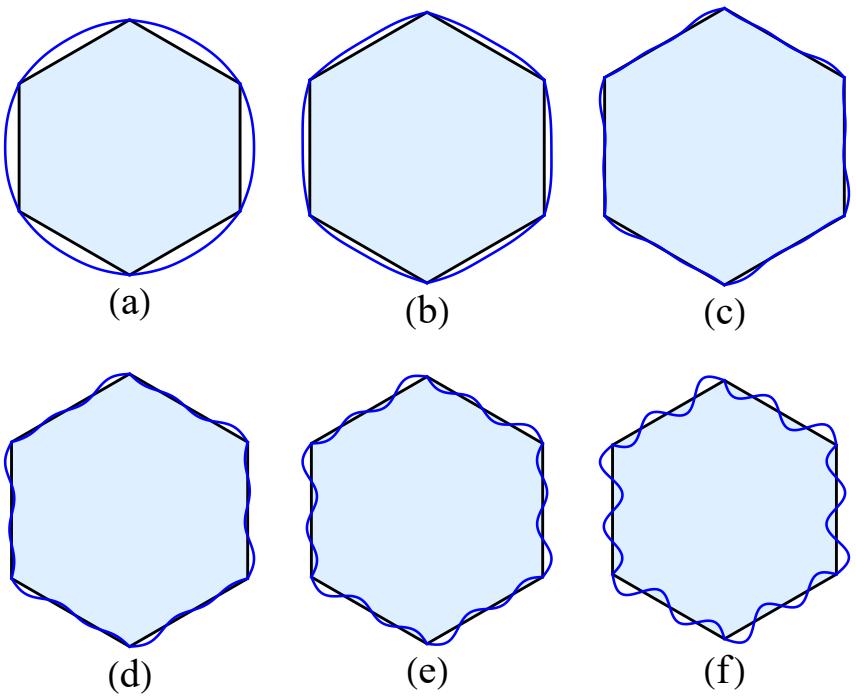}}
		\caption{{\it Different shape indices of curved-edged cells simulated and compared to a regular hexagon.} The target shape index of each shape is (a) 3.55 (b) 3.6 (c) $\frac{2\sqrt{6}}{3^{1/4}}\simeq3.72$ (d) 3.8 (e) 4 (f) 4.5.}
		\label{fig:sc-si-index}
	}
\end{figure*}
\subsection{Transition from convex to non-convex for a single cell} \label{sec-si}
As the target shape index is typically a parameter of interest, we begin with a regular hexagon and then increase the target shape index and look for minimal-energy states. For a regular hexagon, the cell shape index $p_\alpha$ (with $a_\alpha=1$) is as follows:
\begin{align}
p_\alpha &=\frac{6l^{(CE)}}{\sqrt{6a^{(CE)}}} = \frac{\sqrt{6}l^{(CE)}}{\sqrt{a^{(CE)}}} = \sqrt{6} p^{(CE)},
\end{align}
where $l^{(CE)}$ and $a^{(CE)}$ represent the side length and area of one of the six wedges in the curved-edge model. We also consider the energy function from Equation \ref{scaled_e_tot} with $a_\alpha=1$ (equivalently, the area term is negligible). Thus, the energy takes the form:

\begin{align}
E(p) &= K_{SI}\bigg(\sqrt{6}p^{(CE)}-p_0\bigg)^2 \nonumber\\
&=K_{SI}\bigg(\frac{\sqrt{6}\cdot l^{(CE)}}{\sqrt{\frac{\sqrt{3}}{4}+\delta a^{(CE)}}}-p_0\bigg)^2,\label{eq:hex-pie-si}
\end{align}
where $K_{SI}=\frac{K_P}{(K_A A_0)}$, $l^{(CE)}$ is the arc length of a wedge, $\delta a^{(CE)}$ is the area between the curved edge and the straight edge (as shown in Fig. \ref{fig:vm-cell}(b)), and $p_0$ represents the target shape index. 

In Fig. \ref{fig:sc-si-index} we show the results of minimizing $\min{(E(p)+\mathcal{R}_2^{(CE)})}$ with $K_{SI}=10$, $K_{R2}=0.001$ and each target Fourier coefficient $c_t$ for $K_{R2}\cdot\sum_i(c_i-c_t)^2$ is given by $c_t=0.01$ in $\mathcal{R}_2^{(CE)}$. Note that we also fix the vertices in the curved edge minimization to preserve the six-fold symmetry. To test the robustness of our results, we vary $K_{R2}$ from $0.001$ to $0.1$, as well as vary $c_t$, to show that there are predominantly changes in the third decimal place for the cell shape index as indicated in Table \ref{table:error}. As we increase the target cell shape index, the cells transition from convex to non-convex shapes. Such a transition does not occur in the straight edged model for a hexagon. Note that the convex-nonconvex transition occurs slightly below the shape index of a regular hexagon (with unit area). As  $K_{R2}\rightarrow 0$, this transition takes place at the regular hexagon.  However, again, for some range of $K_{R2}$, the solutions are robust. Moreover, for convex, regular polygons, as the shape index increases, the number of edges decreases with the triangle being the minimal number of edges. See Table \ref{table:6}. With a curved edged model, to increase their shape index, regular shapes can either decrease the number of edges or transition to nonconvex shapes to increase their shape index. Of course, cells can also become irregular to increase their shape index as will be discussed in the next subsection. 

\begin{table*}\caption{Measured shape index $p_\alpha$ with different coefficients (Equation \ref{eq:hex-pie-si} with $p_\alpha=\sqrt{6}p^{(CE)},~ p_0=3.55$).}	
\centering
	\begin{tabular}{c c c c c c}
		\hline\hline                        
	($K_{SI}=10$)& $c_t=0.01$	&$c_t=0.02$	&$c_t=0.03$	&$c_t=0.04$	&$c_t=0.05$\\ [0.5ex]
		\hline                 
$K_{R2}=0.001$	&3.54995	&3.55003	&3.55002	&3.55002	&3.55002\\
$K_{R2}=0.01$	&3.55027	&3.55024	&3.55022	&3.55020	&3.55019\\
$K_{R2}=0.1$	&3.55216	&3.55199	&3.55183	&3.55170	&3.55158\\	[1ex]  \hline
\end{tabular}\label{table:error}
\end{table*}

\begin{table*}\caption{Shape index}	\centering
	\begin{tabular}{c c c c c c c}
		\hline\hline                        
		& Circle & Decagon & Octagon & Hexagon & Square & Triangle\\ [0.5ex]
		\hline                 
		Number of edges&-&12&8&6&4&3\\
		Shape index&3.54491&3.58630&3.64072&3.72242&4&4.55901\\	[1ex]      
		\hline
	\end{tabular}\label{table:6}
\end{table*}

\subsection{Expansion of the zero-energy solution space}
\begin{figure*}
	\captionsetup{singlelinecheck = false, justification=raggedright}
	\centering
 \begin{tabular}{c c c c}
	\begin{tikzpicture}[scale=2.5]
		\draw[-,thick,fill=white](0,0)--(1,0)--(1,1)--(0,1)--cycle;
		\draw[thick,<->](1.1, 0.2)--(1.1, 0.9);
		\draw[thick,<->](-0.1, 0.2)--(-0.1, 0.9);
		\draw[thick,<->](0.1, 1.1)--(0.8, 1.1);
		\draw[thick,<->](0.1, 0.1)--(0.8, 0.1);
        \draw[thick,blue] plot [smooth, tension=1] coordinates {(0,1) (0.2,1.25)(0.5,1.35) (0.8,1.25) (1,1)};
		\node[label={0:{\color{blue}\large$l$}},inner sep=2pt] at (0.8,1.3) {};
		\node[label={0:{\large$l_1=1$}},inner sep=2pt] at (0.2,1.2) {};
		\node[label={0:{\large$l_1$}},inner sep=2pt] at (0.4,0.2) {};
		\node[label={0:{\large$l_2=1$}},inner sep=2pt] at (-0.6,0.5) {};
		\node[label={0:{\large$l_2$}},inner sep=2pt] at (1.1,0.5) {};
		\node[label={0:{\large$\theta$}},inner sep=2pt] at (0.7,0.25) {};
		\node[label={0:{\large$\theta$}},inner sep=2pt] at (0.1,0.75) {};
		\node[label={0:{\large$\pi-\theta$}},inner sep=2pt] at (0.4,0.75) {};
		\draw[-](0.8, 0) arc[start angle=180, end angle=90,radius=0.2cm];
		\draw[-](0.8, 1) arc[start angle=-180, end angle=-90,radius=0.2cm];
		\draw[-](0.2, 1) arc[start angle=0, end angle=-90,radius=0.2cm];
		\node[label={90:{\color{blue}\large$O$}},inner sep=1pt] at (-0.1,0) {};
		\draw[fill=blue](0, 0) circle[radius=0.03];
	\end{tikzpicture}
      & 
		\addheight{	\includegraphics[width=3.5cm,valign=b]{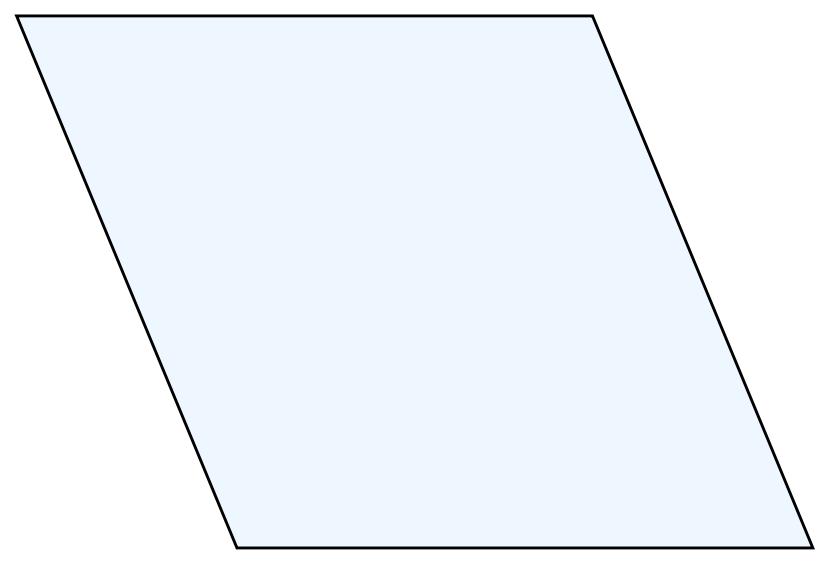}}
        &
		\addheight{	\includegraphics[width=3cm,valign=b]{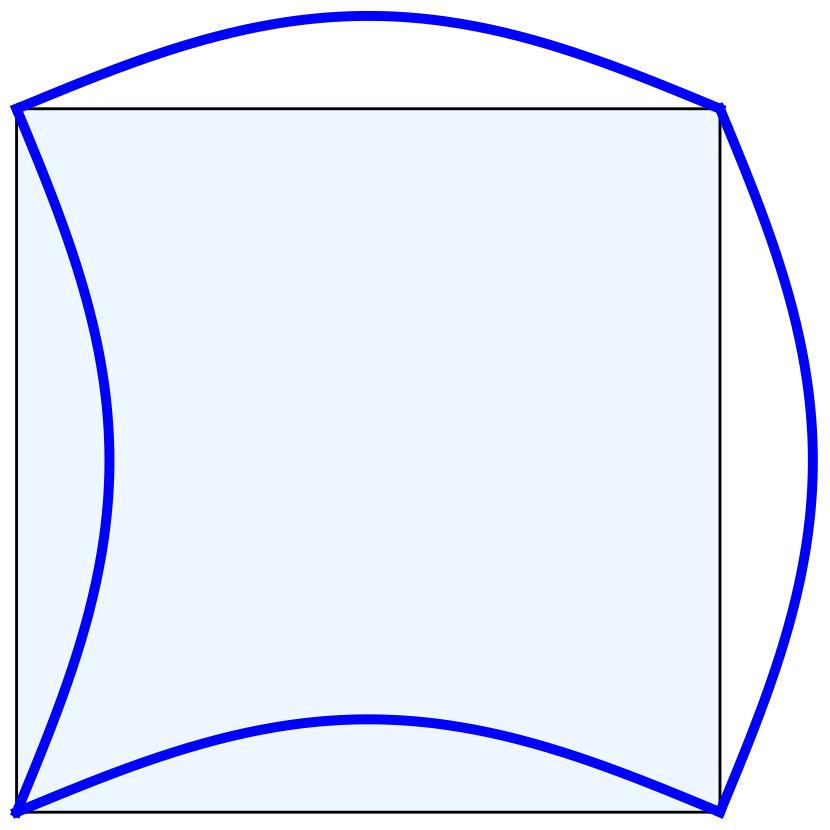}}
        & 
		\addheight{	\includegraphics[width=3cm,valign=b]{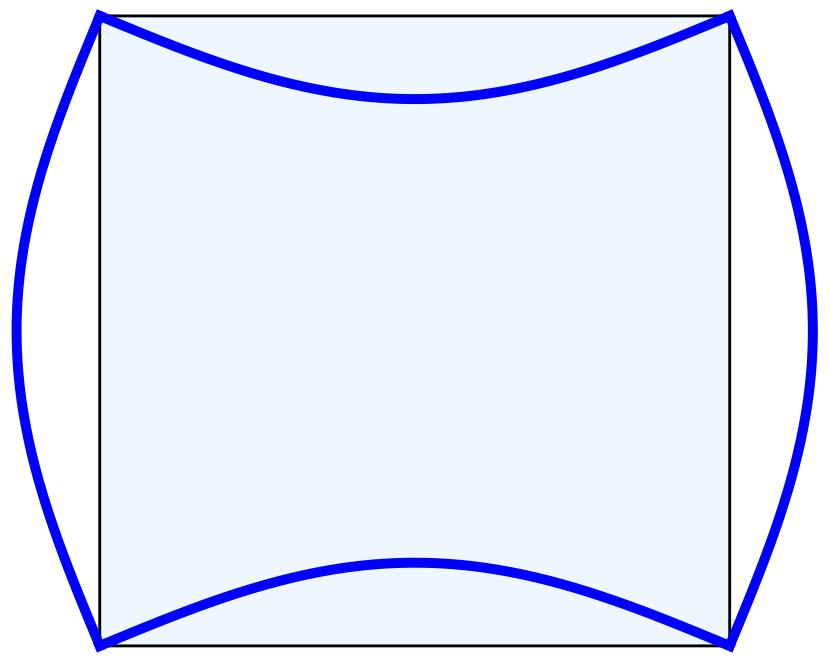}}\\
  (a)&(b)&(c)&(d)\\
      & 
		\addheight{	\includegraphics[width=3cm,valign=b]{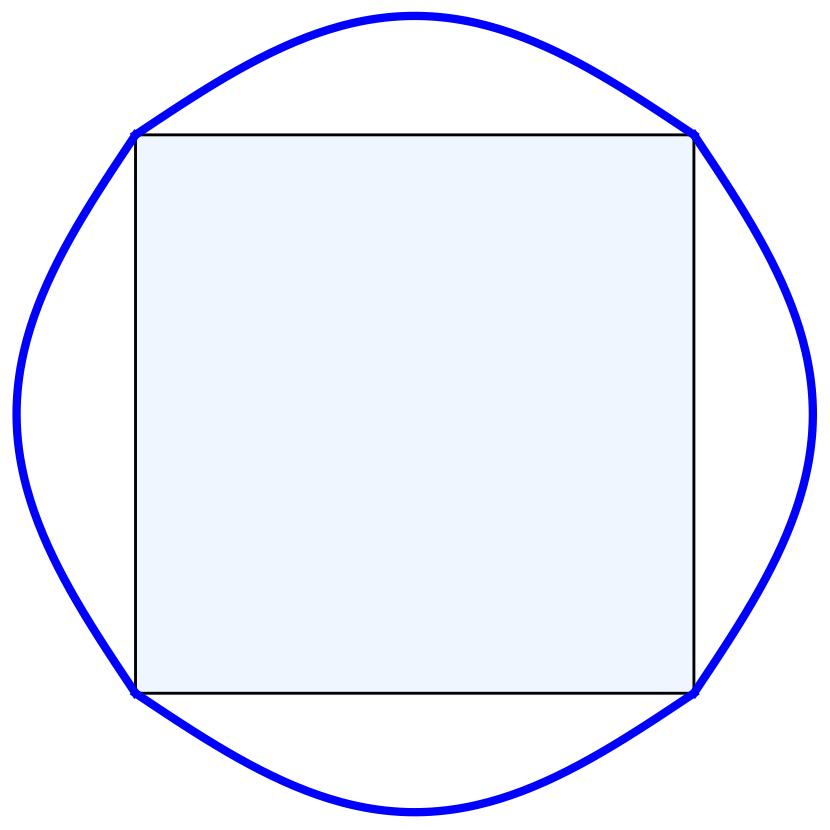}}
        &
		\addheight{	\includegraphics[width=3cm,valign=b]{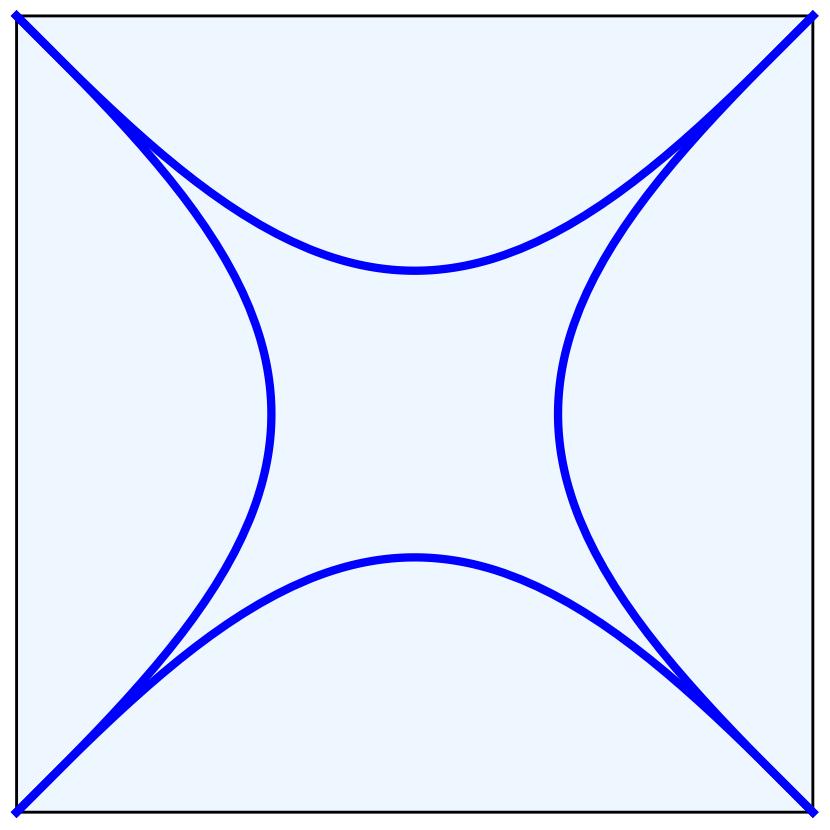}}
        & 
		\addheight{	\includegraphics[width=3cm,valign=b]{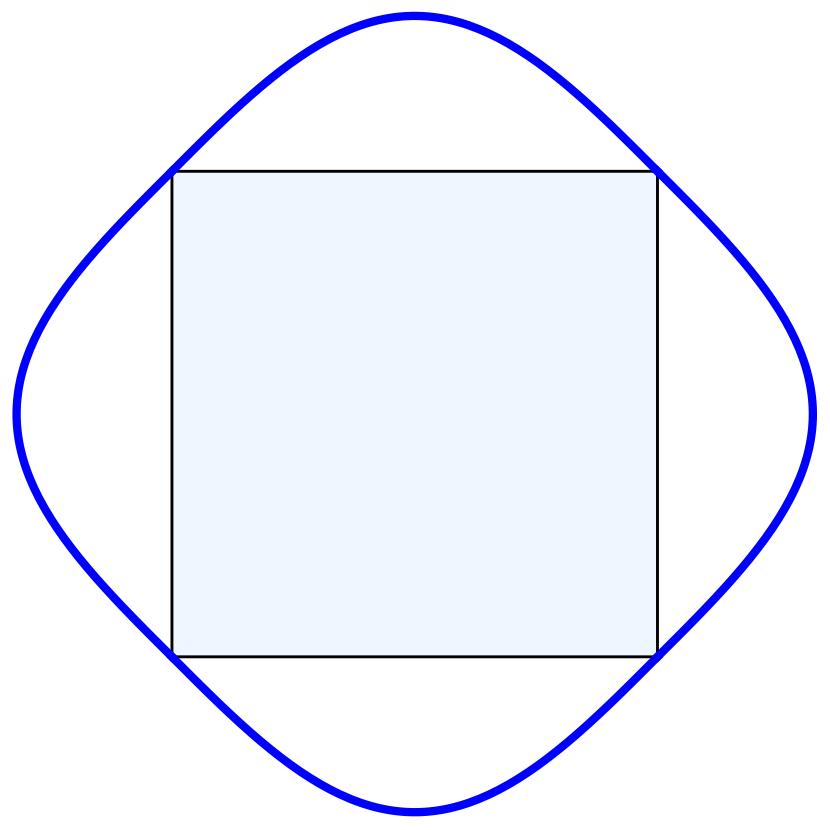}}\\
  &(e)&(f)&(g)\\
 \end{tabular}
	\captionsetup{type=figure}
	\captionof{figure}{{\it A simple rectangular cell model, different representations of cells with the same shape index $p_\alpha\simeq4.17$ (a)-(d) as well as limits on the shape index (e)-(g) for the regular case.} (a) The model can be described with two different edge lengths $l_1,l_2$ and an angle $\theta$. (b)-(d) Different representations of cells with the same shape index $p_\alpha\simeq4.17$. (b) $\theta=\frac{1.5 \pi}{4}$, (c) and (d) $\theta=\frac{\pi}{2},c_1=0.132$ for two edges ($c_1=-0.132$ for the other two edges). (e)-(f) Limits on the shape index for the regular model. (e) The minimum shape index with $c_1=0.213$, (f) The maximum perimeter based on the square with $c_1=-0.32$, and (g) the convex counterpart of (f)}.
	\label{fig:simple-rec}
\end{figure*}

Using the energy functional in Eq. 2.3 for a single cell, it has been shown that there exists a compatibility-incompatibility transition as the target shape index increases where the shapes eventually become zero-energy solutions, i.e., their shapes are compatible with the energy~\cite{Staddon2023}. For instance, for six-sided polygons, the location of the transition occurs at the target shape index of the regular hexagon. This analysis has been recently extended to three dimensions to also demonstrate such a transition ~\cite{kim2023mean}. For the curved edge model, its additional degrees of freedom thus allowing for greater compatibility and, therefore, an expansion of the zero-energy solution space. To more concretely illustrate this point, we will focus on a simplified model using a single sine curve, i.e., $c_2=c_3=c_4=0$. This example, depicted in Fig. \ref{fig:simple-rec}, captures the essence of curved-edge behavior while remaining tractable for calculations. For the straight-edged model, the compatibility-incompatibility transition occurs at $p_0=4$. 

Using a single sinusoid parameterized by the coefficient $c_1$, we can describe points along a curved edge.
\begin{align}
	(x,y)&=(f(t),g(t))\quad\text{for $t\in[0,1]$},\\
	f(t)&=t,\\
	g(t)&=c_1\sin\pi t.
\end{align}
We can also compute the edge length of curved-edged component by $l$
\begin{align}
l=&\int\sqrt{{f'(t)}^2+{g'(t)}^2}dt=\int_0^1\sqrt{1+c_1^2\pi^2\cos\pi t^2}dt\nonumber\\
=&\sqrt{1+c_1^2\pi^2}\int_0^1\sqrt{1-\frac{c_1^2\pi^2}{1+c_1^2\pi^2}\sin\pi t^2}dt\nonumber\\
=&2\frac{\sqrt{1+c_1^2\pi^2}}{\pi}El\bigg(\frac{{c_1^2\pi^2}}{1+c_1^2\pi^2}\bigg)\label{1arclen}.
\end{align}
 This allows us to compute the side length of the curved segment using an integral expression (Equation \ref{1arclen}), which involves the complete elliptic integral of the second kind (denoted by $El$). Assuming equal side lengths for the edges and focusing solely on angle changes, the perimeter is directly proportional to the side length. Interestingly, if we fix the area to $1$, both the perimeter and the shape index increase as we adjust the amplitude of the sine function (represented by $c_1$).

If we consider the area of the straight-edged polygon as $a^{(SE)}$, then we obtain
\begin{align}
a^{(SE)}=l_1 l_2\sin\theta.
\end{align}
The fixed perimeter of $2l_1+2l_2=4$ allows for calculation of the desired angle using the shape index. For the straight-edged polygon case, there are two solutions when $\theta\neq\frac{\pi}{2}$, and a single solution when $\theta=\frac{\pi}{2}$. Varying the values of $l_1$ and $l_2$ can yield additional solutions, but a single representation (up to global rotations) for $\theta=\frac{\pi}{2}$ remains when $l_1$ and $l_2$ are known. The curved-edged approach offers a larger range of shapes when changing the shape index of a square ($\theta=\frac{\pi}{2}$) as compared to the straight-edged case, where the shape index and angle are determined by a fixed perimeter. Fig. \ref{fig:simple-rec}(b)-(d) showcases this range, demonstrating different curved-edge representations for squares with the same shape index $p_\alpha\simeq4.17$. Thus, the zero-energy solution space is now higher dimensional as it now depends on $c_1$ as well. 

With the curved edges, the compatible-incompatible transition for a single cell now occurs near the shape index of the circle (relaxing the tiling condition, though one can consider cells of different concavities in a tiling), which is $c_1=0.213$ with $p_\alpha=3.55562$, as shown in Fig. \ref{fig:simple-rec}(e). We can also consider the limit of the CE model from regular SE-based vertex coordinates for this four-sided polygon when $\theta=\frac{\pi}{4}$, that is, when the edges potentially overlap with minimum area. If we use a single sine function, we have $c_1=0.32$ with shape index $11.32360$ as the limit, which can be seen in Fig. \ref{fig:simple-rec}(f). From its convex counterpart in Fig. \ref{fig:simple-rec}(g), i.e. rearranging curves to be outside the polygon, we can easily see that the extremum occurs closest to the shape of the extended square.  Thus, the compatibility-incompatibility transition point no longer occurs at $p_0=4$ but at a shape index much closer to the circle.

\subsection{A tiling of hexagonal shapes}
\begin{figure*}
	\captionsetup{singlelinecheck = false, justification=raggedright}
	\begin{center}
		\begin{tabular}{c c}
			\addheight{	\includegraphics[width=4.5cm]{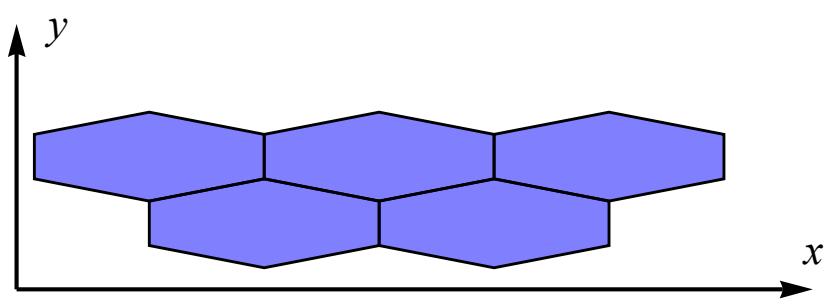}
			}&
			\addheight{\includegraphics[width=4.5cm]{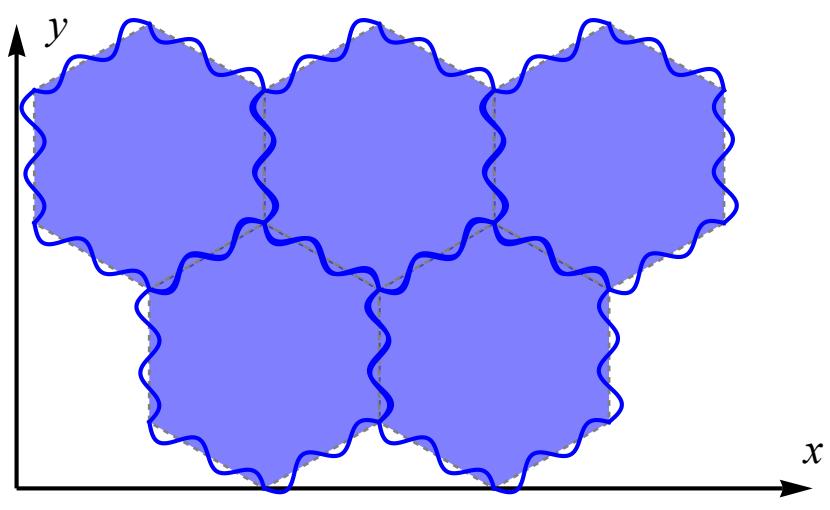}
			} \\
			(a) Shape index $p_\alpha\simeq4.5$&
			(b) Shape index $p_\alpha\simeq4.5$\\
			\addheight{\includegraphics[width=4.5cm]{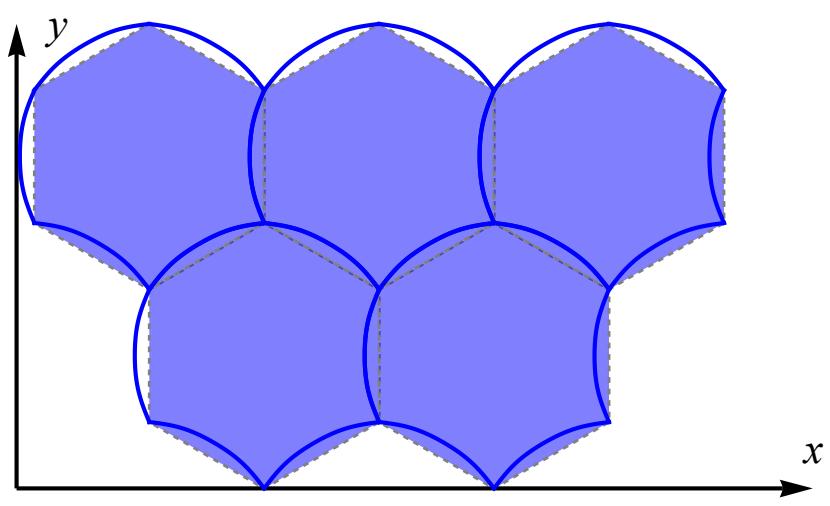}
			}&
			\addheight{\includegraphics[width=4.5cm]{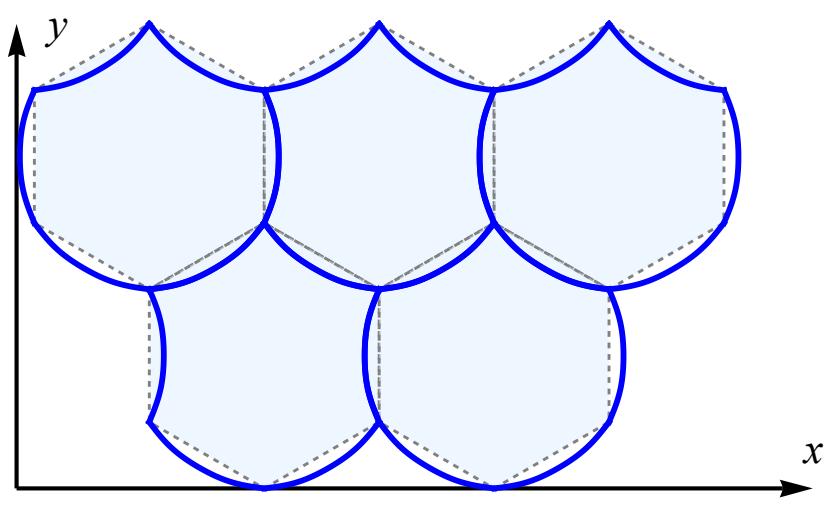}
			} \\
			(c) Ordered packing of a single cell&
			(d) Ordered packing of two cells\\
		\end{tabular}
		\captionsetup{type=figure}
		\captionof{figure}{{\it Tiling of SE and CE shapes for the same shape index and example of circular tiling.} (a) hexagonal cell array with shape index $p_\alpha\simeq4.5$. (b) Curved-edge model with sixfold symmetry with $p_\alpha\simeq4.5$. (c) Three edges are concave and others are convex in orientation relative to the hexagon. The shape index of this cell is $p_\alpha=3.89811$. (d) Two different cells so that, on average, three edges are concave and others are convex with $p_\alpha=3.83190$ and $3.9679$.}
		\label{fig:hex-tile}
	\end{center}	
\end{figure*}

Consider a tiling of ordered polygons as shown in Fig. \ref{fig:hex-tile}(a). Here, we have focused on a specific change for a given deformation, which is one of many possible shape changes, at least in the compatible regime. The polygons indeed remain hexagonal, though their shape index has increased as the polygons have become irregular with the distances between some vertices decreasing and between other vertices increasing  (and the area remaining fixed). However, suppose there is a change in shape index resulting from the stretching of the edges while maintaining the total area without an overall anisotropy developing, as shown in Fig. \ref{fig:hex-tile}(b). To do so in the straight-edge model becomes complicated, requiring additional vertices on the edges to capture the curved edges, but not with the curved-edge model. 

We can also create an ordered array of cells with two different shape indices from a single ordered straight-edged cell array. Motivated by circular cross-sectional cell shapes in the mouse ear epidermis~\cite{Ishimoto2014}, consider the case of a circular-edged tile emerging from a regular hexagonal array, as shown in Fig. \ref{fig:hex-tile} (c)-(d).
As the area of individual cells decreases due to the curved edges, their opposite edges morph into concave shapes. Interestingly, despite these variations, the average cell area for this circular arrangement approaches that of the original hexagon. From Equation \ref{eq:hex-pie-si}, we have:
\begin{align}
	p_\alpha=\sqrt{6}p^{(CE)}=\frac{6\cdot l^{(CE)}}{\sqrt{6\cdot a^{(CE)}}}=\frac{\sqrt{6}\cdot l^{(CE)}}{\sqrt{\frac{\sqrt{3}}{4}+\delta a^{(CE)}}}
\end{align}
While each cell retains the identical perimeter of $2\pi$ (based on the initial hexagonal array with radius $r=1$), their areas differ slightly due to the curved edges. We quantify this difference with $\pm6\delta a^{(CE)}$, where $\delta a^{(CE)}$ represents the area beneath a single curved segment. For a circle with a target shape index of $p_\alpha=3.54491$ as in Fig. \ref{fig:sc-si-index} (a), applying the shape index equations with $l^{(CE)}=\frac{2\pi}{6}$ ($\frac{1}{6}$ times perimeter of the unit circle) yields a required $\delta a^{(CE)}$ of approximately $0.09059$.

While neighboring cells in this circular pattern can adopt different shape indices based on their curved edge orientations, as illustrated by comparing Fig. \ref{fig:sc-si-index} (a) and (b), this freedom ultimately results in a slightly higher average shape index for the entire circular array compared to the hexagonal tiling. This occurs even though the area remains the same, because the increased perimeter of the circular array outweighs the unchanged area in terms of affecting the shape index calculation. However, while the straight-edge solution has the lower energy in this case, it may not be compatible with the symmetry and/or boundary conditions of the system at hand. More generally, energetically preferred shapes depend on the relative weight of the area and perimeter terms in the energy equation (Equation~\ref{scaled_e_tot}). To be concrete, consider a unit circle and a regular hexagon of unit edge length with the desired shape of a circle:
\begin{equation}
e_\alpha = (a_\alpha - 1)^2 + K_{SI}(p_\alpha - \frac{2\pi}{\sqrt{\pi}})^2.
\end{equation}
For the regular hexagon, this becomes:
\begin{equation}
e_\text{hex}=(\frac{3\sqrt{3}}{2\pi}-1)^2 + K_{SI}(\frac{6}{\sqrt{\pi}} - \frac{2\pi}{\sqrt{\pi}})^2.
\end{equation}
Fig.~\ref{fig:hex-tile} (c) has the same area term as the hexagon, but a different perimeter term:
\begin{equation}
e_\text{\ref{fig:hex-tile}(c)}=(\frac{3\sqrt{3}}{2\pi} - 1)^2 + K_{SI}(\frac{2\pi}{\sqrt{\pi}}-\frac{2\pi}{\sqrt{\pi}})^2.
\end{equation}
Fig.~\ref{fig:hex-tile} (d) has a slightly modified area term depending on the cell type (represented by $\pm$):
\begin{equation}
e_\text{\ref{fig:hex-tile}(d)}=(\frac{\frac{3\sqrt{3}}{2}\pm 0.09059 }{\pi} -1)^2 + K_{SI}(\frac{2\pi}{\sqrt{\pi}}-\frac{2\pi}{\sqrt{\pi}})^2.
\end{equation}
While the dimensionless energy $e_\alpha$ ultimately depends on the specific values of $K_{SI}$ for the hexagonal case, the dimensionless perimeter term from $(P_\alpha - 2\pi)^2\frac{1}{\pi}=(p_\alpha - \frac{2\pi}{\sqrt{\pi}})^2$ vanishes for Fig.~\ref{fig:hex-tile} (c)-(d), which can make the shapes shown in Fig.~\ref{fig:hex-tile}(c) the energetically favorable choice.

\section{Multicellular response to applied stresses}
For the first case, we will consider a line of cells in the presence of applied, anisotropic stress. In the second case, inspired by recent experiments~\cite{rigato2022}, we will explore how a group of cells can be compressed by surrounding cells. We study this latter phenomenon in both "ordered" and "disordered" scenarios.

\subsection{Curved-edges induced by anisotropic stress}\label{local-def}
\begin{figure*}
		\captionsetup{singlelinecheck = false, justification=raggedright}
        \begin{center}
        \begin{tabular}{c c c}
       \addheight{\includegraphics[width=7cm,valign=t]{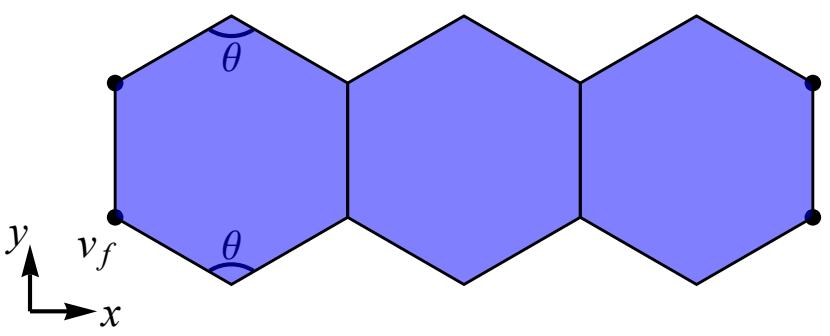}}&
			\addheight{	\includegraphics[width=3cm,valign=t]{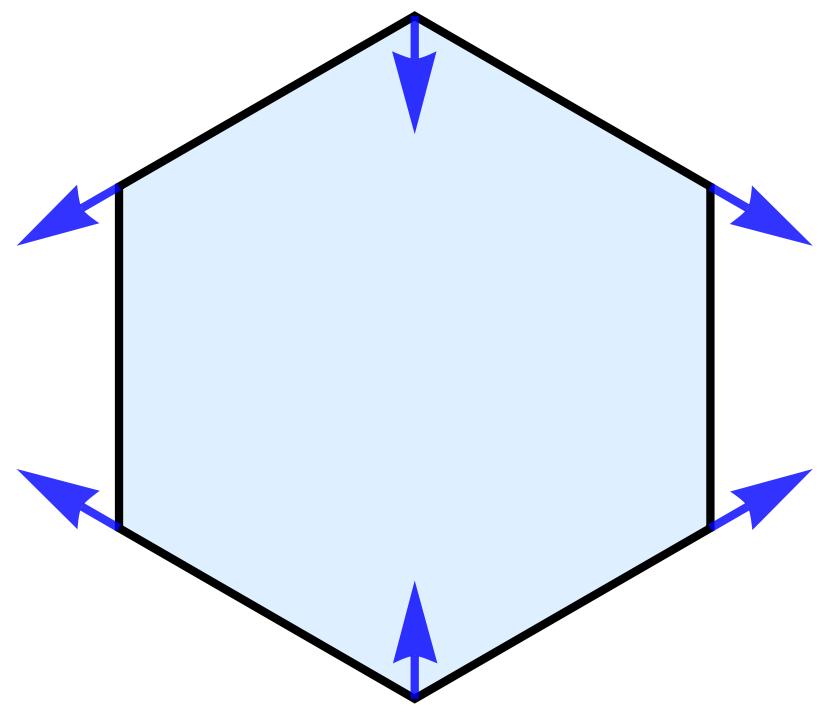}
			}&
			\addheight{\includegraphics[width=2.5cm,valign=t]{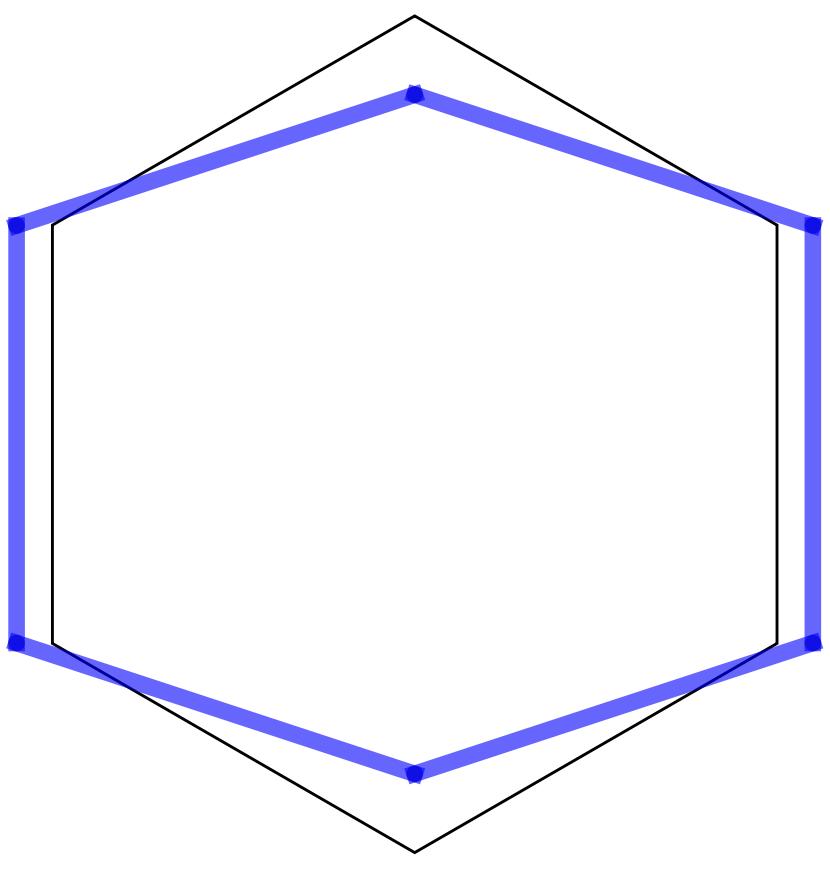}
			} \\
        (a) &  			
        (b) &
		(c) \\
        \end{tabular}            
        \end{center}
			\caption{{\it A basic model for evaluating local deformations under anisotropic stress.} Let $\mathbf{v}_f$ denote the fixed vertices at the end of the hexagonal cell array, marked by black colored points. Angle $\theta$ indicates the top/bottom angles that can convert hexagonal cells into rectangular shapes. The hexagon is changing its angle due to a uniform anisotropic stress~\cite{Lin2022}. Blue arrows in (b) represent the forces applied at each vertex. The blue hexagon in (c) illustrates the change in shape in response to the applied stress.}
			\label{fig:single-model}
	\end{figure*}

 Imagine a line of cells under "anisotropic" stress, meaning that each cell experiences pressure from different directions with different intensities. This non-uniform force field may cause "stress-induced sub-cellular buckling", depending on the boundary conditions. As depicted in Fig. \ref{fig:single-model} (b)-(c), each cell in the line experiences a unique stress pattern that mimics local anisotropic forces but does not perfectly mimic stretching \cite{Sato2021}.
As shown in Fig. \ref{fig:single-model}, we have developed a simple energy function that adjusts the top and bottom angles of a hexagon to induce the transformation from a hexagon into a rectangle, and vice versa. This transformation includes a change in the length of corresponding edges (See Fig. \ref{fig:single-model} (c)).
	\begin{align}
		\mathcal{E}_\theta=&\frac{K_T}{2}\sum_{\alpha}(\theta_{\alpha}-\theta_{t})^2.
	\end{align}
 with $\theta$ as denoted in Fig. \ref{fig:single-model}(a). To drive the cell shape change from hexagon to rectangle, we minimized the top and bottom angles of each hexagon, as opposed to all angles, as our chosen target angle ($\theta_t=\pi$) induces changes in all angles of the polygons given the other energetic contributions. To study the stress-induced deformations, we implemented independent optimizations for the straight-edged (SE) and curved-edged (CE) models. 

\begin{figure*}
	\captionsetup{singlelinecheck = false, justification=raggedright}
	\begin{center}
		\begin{tabular}{c c}
			%		\hline
			\addheight{	\includegraphics[width=5cm]{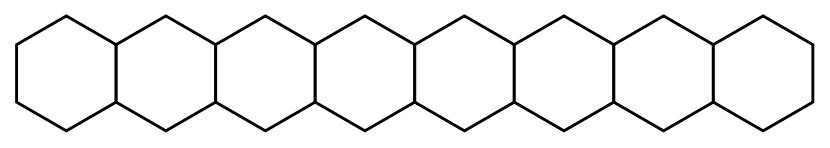}
			}&
			\addheight{\includegraphics[width=5cm]{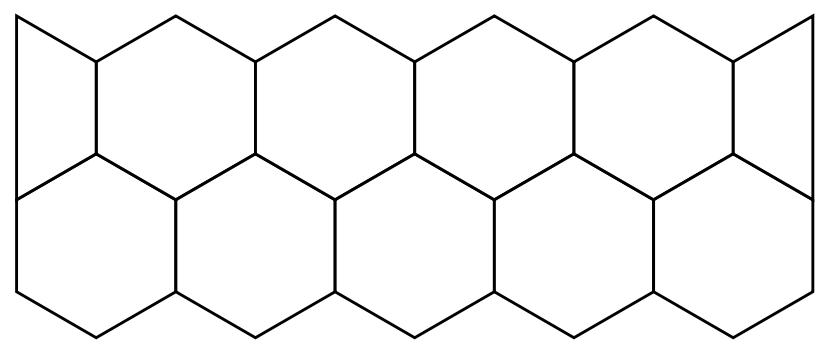}
			} \\
			(a) Single row cell array (initial state).&
			(b) Double row cell array (initial state).\\
			\addheight{	\includegraphics[width=5cm]{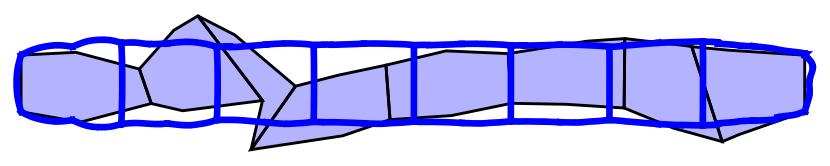}
			}&
			\addheight{\includegraphics[width=5cm]{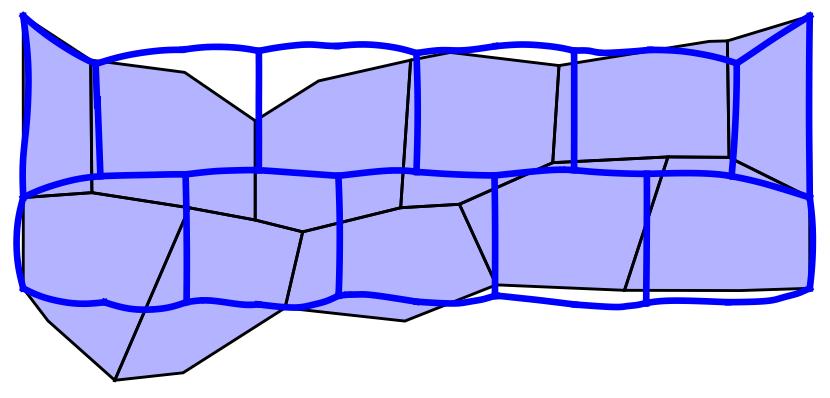}
			} \\
			(c) Single row cell array (final state).&
			(d) Double row cell array (final state).\\
		\end{tabular}
		\captionsetup{type=figure}
		\captionof{figure}{{\it Sub-cellular buckling simulation of single and double array cells for a fixed boundary.} (a)-(b) Initial cell shape, (c)-(d) blue lines represent cells with curved edges and light blue shaded cells represent cells with straight edges.}
		\label{fig:double-bending}
	\end{center}	
\end{figure*}

The energy function for both models is given by $E=\mathcal{E}_\theta+\mathcal{A}+\mathcal{P}+\mathcal{R}$, parametrized by $\theta_t$, $A_0$, $P_0$, $x_t$, and $y_t$, directed both straight-edged (SE) and curved-edged simulations (CE). While $A_0$, $x_t$, and $y_t$ corresponds to those of a regular hexagon, $P_0$ took non-uniform values between $6$ and $6.24$ with shape index to be $3.79549$ in Fig. \ref{fig:double-bending}(a) and $3.75973$ in Fig. \ref{fig:double-bending}(b) on average, except for two half-hexagonal boundary cells in Fig. \ref{fig:double-bending}(b). This range of $P_0$ is used to help induce global buckling. Also note that we have fixed boundary vertices and boundary cell walls. Control parameters are identical for both models: $K_T=2$, $K_A=1$, $K_P=3$ (or $K_A'=1$, $K_P'=3$), and $K_{R1}=1$. The resistance vertex function is implemented to prevent vertex flipping. In the CE model, we used additional parameters, $K_{R2}=0.01$ and $c_t=0.05$.		

Examining Fig.~\ref{fig:double-bending}(c) and~\ref{fig:double-bending}(d), we observe a notable difference in chain shape. The curved model shows a less deformed chain shape to reach the target angle of $\theta_t = \pi$ globally as compared to the SE model. This configuration is due to the increased degrees of freedom provided by the curved edges. These added degrees of freedom allow the cells to energetically conform to the desired $\theta_t = \pi$ with smaller deformations at larger length scales over multiple cells such that sub-cellular buckling emerges as oppose to multi-cellular buckling. 

\begin{table*}
\caption{Average shape index of individual cells (from the Equation $p_\alpha=\frac{P_\alpha}{\sqrt{A_\alpha}}$).}
\centering
\begin{tabular}{c c c c}
\hline\hline
&$p_0$& $p_\alpha$ of SE models & $p_\alpha$ of CE models \\ [0.5ex]
\hline
Single row cell array (Fig.~\ref{fig:double-bending}c) & 3.79549 &4.35291 & 3.81628 \\
Double row cell array (Fig.~\ref{fig:double-bending}d) & 3.75973&4.11177 & 3.92456 \\
\hline
\end{tabular}
\label{table:7}
\end{table*}

Beyond the impact of curved edges, confinements plays a crucial role in shaping cell morphology. The availability of space at the top and bottom can introduce distinct constraints, even within the same model. For example, comparing Figures~\ref{fig:double-bending}(c) and~\ref{fig:double-bending}(d), we see that the restructuring of the cells occurs in double arrays to satisfy $\theta_t = \pi$, going beyond top/bottom flattening. This suggests that cells in confined spaces adapt their arrangements to meet target conditions, highlighting the interplay between shape flexibility and confinement. Furthermore, the curved-edged model is expected to have lower energy costs associated with shape changes due to its proximity to the target shape index of $3.79549$ (Fig.~\ref{fig:double-bending}(a)) and $3.75973$ (Fig.~\ref{fig:double-bending}(b)). We note that due to the two boundary tetragonal cells in Fig.~\ref{fig:double-bending}(b), there is a slight increase in the shape index compared to the hexagonal one for Fig.~\ref{fig:double-bending}(d). Without these two cells, we obtain $p_\alpha=4.04877$ and $3.81244$ for the straight-edged and curved-edged models in Fig.~\ref{fig:double-bending}(d).
 
\subsection{Compression-induced curved edges}\label{stress-def}
We now focus on compression of a group of cells in both \textit{disordered} and \textit{ordered} scenarios surrounded by boundary cells. For the ordered case, we modified the target parameters in a way that mimics the swelling of boundary cells \cite{ben2016onset}. This mechanism uses vertex position data from the previous minimized state to evaluate and determine the next minimized state. Simulations details are provided in the Appendix. For both the ordered and disordered cases, we utilize a simultaneous optimization approach. We define constraints governing the relative increase or decrease in area and perimeter compared to the initial state (Fig. 1C, refer also to Fig. \ref{fig:disordered-cell}). Simulations were conducted using the Mathematica software \cite{Mathematica}, specifically employing the NMinimize function for optimization. 

	\begin{figure*}
		\captionsetup{singlelinecheck = false, justification=raggedright}
		\begin{center}
			\begin{tabular}{c c}
				%		\hline
				\addheight{\includegraphics[height=3cm]{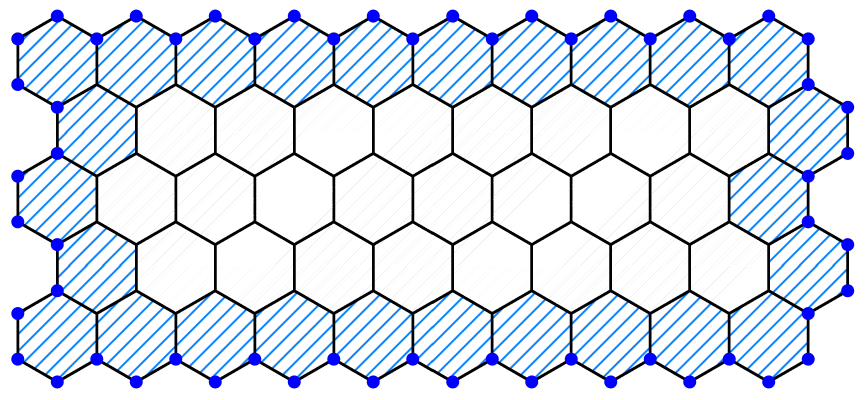}}
				&\addheight{	\includegraphics[height=4cm]{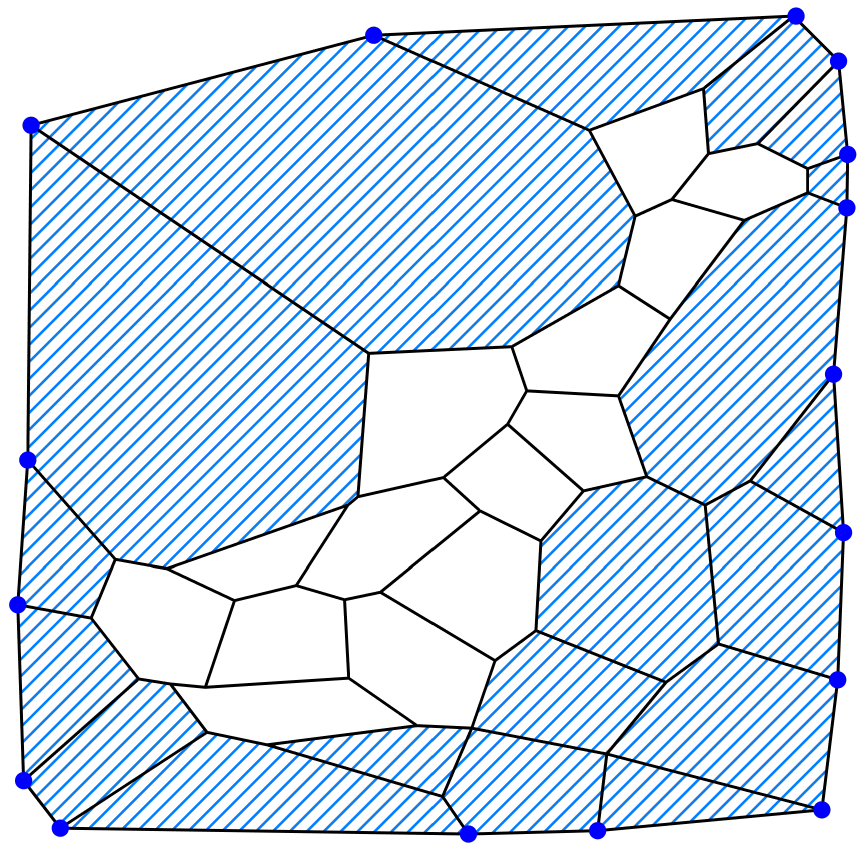}
				} \\
				(a) Ordered cell array &
                (b) Disordered cell array\\
			\end{tabular}
			\captionsetup{type=figure}
			\captionof{figure}{{\it Ordered/disordered cell array used in the simulations.} Interior cells are painted white, while boundary cells are represented by blue diagonal stripes. For the ordered (a) scenario, we conduct simulations on 24 interior cell arrays to investigate sub-cellular buckling caused by compression by the boundary cells. In the case of disordered cell array in (b), we utilized a graph inspired by experiments reported in Ref.~\cite{rigato2022} (See Fig. \ref{fig:disordered-cell}).}
			\label{fig:vm-ref}
		\end{center}	
	\end{figure*}

\subsubsection{Ordered Case}
To study the ordered system under compression by a group of surrounding cells that are swelling, we 
\begin{enumerate}
    \item Increase the target boundary cell area by 5\%.
    \item Decrease the target boundary cell perimeter by 5\%.
\end{enumerate}
These percentage changes are similar to the disordered case, which is directly relevant to experiments~\cite{rigato2022}. States 1-5 refer to this increase in area/decrease in perimeter for each consecutive step. The ordered simulation evaluates two different sub-cellular buckling mechanisms in terms of deformation type A and B. In deformation type A, again, we imagine a cell contracting while maintaining its edge length, resulting in the curved edge length remaining the same as the initial edge length, but the straight edge length becoming smaller. On the other hand, deformation type B involves stretching the side length while maintaining the distance from vertex to vertex, so the straight edge length remains the same while the curved edge length increases. 
\begin{enumerate}
    \item Deformation type A (Fig. \ref{fig:vertex-model-type}(b))
      \begin{itemize}
            \item Cells maintain their initial edge length while contracting their area.
            \item Shape index remains relatively stable, indicating that the SE behavior is energetically favorable.
       \end{itemize}
       \item Deformation type A+B  (Fig. \ref{fig:vertex-model-type}(b)+(c))
      \begin{itemize}
        \item Side length increases to approximately 1.4 during cell contraction, mirroring the disordered case.
        \item Significant rise in shape index, suggesting a transition towards CE behavior.
        \end{itemize}
\end{enumerate}
Note the combination of deformation type A+B is not equivalent to deformation type C, the latter of which does not conserve the number of vertices. 

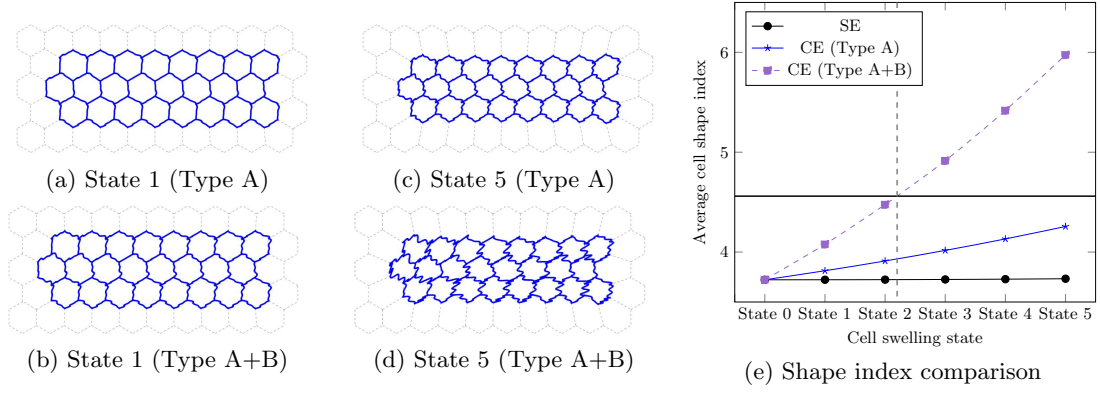
\begin{figure*}
		\captionsetup{singlelinecheck = false, justification=raggedright}
		\begin{center}
  \begin{tabular}{c c c}
			\begin{tabular}{c}
				\addheight{	\includegraphics[width=4cm]{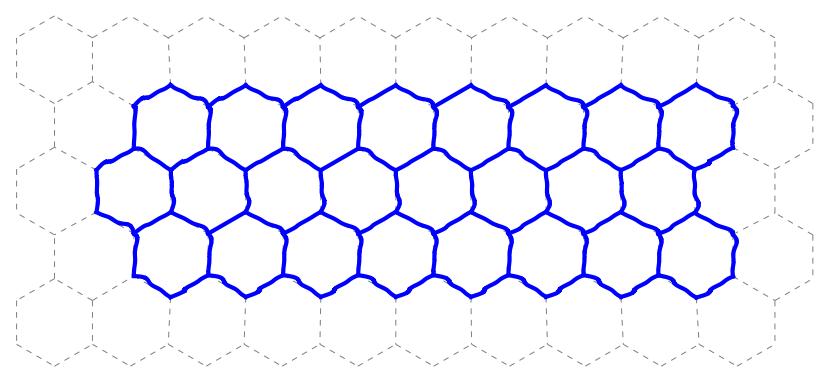}
				}\\
				(a) State 1 (Type A)\\
				\addheight{\includegraphics[width=4cm]{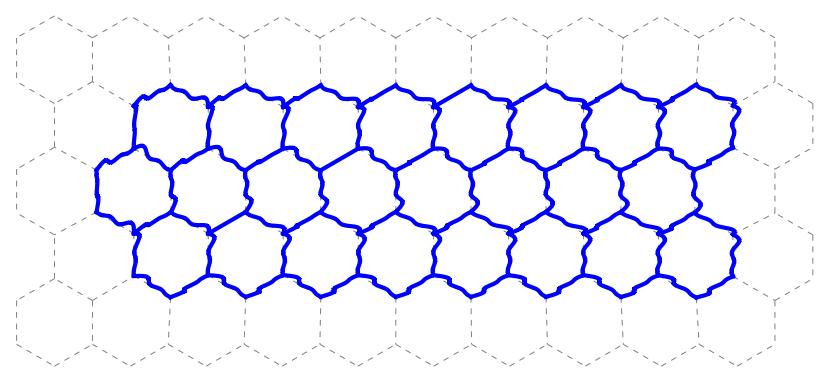}
				} \\
				(b) State 1 (Type A+B)\\
			\end{tabular}&
   \begin{tabular}{c}
				\addheight{	\includegraphics[width=4cm]{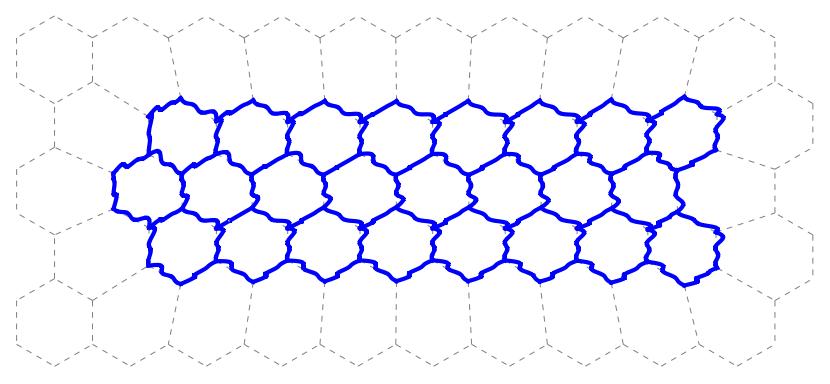}
				}\\
				(c) State 5 (Type A)\\
				\addheight{\includegraphics[width=4cm]{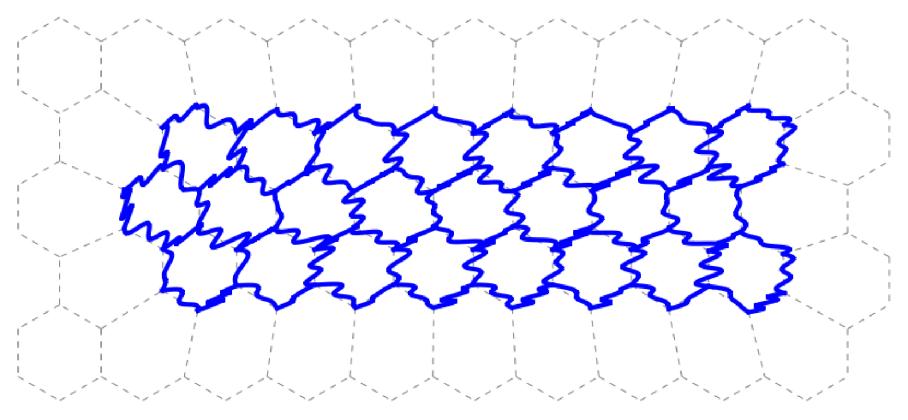}
				} \\
				(d) State 5 (Type A+B)\\
   \end{tabular}&
   \begin{tabular}{c}
   			\begin{tikzpicture}[thick,scale=0.7, every node/.style={transform shape}]
				\begin{axis}[
					xlabel=Cell swelling state,
					ylabel=Average cell shape index,
					legend pos=north west,
					xmin=-0.5, xmax=5.5,
					ymin=3.5, ymax=6.5,
					xtick={-0.5,0,1,2,3,4,5},
					xticklabels={,State 0,State 1,State 2,State 3,State 4,State 5,},  
					]
					\addplot[smooth,mark=*,black] plot coordinates {
						(0, 3.72242)
                        (1,3.722724167)
                        (2,3.723738333)
                        (3,3.725619583)
                        (4,3.728576667)
                        (5,3.73287625)
                    };
					\addlegendentry{SE}
					\addplot[smooth,color=blue,mark=star]
					plot coordinates {
						(0, 3.72242)
                        (1,3.81293)
                        (2,3.910881667)
                        (3,4.0166775)
                        (4,4.131634583)
                        (5,4.256133333)
                    };
                    \addlegendentry{CE (Type A)}
					\addplot[smooth,color=amethyst,mark={square*}, dashed]
					plot coordinates {
						(0, 3.72242)
                        (1,4.077920833)
                        (2,4.475821667)
                        (3,4.91530875)
                        (4,5.418001667)
                        (5,5.974689583)
                        };
                    \addlegendentry{CE (Type A+B)}
					\addplot[mark=none, black,thick] coordinates {(-0.5,4.56) (5.5,4.56)};
					\addplot[mark=none, gray,dashed,thick] coordinates {(2.2,0) (2.2,6.5)};
				\end{axis}
			\end{tikzpicture}
            \\
            (e) Shape index comparison\\
            \end{tabular}
  \end{tabular}
			\captionsetup{type=figure}
			\captionof{figure}{{\it Simulation results and shape index comparison for ordered collection of cells.} In (a), (b), and (c) the cell edges show wavy shapes and the cell shape indices are less than $p_\alpha\sim4.56$, still not significant compared to (d) where the cell shape index is greater than $p_\alpha\sim4.56$. (e) The black horizontal line represents shape index of the regular triangle. A transition point $p_\alpha\sim4.56$ is represented by dashed gray vertical line.}
			\label{fig:ordered-cell-CE}
		\end{center}	
	\end{figure*}

Our simulations show that both area contraction and perimeter expansion are necessary for curved edge morphologies. The transition from SE to CE behavior, as observed in deformation type A+B, begins at state 2 when the SE shape index limit ($p_\alpha\sim4.56$ for regular triangles) is used as the transition point, which is represented in Fig. \ref{fig:ordered-cell-CE} (e).
	\subsubsection{Disordered Cell Simulation}
\begin{figure*}
		\captionsetup{singlelinecheck = false, justification=raggedright}
		\begin{center}
			\begin{tabular}{c c c}
				\addheight{\includegraphics[height=3.5cm]{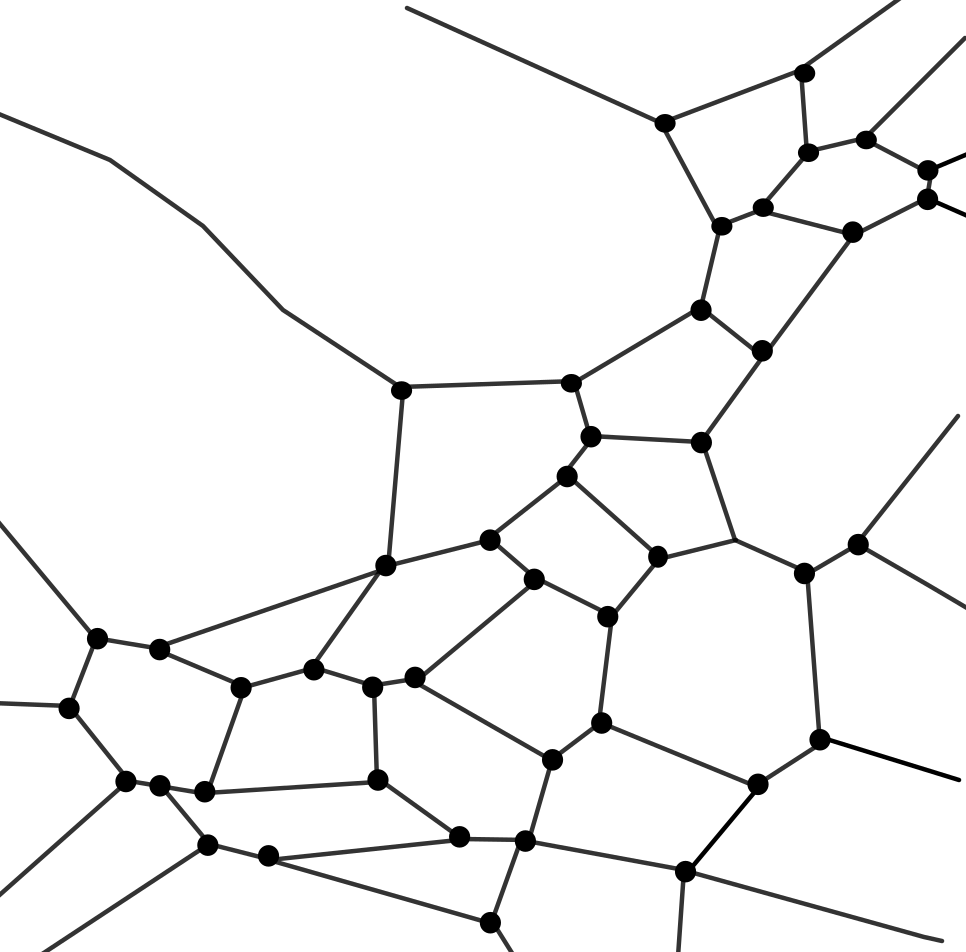}
				}&\addheight{	\includegraphics[height=3.5cm]{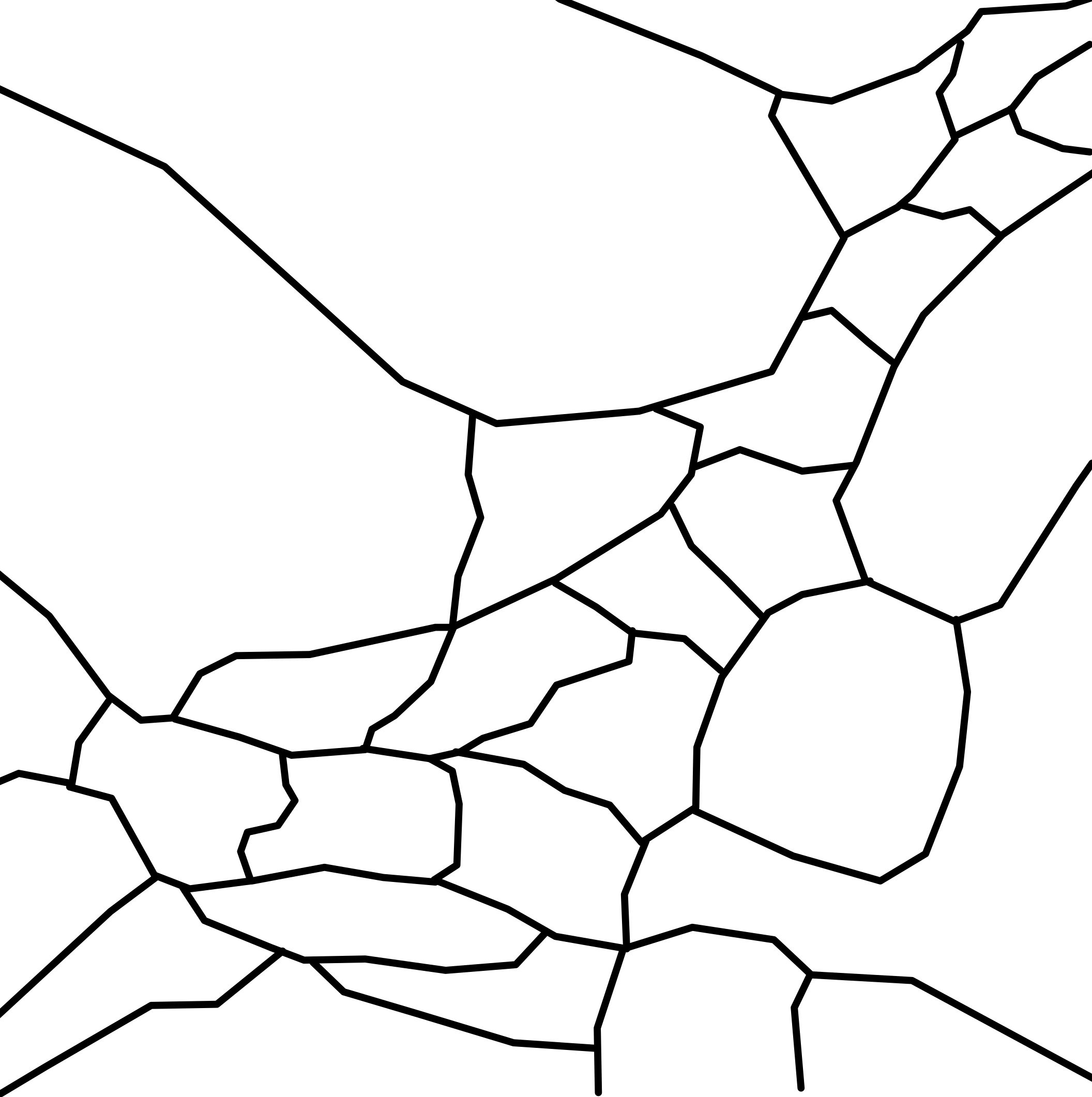}
				}&\addheight{\includegraphics[height=3.5cm]{Fig1_A_drawing02W.png}}\\
				(a) Vertex model for (b)&
				(b) Drawings of Fig. 1A in the experiments&
				(c) Drawings of Fig. 1D in the experiments\\
				\addheight{	\includegraphics[height=3.3cm]{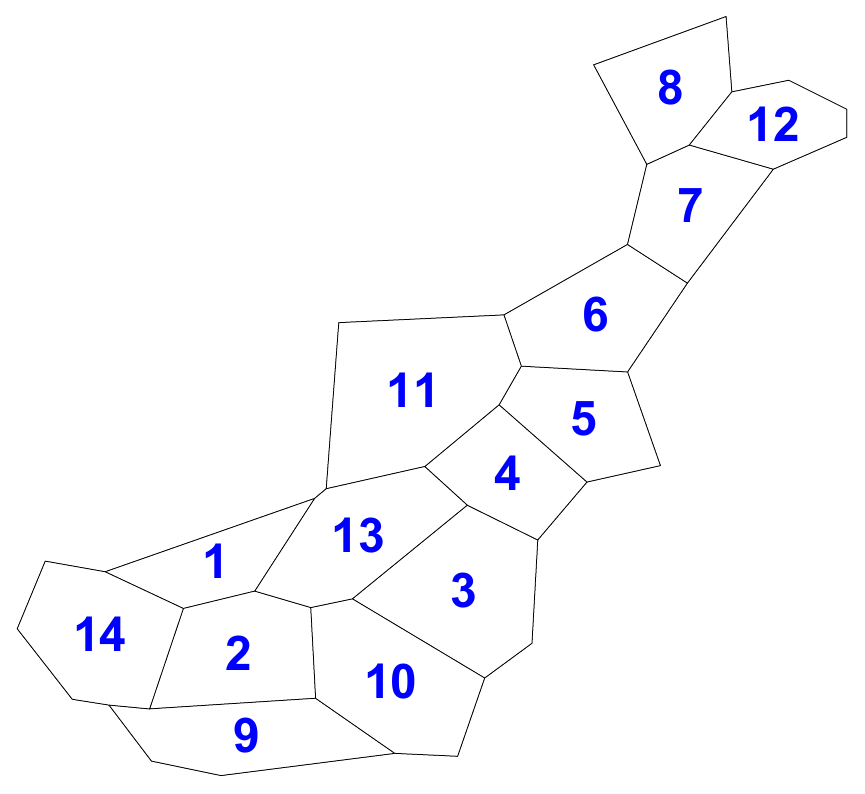}
				}&\addheight{	\includegraphics[height=4cm]{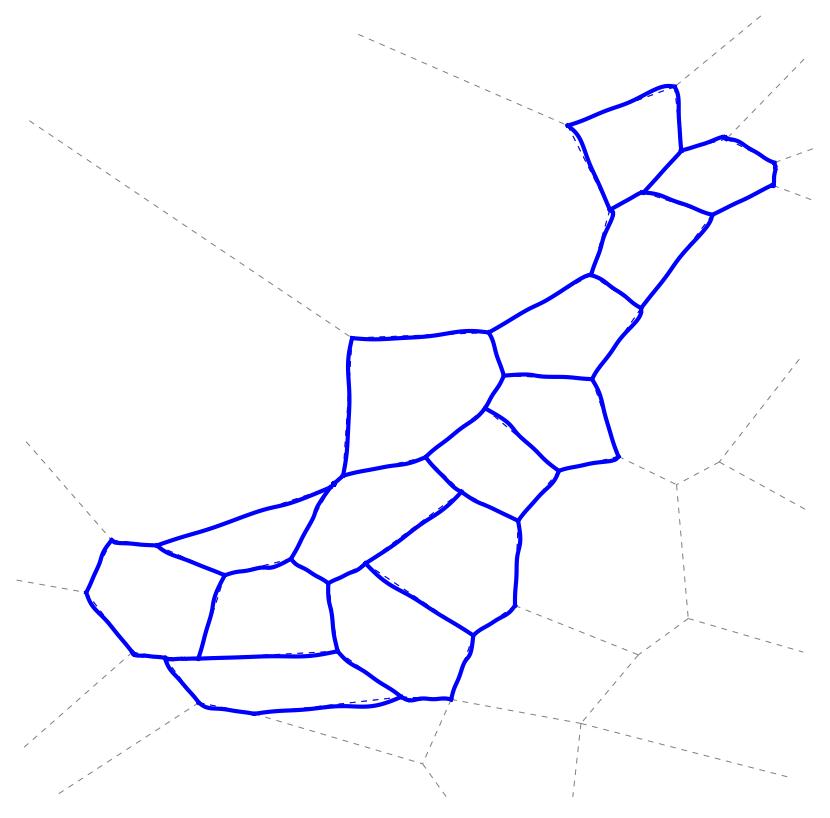}
				}&\addheight{\includegraphics[height=4cm]{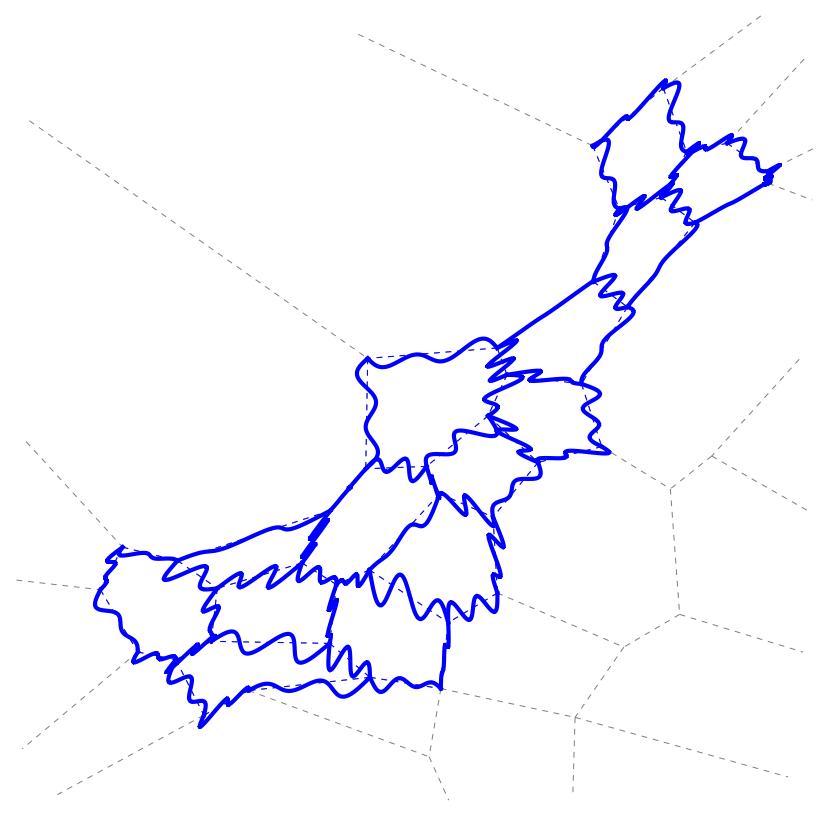}
				}\\
				(d) Cell ID &	(e) State 0 (corresponded to (b)) &	(f) State 3 (corresponded to (c))\\
			\end{tabular}
			\captionsetup{type=figure}
			\captionof{figure}{(a)-(c) Extracted from the experimental data published in bioarchiv. of Rigato {\it et al}. (d)-(f) Disordered cell simulation results. (a) straight-edged cell vertex model for (b). (b)-(c) Drawings from the Fig. 1A. (d) Cell id of each polygons. (e),(f) Simulation results of state 0,3 using conditions in Table. \ref{table:2}.}
			\label{fig:disordered-cell}
		\end{center}	
	\end{figure*}

As mentioned previously, in development, larval epithelial cells and histoblasts compete for limited space on the growing larval body surface (even in the absence of cell division)~\cite{rigato2022}. Histoblasts undergo a remarkable morphological transformation from convex polygons into cells with curved edges, as shown in Fig. \ref{fig:disordered-cell}(a)-(c), as surrounding larval epithelial cells swell to compress the histoblasts. Our simulation focuses on the two-dimensional aspects of this process, where readily available information on cell area and perimeter allows one to analyze the shape changes observed in the experiments. Given to the lack of specific data for individual boundary and histoblasts, we selected parameters by comparing figures from Ref.~\cite{rigato2022} and opted for a smaller value for the scaling factor ($s$) than the reported average given sample-to-sample fluctuations. Table \ref{table:2} in the Appendix details the chosen conditions and parameter values.

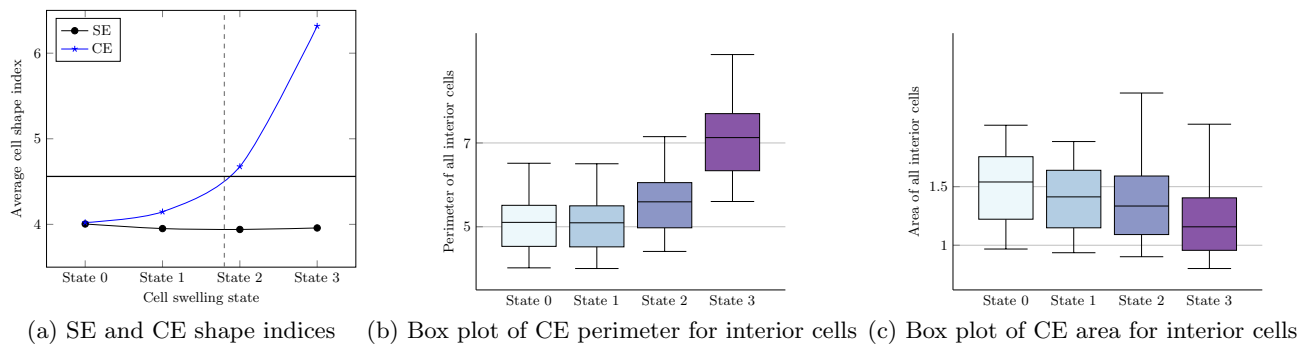
\begin{figure*}
	\captionsetup{singlelinecheck = false, justification=raggedright}
		\centering
        \begin{tabular}{c c c}
			\begin{tikzpicture}[thick,scale=0.6, every node/.style={transform shape}]
				\begin{axis}[
					xlabel=Cell swelling state,
					ylabel=Average cell shape index,
					legend pos=north west,
					xmin=-0.5, xmax=3.5,
					ymin=3.5, ymax=6.5,
					xtick={-0.5,0,1,2,3},
					xticklabels={,State 0,State 1,State 2,State 3,},   % <---
					]
					\addplot[smooth,mark=*,black] plot coordinates {
                        (0,4.004329167)
                        (1,3.9507325)
                        (2,3.9403475)
                        (3,3.957995833)
    				};
					\addlegendentry{SE}
					\addplot[smooth,color=blue,mark=star]
					plot coordinates {
                        (0,4.021421667)
                        (1,4.147305833)
                        (2,4.6755275)
                        (3,6.316833333)
                    };
					\addlegendentry{CE}
					\addplot[mark=none, gray,thick,dashed] coordinates {(1.8,3.5) (1.8,6.5)};
					\addplot[mark=none, black,thick] coordinates {(-0.5,4.56) (3.5,4.56)};
				\end{axis}
			\end{tikzpicture}
            &
				\begin{tikzpicture}[thick,scale=0.6, every node/.style={transform shape}]
					\pgfplotstableread[col sep=comma]{disordered_scperi.csv}\csvdata
					\pgfplotstabletranspose\datatransposed{\csvdata} 
					\begin{axis}[
						boxplot/draw direction = y,
						axis x line* = bottom,
						axis y line* = left,
						enlarge y limits,
						ymajorgrids,
						xtick = {1, 2, 3, 4},
						xticklabel style = {align=center, font=\small, rotate=0},
						xticklabels = {State 0, State 1, State 2, State 3},
						xtick style = {draw=none},
						ylabel = {Perimeter of all interior cells},
						ytick = {5,7}
						]
						\foreach \n in {1,...,4} {
							\addplot+[boxplot,draw=black,fill] table[y index=\n] {\datatransposed};
						}
					\end{axis}
				\end{tikzpicture}
            &
				\begin{tikzpicture}[thick,scale=0.6, every node/.style={transform shape}]
					\pgfplotstableread[col sep=comma]{disordered_scarea.csv}\csvdata
					\pgfplotstabletranspose\datatransposed{\csvdata} 
					\begin{axis}[
						boxplot/draw direction = y,
						axis x line* = bottom,
						axis y line* = left,
						enlarge y limits,
						ymajorgrids,
						xtick = {1, 2, 3, 4},
						xticklabel style = {align=center, font=\small, rotate=0},
						xticklabels = {State 0, State 1, State 2, State 3},
						xtick style = {draw=none}, 
						ylabel = {Area of all interior cells},
						ytick = {1,1.5}
						]
						\foreach \n in {1,...,4} {
							\addplot+[boxplot,draw=black,fill] table[y index=\n] {\datatransposed};
						}
					\end{axis}
				\end{tikzpicture}
        \\
        (a) SE and CE shape indices &
   (b) Box plot of CE perimeter for interior cells&
   (c) Box plot of CE area for interior cells\\
        \end{tabular}
   \caption{{\it Perimeter and area of all interior CE cells and comparison of SE and CE shape indices for disordered cells} (*excluding cell ID 1 and 9). The solid black horizontal line represents a shape index of $4.56$, which is the shape index of a regular triangle. A dashed gray vertical line represents the transition point from SE to CE dominated cells.}
		\label{fig:disordered-cell-SI}
	\end{figure*}

Fig. \ref{fig:disordered-cell} depicts our findings. Panel (a) shows the straight-edged cell vertex model, which serves as the starting point for our analysis. Panels (b) and (c) feature drawings adapted from Fig. 1A/1D in the experiments, showcasing the configuration of cells. Panel (d) displays the assigned cell IDs for each polygon. Finally, panels (e) and (f) present the simulation results for states 0 and 3, respectively, achieved using the conditions specified in Table \ref{table:2}. Each state denotes a further increment in time as the boundary cells swell. With our parameterization choice given in Eqns. \ref{eq08}-\ref{eq09}, we observe anisotropically-curved edges in Fig.\ref{fig:disordered-cell} (f), which are similar to those depicted in Fig.\ref{fig:disordered-cell} (c).
Fig. \ref{fig:disordered-cell-SI} (a) shows the average cell shape indices for the disordered system for each boundary cell swelling state. We exclude cell IDs as 1 and 9 due to their significantly higher shape index ($>5$) as compared to other cells ($\sim4$).
To compare our results with those of the experiments, we collect data for both area and perimeter of curved-edged (CE) cells. Fig. \ref{fig:disordered-cell-SI} aligns well with the general trends observed in Figures 1F and 1G of the experimental average values.

Our analysis also sheds light on the transition point from straight-edged (SE) to curved-edged (CE) behavior. For regular polygons, the triangle shape index is the highest. As the shape index increases, the number of edges decreases while the number of edges per vertex increases, aligning with the findings of Lin {\it et al.} \cite{Lin2023}. We can expect a transition to a curved-edge dominated regime when the shape index exceeds a critical value, approximately $p_\alpha\sim 4.56$ for a straight-edged cell limit. At this point, cell edge buckling, similar to those observed in state 2 of Fig. \ref{fig:ordered-cell-CE} (e), becomes likely. Alternative criteria can be used for conditions where the initial average cell shape index for SE cells exceeds 4.56. For example, the region dominated by wavy cell edges can be identified by calculating the difference in the average shape index. These conditions apply to both area decreasing and perimeter increasing. 

To facilitate further comparison with the results of the experiments, we converted the dimensionless shape index $p_\alpha$ to circularity $C$ using the following equation:
\begin{align*}
C &= \frac{4\pi a}{p^2}= \frac{4\pi}{{p_\alpha}^2}
\end{align*}
For state 0, the circularity is $C = 0.77472$, and for state 3, it becomes $C = 0.31290$ (again, excluding cell IDs 1 and 9). While similar trends to the experiment are observed, we do not yet have quantitative agreement with experiments. Since we do not yet have information distinguishing changes between the boundary cells and the histoblasts, as evidenced by the differences between Fig. \ref{fig:disordered-cell}(c) and (f), it is clear that the boundary cells are not identical (especially top/bottom and bottom left). We will incorporate such details in future work to ultimately yield a quantitative comparison with experiments. 

	\section{Discussion}
We have introduced a curved-edge vertex model for curved cell-cell interfaces in tissues at the subcellular scale using a parametric function to quantity the edge between two vertices. We can now explore non-convex cell shapes in a tiling without adding more vertices explicitly. From a cell shape index perspective, there are now two ways to increase the cell shape index, either by remaining convex and becoming more elongated or by remaining globular and becoming more non-convex. Which shape change pathway a cell takes depends on a multitude of factors, including the morphology of the underlying cytoskeleton.  For instance, in cell monolayer experiments discovering that an upregulation of RAB5A induced an unjamming transition, the cells morphed from non-convex to convex shapes~\cite{Malinverno2017}. As the cells become convex, they can then perhaps undergo T1 transitions at very little energetic cost and so the system fluidizes, provided the cell shape index is high enough~\cite{Bi2015}. In other words, perhaps cell nonconvexity helps put a brake on cell fluidization. Therefore, in addition to the target cell shape index parameter, topology of the tiling~\cite{Yan2019}, and applied shear~\cite{Huang2022} as drivers of a rigidity transition, we perhaps should add yet another axis of the fraction of convex cells to the rigidity transition phase diagram.  Interestingly, recent work has demonstrated that when a polygon, consisting of two-body springs and an area spring constraint, is subjected to expansive strain, the convexity of the polygon is a necessary condition to generate a state of self-stress, or rigidity, while a cyclic configuration of the polygon is a sufficient condition for the self stress~\cite{Gandikota2022}. 

While cellular nonconvexity may place a brake on cell fluidization, how then can cells adjust in the presence of applied compression? Our results indicate that as cells become increasingly nonconvex, they can respond to applied compression without giving up their area by becoming more like jigsaw puzzle pieces. This phenomenon is presumably widely observed in plant cells because plant cells, which contain cell walls, are not motile. As depicted in Fig. \ref{fig:1} (left), Bidhendi, {\it et al.} \cite{Bidhendi563403} established that the cell wall shaping process relies on spatially confined, feedback-augmented stiffening of the cell wall in the periclinal walls. Moreover, Belteton, {\it et al.} \cite{Belteton2021} demonstrated that tensile forces and the microtubule–cellulose synthase systems dictate the patterns of interdigitated growth. Finally, experiments in drosophila~\cite{rigato2022}, demonstrate similar behavior in which a group of cells eventually develops significant nonconvexity to accommodate the compression by surrounding swelling cells. It would be interesting to search for other examples of this convex to nonconvex transition in other instances in animal tissues. 

\begin{figure*}
		\captionsetup{singlelinecheck = false, justification=raggedright}
		\begin{center}
  \begin{tabular}{C{7.5cm}C{7.5cm}}
   			\addheight{\includegraphics[height=3.5cm]{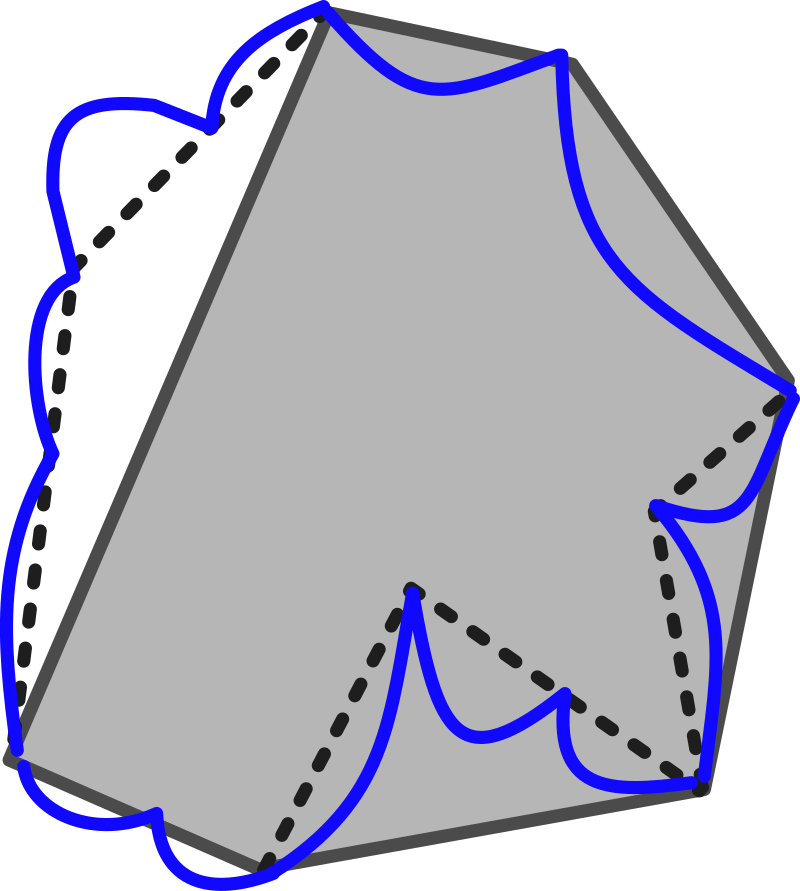}
      }&
    		\addheight{\includegraphics[height=3.5cm]{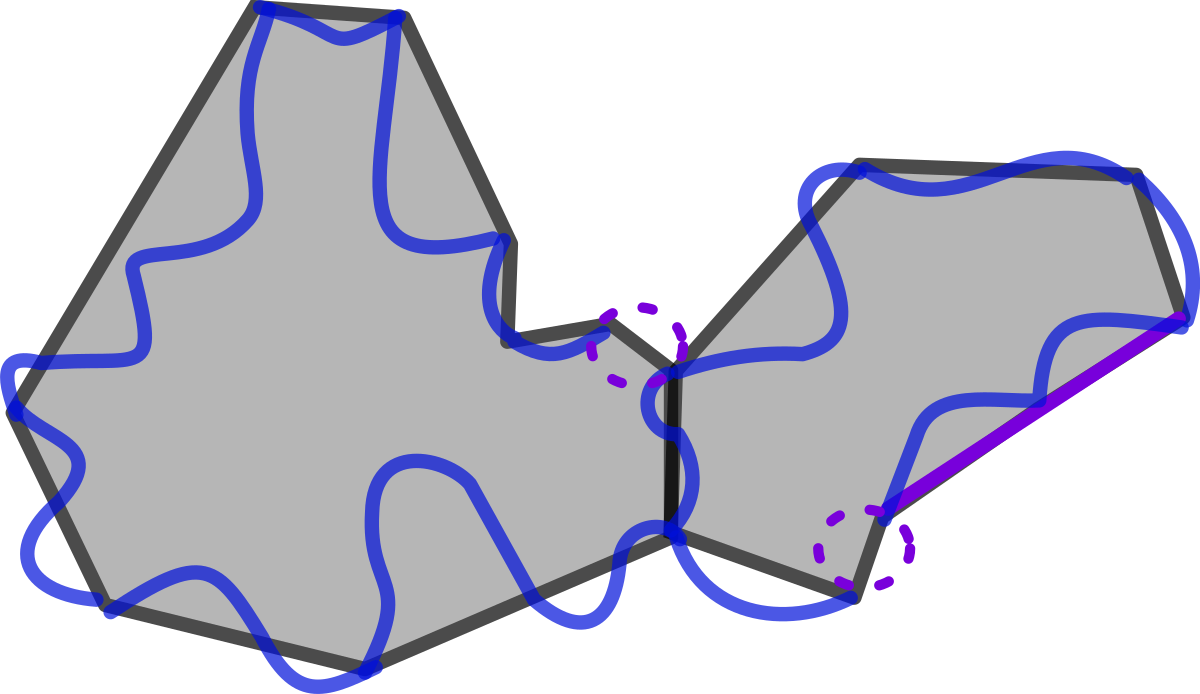}
      }\\
                (a) Cell cross-sectional shape reconstruction with the help of hidden edges. &  (b) Cell cross-sectional shape for different functions of the curved edges. \\
				\addheight{	\includegraphics[height=5cm]{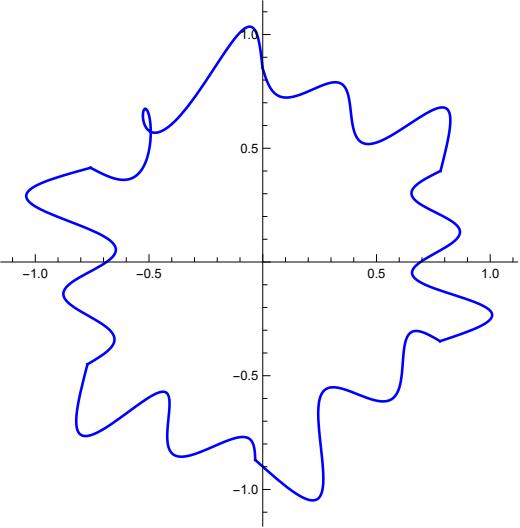}
				}&
				\addheight{	\includegraphics[height=5cm]{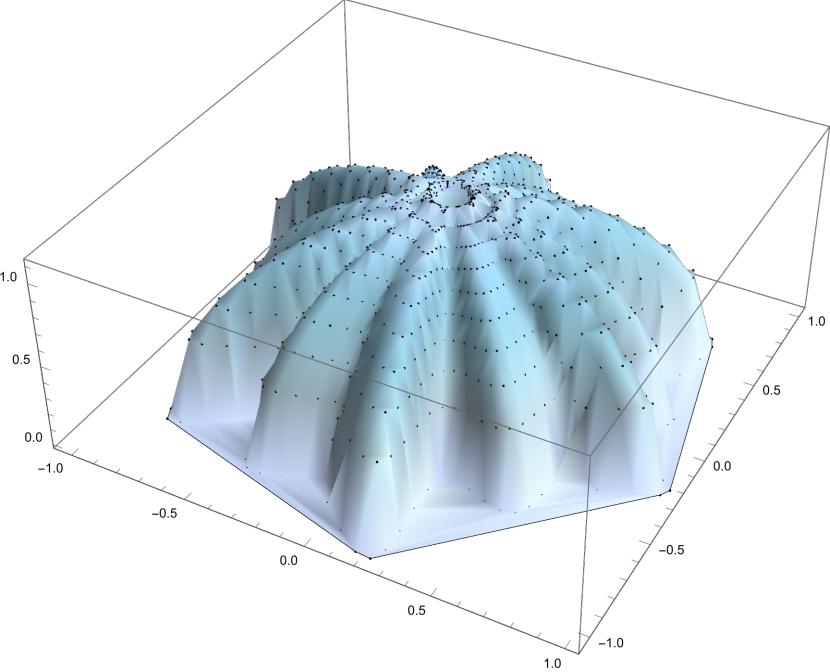}
				} \\
				(c) Planar view &
				(d) Three-dimensional representation using (c) as a boundary\\

  \end{tabular}
			\caption{{\it Possible extensions of the curved-edge model.} (a) Cell reconstruction with the help of hidden edges. Black solid lines filled with gray shade represent SE components and dashed lines depict hidden edges. (b) Cell reconstruction for different functions of the curved edges. Two dashed purple circles represent edges with zero curvature, and two purple lines show two straight edges forming a line similar to deformation type C in Fig. \ref{fig:vertex-model-type}. (c) A single-cell example with a singular point, (d) the three-dimensional representation using the bounded spherical balloon equation in \cite{kimthesis}. The singular boundary point is ignored in the calculation.}
			\label{fig:extended-model-examples}
		\end{center}	
	\end{figure*}

We envision future extensions to the model we have introduced here that remain distinct from other approaches to capture edge cell curvature \cite{boromand2018jamming,vetter2023polyhoop}. For example, if concave polygons are to be avoided for technical purposes, the CE model can be extended by adding more vertex points, and different curve functions can be selected for the CE model to improve the representation of cellular shape profiles. Simply, invisible vertex points can be added on each edge. Fig. \ref{fig:extended-model-examples} (a) has hidden edges (for invisible vertex points) to simulate cells with lobes. Representing lobes in the simulations may involve folding certain edges. By implementing invisible vertex points, cell profiles can be accurately represented and convex polygons can be maintained for the SE model (gray shaded polygon in Fig. \ref{fig:extended-model-examples} (a)). The possibility of setting angles between hidden edges to zero enables the representation of a wider range of shapes. Fig. \ref{fig:extended-model-examples} (a) shows an example with hidden edges inspired by Fig. 2 a in \cite{Malinverno2017}, Fig. \ref{fig:extended-model-examples} (b) is the demonstration of the complicated cell structure inspired by Fig. \ref{fig:1} left and Fig. 1 in \cite{koab250}. 
 
    Because we are using parameterized curves at two fixed ends that correspond to the edges of the straight polygon, we may get unintended results during the optimization. For example, when implementing sinusoids as the parametric function, there can be overlaps resulting from adjacent edges, as shown in Fig. \ref{fig:extended-model-examples} (c). These points of overlap can be a source of error in the evaluation of the correct perimeter and area. In addition, if we look in three dimensions, these points will generate several boundary points, therefore some of them will have to be recalculated or omitted, as shown in the Fig. \ref{fig:extended-model-examples} (d). To achieve more accurate cell profiles, one may consider using different coefficients for sinusoidal harmonics, such as $(c_i)^2$ instead of $c_i$, or limiting $c_i$ to be positive or small enough, will presumably avoid such singular points. Another approach to avoid singular points is by imposing constraints on the slope of the two endpoints. Moreover, instead of sine functions, other trigonometric functions, such as elliptic functions, hypergeometric functions, or B\'ezier type functions \cite{brander2018bezier} can be chosen to generate different edge profiles. 

Finally, our analysis has so far been restricted to two dimensions. However, we can model the three-dimensional shape of a nonconvex  cell using two-dimensional representations of the cell's boundary. Re-optimization with constraints can then be performed in two-dimensional or three-dimensional computations, as shown in Fig. \ref{fig:extended-model-examples} (d) serving as initial parameters. By conceptualizing the shape of a cell as two-dimensional surface sheets, we can perform minimum energy calculations~\cite{Mimura2023,Murisic2015} to ultimately study the three-dimensional aspects of cellular nonconvexity at the subcellular scale.

	\appendix
	\section{Appendix}
\renewcommand{\thefigure}{A\arabic{figure}}
\setcounter{figure}{0}
\subsection{More on the methodology}
This section will discuss some more of the technical details for our curved-edge model.  Let us first address the issue of how many Fourier coefficients to include in the computation from a constraint counting perspective. For the cellular packings we considered, there are approximately two-thirds of the number of vertices where the number of edges is equal to the number of vertices. Suppose we have a total of $N$ number of packed cell arrays with boundaries. In general, the relation between the number of cells $N$, the number of edges $E$ and the number of vertices $V$ is as follows:
\begin{align}
V&=\alpha E,\\
N&=E-V+1\quad&\text{(for open boundary)},\\
\text{ or }N&=E-V\quad&\text{(for periodic boundary)},\\
N&=(1-\alpha)E+1\quad&(\text{or }N=(1-\alpha)E)\label{eq-con},
\end{align}
where $\alpha$ is the ratio of the total number of vertices to the number of edges. For example, $\alpha=\frac{2}{3}$ for a cell packing with periodic boundary conditions. From the energy functional (without the resistance contributions), there are $2N$ constraints. Using Eqn. A4, we obtain the number of constraints  by comparing  the number of degrees of freedom in terms of $E$, and we can deduce  whether the system is marginal, under-constrained, or over-constrained. For our case, if we ignore boundary edges for the open boundary case, we have about $E'=\frac{2}{3}E$ number of shared edges. This number can be adjusted by the ratio of boundary edges to interior edges. Similarly, we have $V'=\frac{2}{3}E'$ number of shared vertices between cells. So the number of degrees of freedom is $(\frac{4}{9}\cdot 2+\frac{2}{3}\cdot4)E=\frac{32}{9} E$, with the second term accounting for the 4 Fourier coefficients. By inserting $\alpha=\frac{2}{3}$ into the equation \ref{eq-con}, $N\sim\frac{1}{3}E$ such that $2N=\frac{2}{3}E<\frac{32}{9} E$ and so the system is under-constrained. Including the resistance function as constraints may result in a slightly over-constrained system.  Still, there are fewer unknowns for the CE model as compared to adding more SE vertices $V^*$ because the number of unknowns is increased to $2(V^*+V)$ even though the number of constraints remains the same.

\begin{figure*}
		\captionsetup{singlelinecheck = false, justification=raggedright}
		\begin{center}
  \begin{tabular}{c c}
   			\addheight{\includegraphics[height=3.5cm]{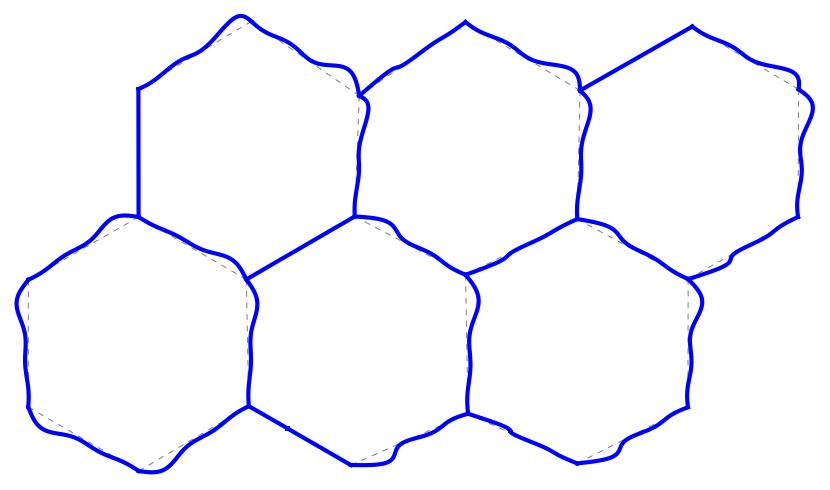}
      }&
    		\addheight{\includegraphics[height=3.5cm]{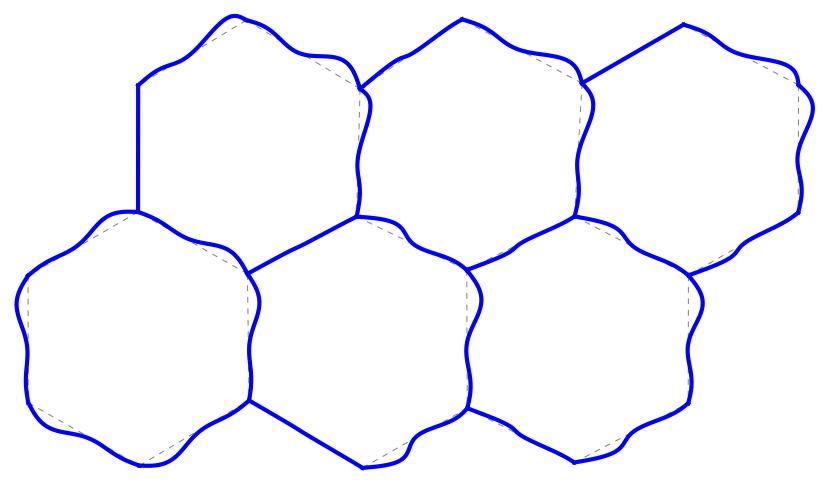}
      }\\
                (a) four sinusoidal harmonics, &  (b) three sinusoidal harmonics, \\
				\addheight{	\includegraphics[height=3.5cm]{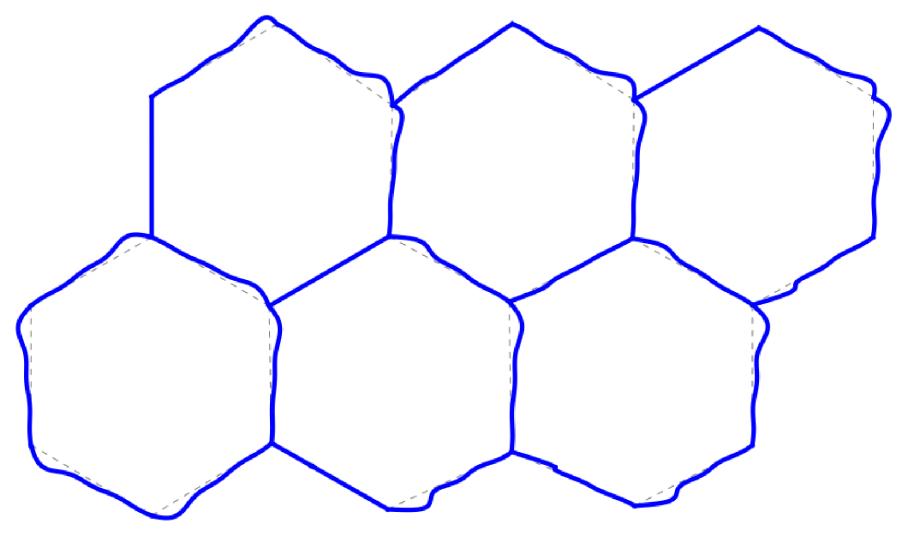}
				}&
				\addheight{	\includegraphics[height=3.5cm]{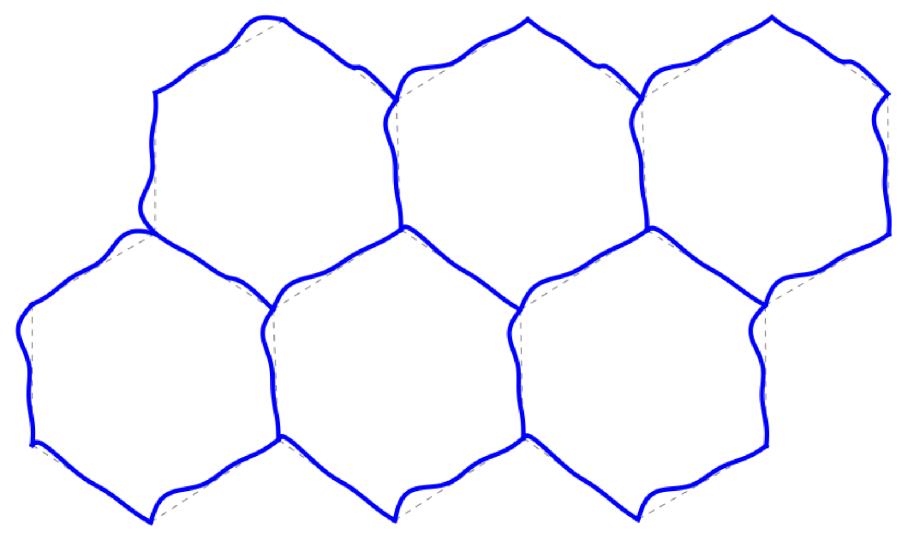}
				} \\
				(c) five sinusoidal harmonics, &
				(d) four sinusoidal harmonics using Equation \ref{eq08b} and \ref{eq09b}.\\

  \end{tabular}
			\caption{{\it Exploring different numbers of Fourier coefficients as well as the curve parameterization in the CE model.} }
			\label{fig:6cell-methods2}
		\end{center}	
	\end{figure*}

Next, we check how the number of Fourier coefficients affect our results, at least for one cell packing. Fig. \ref{fig:6cell-methods2} (a)-(c) illustrates the minimal energy configurations for 3,4 and 5 Fourier coefficients. It is evident that smaller numbers may result in rounded profiles. However, the results of the four and five frequency simulations showed minimal differences. The measured $P^{(CE)}_{\alpha}$ respectively were $6.20037$ (3 harmonics)/ $6.20103$ (4 harmonics)/ $6.20144$ (5 harmonics), i.e., a change in the third decimal place.

The parameterization implemented Eqns. \ref{eq08} and \ref{eq09} have several singularities. Here, we introduce another parameterization that may be useful for other cell packing geometries. The parameterization is:
\begin{align}
    	\textbf{v}_{k}(t)&=\textbf{v}_{\alpha,i}(t)=(f_{\alpha,i}(t),g_{\alpha,i}(t)),\\
		f_{\alpha,i}(t)&=(1 - t) x_{\alpha,i} + 
		t x_{\alpha,i+1} \nonumber\\
 & - \frac{c_{\alpha,i}(t)(x_{\alpha,i} -
			x_{\alpha,i+1})}{\sqrt{
				(x_{\alpha,i} - x_{\alpha,i+1})^2 + (y_{\alpha,i} - y_{\alpha,i+1})^2}}
		,\label{eq08b}\\
		g_{\alpha,i}(t)&=(1 - t) y_{\alpha,i} + 
		t y_{\alpha,i+1}\nonumber\\
 & + \frac{c_{\alpha,i}(t)(y_{\alpha,i} - 
			y_{\alpha,i+1})}{\sqrt{
				(x_{\alpha,i} - x_{\alpha,i+1})^2 + (y_{\alpha,i} - y_{\alpha,i+1})^2}}.\nonumber\label{eq09b}\\
\end{align}
Using this alternate parameterization does not significantly affect the minimized cell structures (see Figs. \ref{fig:6cell-methods2} (a) and (d)). It is worth noting that (d) does not have perfectly flat edges at certain angles.

\subsection{Simulation conditions for compression-induced curved edges}
\begin{figure}[H]
		\captionsetup{singlelinecheck = false, justification=raggedright}
		\centering{
			\resizebox{70mm}{!}{\includegraphics{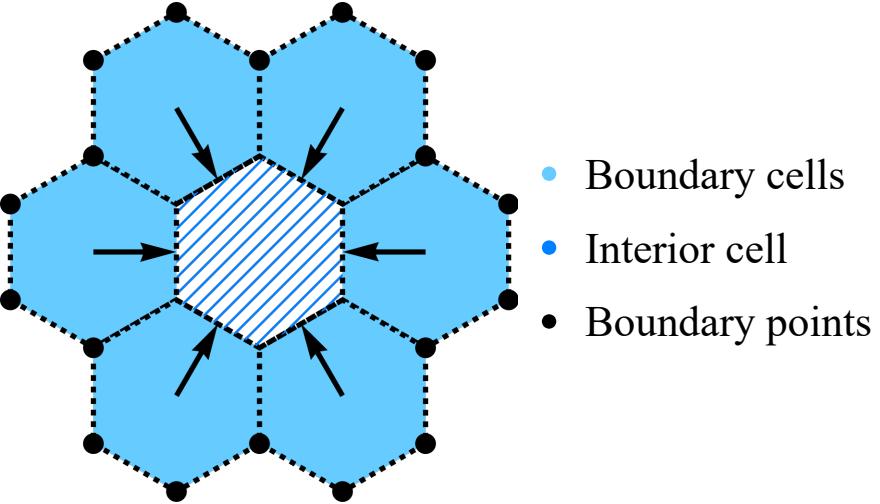}}
			\caption{A model for swelling of the boundary cells. Fixed boundary points are symbolized as black points. Boundary cells are marked with blue shading, and interior cells are represented by blue diagonal stripes, surrounded by boundary cells. Black arrows indicate the direction of swelling.}
			\label{fig:vm-cell-array}
		}
	\end{figure}

To study the deformation of the interior cells as a result of swelling boundary cells, we establish a scaling factor $s$ for the $m$-th state of cell $\alpha$ as stated below.
	\begin{align}
		A_0^{(SE)}(m)&=s^{(SE)}_{a}\cdot A_0,\\
		P_0^{(SE)}(m)&=s^{(SE)}_{p}\cdot P_0,\\	
		A_0^{(CE)}(m)&=s^{(CE)}_{a}\cdot A_0,\\	
		P_0^{(CE)}(m)&=s^{(CE)}_{p}\cdot P_0.	
	\end{align}
	where $A_0$ and $P_0$ are the area and perimeter of the unit-length regular hexagon. Both straight-edged cell's characteristics and curved-edged cell's characteristics are determined by its area, perimeter, and the cell index $\alpha$. If $s^{(SE)}_{p}=\sqrt{s^{(SE)}_{a}}$, then the target shape index of for the $m$-th state $p_0^{(SE)}(m)$ will be
	\begin{align}
		p_0^{(SE)}(m)=&\frac{P_0^{(SE)}(m)}{A_0^{(SE)}(m)}=\frac{s^{(SE)}_{p}\cdot P_0}{\sqrt{s^{(SE)}_{a}\cdot A_0}},\\
  =&\frac{s^{(SE)}_{p}\cdot P_0}{\sqrt{s^{(SE)}_{a}}\cdot\sqrt{A_0}}=\frac{P_0}{\sqrt{A_0}}.\label{eq:si}
	\end{align}
	Thus, the shape index of SE remains unchanged. Then we can define the following energy functions in the form of $\mathcal{W}^{(*)}(m)=\frac{K_W}{2}\sum_{\alpha}(W^{(*)}_{\alpha}(m)-W^{(*)}_{0}(m))^2$,
	\begin{align}
		\mathcal{A}^{(SE)}(m)=&\frac{K_A}{2}\sum_{\alpha}(A^{(SE)}_{\alpha}(m)-s^{(SE)}_{a}A_{\alpha,0})^2,\\
		\mathcal{P}^{(SE)}(m)=&\frac{K_P}{2}\sum_{\alpha}(P^{(SE)}_{\alpha}(m)-s^{(SE)}_{p}P_{\alpha,0})^2,\\
		\mathcal{A}^{(CE)}(m)=&\frac{K_A'}{2}\sum_{\alpha}(\delta a^{(CE)}_{\alpha}(m))^2\rightarrow0\quad \text{(conserved),}\\
		\mathcal{P}^{(CE)}(m)=&\frac{K_P'}{2}\sum_{\alpha}(P^{(CE)}_{\alpha}(m)-s^{(CE)}_{p}P_{\alpha,0})^2,
	\end{align}
	where $\delta a^{(CE)}_{\alpha}(m)$ denotes an area under the curve such that $\delta a^{(CE)}_{\alpha}(m)+s^{(CE)}_a\cdot A_{\alpha,0}=A^{(CE)}_{\alpha}(m)$ (because the area is conserved when simulating curved edges for $s^{(CE)}_a=1$, we get $A^{(CE)}_{\alpha}(m)=A_{\alpha,0}$). We set $K_A=1,K_P=1,K_A'=2,$ and $K_P'=2$. Fig. \ref{fig:vm-cell-array} illustrates the swelling of the boundary cells with immovable boundaries (represented by black points), accompanied by the compression of interior cells. In the ordered scenario, we utilize an 8 x 3 array of inner cells and impose three rows of cells with central cells undergoing vertical compression from top and bottom. For the disordered case, we convert a hand-drawn vector image into a Mathematica graph object \cite{Mathematica}. Fig. \ref{fig:vm-ref} illustrates the use of both disordered and ordered cell arrays in the simulation. 
Since  obtaining precise initial configurations from the experimental images is challenging due to the curved edges, we use the same initial parameters for the different states via the resistance function. However, in the ordered case, to mimic the continuous area expansion of the boundary cells, we implement a feed-forward loop that uses the vertex position data of the previous state to evaluate the next state. Therefore, resistance functions for a state $m$ can be evaluated by
	\begin{align}
		\mathcal{R}^{(SE)}(m)&=\frac{K_{R}}{2}\sum_{i}((x_{i}(m)-x_{i}(0))^2\nonumber\\
        &+(y_{i}(m)-y_{i}(0))^2) \quad\text{(disordered)},\\
		\mathcal{R}^{(SE)}(m)&=\frac{K_{R}}{2}\sum_{i}((x_{i}(m)-x_{i}(m-1))^2\nonumber\\
    &+(y_{i}(m)-y_{i}(m-1))^2) \quad\text{(ordered)},\\
		\mathcal{R}_2^{(CE)}(m)&=\frac{K_{R2}}{2}\sum_{k}\sum_{j=1}^4(c^{(j)}_{k}(m)-0.05)^2.
	\end{align}
	We run simulations for $K_{R}=0.1,K_{R1}=0,$ and $K_{R2}=0.2$. Moreover, $(x_{i}(0),y_{i}(0))$ denotes the initial position for the $i$-th vertex and $(x_{i}(m),y_{i}(m))$ denotes the position at the $i$-th vertex in the m-th state. In order to reduce computation time, we separate the SE and CE contributions for the ordered cases, as depicted in Table \ref{table:1}. We start with SE, using its data as input for both Type A and Type A+B simulations. Thus, a comparison between CE Type A and Type A+B can be made without factoring in the SE cell shape index. The constraints are chosen to correspond the disordered scenario. For example $(1.07)^2\simeq1.145$, which matches state 2 in the disordered scenario (Table. \ref{table:2} and \ref{table:5}), and $(1.07)^5\simeq1.403$, which matches state 5. 
 
	\begin{table}[ht]\caption{Optimization Methods}
		\centering
		\begin{tabular}{l l}
			\hline\hline                        
			& Optimization Methods \\ [0.5ex]
			\hline                 
			Disordered & Simultaneous  (SE and CE)\\
			Ordered (Type A) & Sequential (SE first)\\
			Ordered (Type A+B) & Sequential (SE first)\\[1ex]      
			\hline
		\end{tabular}\label{table:1}
	\end{table}
	\begin{table}[ht]\caption{Ordered Cell Constraints (SE)}
	\centering
	\begin{tabular}{l c c c c c c}
		\hline\hline                        
		& State 0 & State 1 & State 2 & State 3 & State 4 & State 5\\ [0.5ex]
		\hline                 
		Bd. SE area&$s^{(SE)}_a=1$& ${1.05}^1$ & ${1.05}^2$ & ${1.05}^3$&${1.05}^3$&${1.05}^5$\\
		Bd. SE peri.&$s^{(SE)}_p=1$& $\sqrt{{1.05}^1}$& $\sqrt{{1.05}^2}$ & $\sqrt{{1.05}^3}$ & $\sqrt{{1.05}^4}$ & $\sqrt{{1.05}^5}$\\
		Int. SE area&$s^{(SE)}_a=1$&${0.95}^1$ & ${0.95}^2$&${0.95}^3$&${0.95}^4$&${0.95}^5$\\		 
		Int. SE peri.&$s^{(SE)}_p=1$&$\sqrt{{0.95}^1}$&$\sqrt{{0.95}^2}$&$\sqrt{{0.95}^3}$&$\sqrt{{0.95}^4}$&$\sqrt{{0.95}^5}$\\	[1ex]      
		\hline
	\end{tabular}\label{table:3}
\end{table}
 \subsubsection{Tables for Disordered/Ordered simulation conditions}
		\begin{table}[ht]\caption{Disordered Cell Constraints}
		\centering
		\begin{tabular}{l c c c c}
			\hline\hline                        
			& State 0 & State 1 & State 2 & State 3\\ [0.5ex]
			\hline                 
			Bd. SE area &$s^{(SE)}_{a}=1$& $1.01$ & $1.02$ & $1.05$\\
			Bd. SE peri. &$s^{(SE)}_{p}=1$& $\sqrt{1.01}$ & $\sqrt{1.02}$ & $\sqrt{1.05}$\\
			Int. SE area&$s^{(SE)}_a=1$&$0.9$&$0.85$&$0.75$\\		 
			Int. SE peri.&$s^{(SE)}_p=1$&$\sqrt{0.9}$&$\sqrt{0.85}$&$\sqrt{0.75}$\\		 
			Int. CE area&$s^{(CE)}_a=1$&$1$&$1$&$1$\\		 
			Int. CE peri.&$s^{(CE)}_p=1$&$1$&$1.1$&$1.4$\\		 
			Images from Rigato {\it et al.} & Fig. 1A& Fig. 1B& Fig. 1C& Fig. 1D\\
			[1ex]      
			\hline
		\end{tabular}\label{table:2}
	\end{table}
	
\begin{table}[ht]\caption{Ordered Cell (CE, Type A)}
	\centering
	\begin{tabular}{l c c c c c c}
		\hline\hline                        
		& State 0 & State 1 & State 2 & State 3 & State 4 & State 5\\ [0.5ex]
		\hline                 
		Int. CE area&$s^{(CE)}_a=1$&$1$&$1$&$1$&$1$&$1$\\		 
		Int. CE peri.&$s^{(CE)}_{p}=1$&$1$&$1$&$1$&$1$&$1$\\	[1ex]      
		\hline
	\end{tabular}\label{table:4}
\end{table}
\begin{table}[ht]\caption{Ordered Cell (CE, Type A+B)}
	\centering
	\begin{tabular}{l c c c c c c}
		\hline\hline                        
		& State 0 & State 1 & State 2 & State 3 & State 4 & State 5\\ [0.5ex]
		\hline                 
		Int. CE area&$s^{(CE)}_a=1$&$1$&$1$&$1$&$1$&$1$\\		 
		Int. CE peri.&$s^{(CE)}_{p}=1$&${1.07}^1$&${1.07}^2$&${1.07}^3$&${1.07}^4$&${1.07}^5$\\	[1ex]      
		\hline
	\end{tabular}\label{table:5}
\end{table}
\enlargethispage{20pt}
\newpage
\section*{Funding}
JMS acknowledges financial support from NSF-PHY-PoLS-2014192. MBA acknowledges the financial support from ITMO Cancer of Aviesan within the framework of the 2021-2030 Cancer Control Strategy, on funds administrated by Inserm (PCSI 2021,MCMP 2022).
\section*{Acknowledgements}
The authors express their gratitude to G. Scita for permission to use an unpublished image and to Loic LeGoff for permission to use extracted data from experimental figures in \cite{rigato2022}.
\newpage
 
\bibliography{main}

\begin{thebibliography}{65}%
\makeatletter
\providecommand \@ifxundefined [1]{%
 \@ifx{#1\undefined}
}%
\providecommand \@ifnum [1]{%
 \ifnum #1\expandafter \@firstoftwo
 \else \expandafter \@secondoftwo
 \fi
}%
\providecommand \@ifx [1]{%
 \ifx #1\expandafter \@firstoftwo
 \else \expandafter \@secondoftwo
 \fi
}%
\providecommand \natexlab [1]{#1}%
\providecommand \enquote  [1]{``#1''}%
\providecommand \bibnamefont  [1]{#1}%
\providecommand \bibfnamefont [1]{#1}%
\providecommand \citenamefont [1]{#1}%
\providecommand \href@noop [0]{\@secondoftwo}%
\providecommand \href [0]{\begingroup \@sanitize@url \@href}%
\providecommand \@href[1]{\@@startlink{#1}\@@href}%
\providecommand \@@href[1]{\endgroup#1\@@endlink}%
\providecommand \@sanitize@url [0]{\catcode `\\12\catcode `\$12\catcode `\&12\catcode `\#12\catcode `\^12\catcode `\_12\catcode `\%12\relax}%
\providecommand \@@startlink[1]{}%
\providecommand \@@endlink[0]{}%
\providecommand \url  [0]{\begingroup\@sanitize@url \@url }%
\providecommand \@url [1]{\endgroup\@href {#1}{\urlprefix }}%
\providecommand \urlprefix  [0]{URL }%
\providecommand \Eprint [0]{\href }%
\providecommand \doibase [0]{http://dx.doi.org/}%
\providecommand \selectlanguage [0]{\@gobble}%
\providecommand \bibinfo  [0]{\@secondoftwo}%
\providecommand \bibfield  [0]{\@secondoftwo}%
\providecommand \translation [1]{[#1]}%
\providecommand \BibitemOpen [0]{}%
\providecommand \bibitemStop [0]{}%
\providecommand \bibitemNoStop [0]{.\EOS\space}%
\providecommand \EOS [0]{\spacefactor3000\relax}%
\providecommand \BibitemShut  [1]{\csname bibitem#1\endcsname}%
\let\auto@bib@innerbib\@empty
%</preamble>
\bibitem [{\citenamefont {Menton}(1976)}]{MENTON1976}%
  \BibitemOpen
  \bibfield  {author} {\bibinfo {author} {\bibfnamefont {D.~N.}\ \bibnamefont {Menton}},\ }\href {\doibase 10.1111/1523-1747.ep12482234} {\bibfield  {journal} {\bibinfo  {journal} {Journal of Investigative Dermatology}\ }\textbf {\bibinfo {volume} {66}},\ \bibinfo {pages} {283} (\bibinfo {year} {1976})}\BibitemShut {NoStop}%
\bibitem [{\citenamefont {ALLEN}\ and\ \citenamefont {POTTEN}(1976)}]{ALLEN1976}%
  \BibitemOpen
  \bibfield  {author} {\bibinfo {author} {\bibfnamefont {T.~D.}\ \bibnamefont {ALLEN}}\ and\ \bibinfo {author} {\bibfnamefont {C.~S.}\ \bibnamefont {POTTEN}},\ }\href {\doibase 10.1038/264545a0} {\bibfield  {journal} {\bibinfo  {journal} {Nature}\ }\textbf {\bibinfo {volume} {264}},\ \bibinfo {pages} {545} (\bibinfo {year} {1976})}\BibitemShut {NoStop}%
\bibitem [{\citenamefont {Yokouchi}\ \emph {et~al.}(2016)\citenamefont {Yokouchi}, \citenamefont {Atsugi}, \citenamefont {Logtestijn}, \citenamefont {Tanaka}, \citenamefont {Kajimura}, \citenamefont {Suematsu}, \citenamefont {Furuse}, \citenamefont {Amagai},\ and\ \citenamefont {Kubo}}]{Yokouchi2016}%
  \BibitemOpen
  \bibfield  {author} {\bibinfo {author} {\bibfnamefont {M.}~\bibnamefont {Yokouchi}}, \bibinfo {author} {\bibfnamefont {T.}~\bibnamefont {Atsugi}}, \bibinfo {author} {\bibfnamefont {M.~v.}\ \bibnamefont {Logtestijn}}, \bibinfo {author} {\bibfnamefont {R.~J.}\ \bibnamefont {Tanaka}}, \bibinfo {author} {\bibfnamefont {M.}~\bibnamefont {Kajimura}}, \bibinfo {author} {\bibfnamefont {M.}~\bibnamefont {Suematsu}}, \bibinfo {author} {\bibfnamefont {M.}~\bibnamefont {Furuse}}, \bibinfo {author} {\bibfnamefont {M.}~\bibnamefont {Amagai}}, \ and\ \bibinfo {author} {\bibfnamefont {A.}~\bibnamefont {Kubo}},\ }\href {\doibase 10.7554/eLife.19593} {\bibfield  {journal} {\bibinfo  {journal} {eLife}\ }\textbf {\bibinfo {volume} {5}},\ \bibinfo {pages} {e19593} (\bibinfo {year} {2016})}\BibitemShut {NoStop}%
\bibitem [{\citenamefont {Honda}\ and\ \citenamefont {Eguchi}(1980)}]{honda1980much}%
  \BibitemOpen
  \bibfield  {author} {\bibinfo {author} {\bibfnamefont {H.}~\bibnamefont {Honda}}\ and\ \bibinfo {author} {\bibfnamefont {G.}~\bibnamefont {Eguchi}},\ }\href@noop {} {\bibfield  {journal} {\bibinfo  {journal} {Journal of theoretical biology}\ }\textbf {\bibinfo {volume} {84}},\ \bibinfo {pages} {575} (\bibinfo {year} {1980})}\BibitemShut {NoStop}%
\bibitem [{\citenamefont {Honda}\ \emph {et~al.}(1982)\citenamefont {Honda}, \citenamefont {Ogita}, \citenamefont {Higuchi},\ and\ \citenamefont {Kani}}]{Honda1982}%
  \BibitemOpen
  \bibfield  {author} {\bibinfo {author} {\bibfnamefont {H.}~\bibnamefont {Honda}}, \bibinfo {author} {\bibfnamefont {Y.}~\bibnamefont {Ogita}}, \bibinfo {author} {\bibfnamefont {S.}~\bibnamefont {Higuchi}}, \ and\ \bibinfo {author} {\bibfnamefont {K.}~\bibnamefont {Kani}},\ }\href@noop {} {\bibfield  {journal} {\bibinfo  {journal} {Journal of Morphology}\ }\textbf {\bibinfo {volume} {174}},\ \bibinfo {pages} {25} (\bibinfo {year} {1982})}\BibitemShut {NoStop}%
\bibitem [{\citenamefont {Fletcher}\ \emph {et~al.}(2014)\citenamefont {Fletcher}, \citenamefont {Osterfield}, \citenamefont {Baker},\ and\ \citenamefont {Shvartsman}}]{fletcher2014vertex}%
  \BibitemOpen
  \bibfield  {author} {\bibinfo {author} {\bibfnamefont {A.~G.}\ \bibnamefont {Fletcher}}, \bibinfo {author} {\bibfnamefont {M.}~\bibnamefont {Osterfield}}, \bibinfo {author} {\bibfnamefont {R.~E.}\ \bibnamefont {Baker}}, \ and\ \bibinfo {author} {\bibfnamefont {S.~Y.}\ \bibnamefont {Shvartsman}},\ }\href@noop {} {\bibfield  {journal} {\bibinfo  {journal} {Biophysical journal}\ }\textbf {\bibinfo {volume} {106}},\ \bibinfo {pages} {2291} (\bibinfo {year} {2014})}\BibitemShut {NoStop}%
\bibitem [{\citenamefont {Honda}\ and\ \citenamefont {Nagai}(2022)}]{honda2022vertex}%
  \BibitemOpen
  \bibfield  {author} {\bibinfo {author} {\bibfnamefont {H.}~\bibnamefont {Honda}}\ and\ \bibinfo {author} {\bibfnamefont {T.}~\bibnamefont {Nagai}},\ }in\ \href@noop {} {\emph {\bibinfo {booktitle} {Mathematical Models of Cell-Based Morphogenesis: Passive and Active Remodeling}}}\ (\bibinfo  {publisher} {Springer},\ \bibinfo {year} {2022})\ pp.\ \bibinfo {pages} {39--57}\BibitemShut {NoStop}%
\bibitem [{\citenamefont {Farhadifar}\ \emph {et~al.}(2007)\citenamefont {Farhadifar}, \citenamefont {R\"oper}, \citenamefont {Aigouy}, \citenamefont {Eaton},\ and\ \citenamefont {J\"ulicher}}]{Farhadifar2007}%
  \BibitemOpen
  \bibfield  {author} {\bibinfo {author} {\bibfnamefont {R.}~\bibnamefont {Farhadifar}}, \bibinfo {author} {\bibfnamefont {J.-C.}\ \bibnamefont {R\"oper}}, \bibinfo {author} {\bibfnamefont {B.}~\bibnamefont {Aigouy}}, \bibinfo {author} {\bibfnamefont {S.}~\bibnamefont {Eaton}}, \ and\ \bibinfo {author} {\bibfnamefont {F.}~\bibnamefont {J\"ulicher}},\ }\href {\doibase https://doi.org/10.1016/j.cub.2007.11.049} {\bibfield  {journal} {\bibinfo  {journal} {Current Biology}\ }\textbf {\bibinfo {volume} {17}},\ \bibinfo {pages} {2095} (\bibinfo {year} {2007})}\BibitemShut {NoStop}%
\bibitem [{\citenamefont {Staple}\ \emph {et~al.}(2010)\citenamefont {Staple}, \citenamefont {Farhadifar}, \citenamefont {Röper}, \citenamefont {Aigouy}, \citenamefont {Eaton},\ and\ \citenamefont {Jülicher}}]{Staple2010}%
  \BibitemOpen
  \bibfield  {author} {\bibinfo {author} {\bibfnamefont {D.~B.}\ \bibnamefont {Staple}}, \bibinfo {author} {\bibfnamefont {R.}~\bibnamefont {Farhadifar}}, \bibinfo {author} {\bibfnamefont {J.~C.}\ \bibnamefont {Röper}}, \bibinfo {author} {\bibfnamefont {B.}~\bibnamefont {Aigouy}}, \bibinfo {author} {\bibfnamefont {S.}~\bibnamefont {Eaton}}, \ and\ \bibinfo {author} {\bibfnamefont {F.}~\bibnamefont {Jülicher}},\ }\href {\doibase 10.1140/epje/i2010-10677-0} {\bibfield  {journal} {\bibinfo  {journal} {The European Physical Journal E}\ }\textbf {\bibinfo {volume} {33}},\ \bibinfo {pages} {117} (\bibinfo {year} {2010})}\BibitemShut {NoStop}%
\bibitem [{\citenamefont {Okuda}\ \emph {et~al.}(2012)\citenamefont {Okuda}, \citenamefont {Inoue}, \citenamefont {Eiraku}, \citenamefont {Sasai},\ and\ \citenamefont {Adachi}}]{Okuda2012}%
  \BibitemOpen
  \bibfield  {author} {\bibinfo {author} {\bibfnamefont {S.}~\bibnamefont {Okuda}}, \bibinfo {author} {\bibfnamefont {Y.}~\bibnamefont {Inoue}}, \bibinfo {author} {\bibfnamefont {M.}~\bibnamefont {Eiraku}}, \bibinfo {author} {\bibfnamefont {Y.}~\bibnamefont {Sasai}}, \ and\ \bibinfo {author} {\bibfnamefont {T.}~\bibnamefont {Adachi}},\ }\href {\doibase 10.1007/s10237-012-0430-7} {\bibfield  {journal} {\bibinfo  {journal} {Biomechanics and Modeling in Mechanobiology}\ }\textbf {\bibinfo {volume} {12}} (\bibinfo {year} {2012}),\ 10.1007/s10237-012-0430-7}\BibitemShut {NoStop}%
\bibitem [{\citenamefont {Guillot}\ and\ \citenamefont {Lecuit}(2013)}]{guillot2013mechanics}%
  \BibitemOpen
  \bibfield  {author} {\bibinfo {author} {\bibfnamefont {C.}~\bibnamefont {Guillot}}\ and\ \bibinfo {author} {\bibfnamefont {T.}~\bibnamefont {Lecuit}},\ }\href@noop {} {\bibfield  {journal} {\bibinfo  {journal} {Science}\ }\textbf {\bibinfo {volume} {340}},\ \bibinfo {pages} {1185} (\bibinfo {year} {2013})}\BibitemShut {NoStop}%
\bibitem [{\citenamefont {Bi}\ \emph {et~al.}(2015)\citenamefont {Bi}, \citenamefont {Lopez}, \citenamefont {Schwarz},\ and\ \citenamefont {Manning}}]{Bi2015}%
  \BibitemOpen
  \bibfield  {author} {\bibinfo {author} {\bibfnamefont {D.}~\bibnamefont {Bi}}, \bibinfo {author} {\bibfnamefont {J.~H.}\ \bibnamefont {Lopez}}, \bibinfo {author} {\bibfnamefont {J.~M.}\ \bibnamefont {Schwarz}}, \ and\ \bibinfo {author} {\bibfnamefont {M.~L.}\ \bibnamefont {Manning}},\ }\href {\doibase 10.1038/nphys3471} {\bibfield  {journal} {\bibinfo  {journal} {Nature Physics}\ }\textbf {\bibinfo {volume} {11}},\ \bibinfo {pages} {1074} (\bibinfo {year} {2015})}\BibitemShut {NoStop}%
\bibitem [{\citenamefont {Bi}\ \emph {et~al.}(2016)\citenamefont {Bi}, \citenamefont {Yang}, \citenamefont {Marchetti},\ and\ \citenamefont {Manning}}]{Bi2016}%
  \BibitemOpen
  \bibfield  {author} {\bibinfo {author} {\bibfnamefont {D.}~\bibnamefont {Bi}}, \bibinfo {author} {\bibfnamefont {X.}~\bibnamefont {Yang}}, \bibinfo {author} {\bibfnamefont {M.~C.}\ \bibnamefont {Marchetti}}, \ and\ \bibinfo {author} {\bibfnamefont {M.~L.}\ \bibnamefont {Manning}},\ }\href {\doibase 10.1103/PhysRevX.6.021011} {\bibfield  {journal} {\bibinfo  {journal} {Phys. Rev. X}\ }\textbf {\bibinfo {volume} {6}},\ \bibinfo {pages} {021011} (\bibinfo {year} {2016})}\BibitemShut {NoStop}%
\bibitem [{\citenamefont {Kim}\ \emph {et~al.}(2018)\citenamefont {Kim}, \citenamefont {Wang},\ and\ \citenamefont {Hilgenfeldt}}]{kim2018universal}%
  \BibitemOpen
  \bibfield  {author} {\bibinfo {author} {\bibfnamefont {S.}~\bibnamefont {Kim}}, \bibinfo {author} {\bibfnamefont {Y.}~\bibnamefont {Wang}}, \ and\ \bibinfo {author} {\bibfnamefont {S.}~\bibnamefont {Hilgenfeldt}},\ }\href@noop {} {\bibfield  {journal} {\bibinfo  {journal} {Physical review letters}\ }\textbf {\bibinfo {volume} {120}},\ \bibinfo {pages} {248001} (\bibinfo {year} {2018})}\BibitemShut {NoStop}%
\bibitem [{\citenamefont {Krajnc}\ \emph {et~al.}(2018)\citenamefont {Krajnc}, \citenamefont {Dasgupta}, \citenamefont {Ziherl},\ and\ \citenamefont {Prost}}]{krajnc2018fluidization}%
  \BibitemOpen
  \bibfield  {author} {\bibinfo {author} {\bibfnamefont {M.}~\bibnamefont {Krajnc}}, \bibinfo {author} {\bibfnamefont {S.}~\bibnamefont {Dasgupta}}, \bibinfo {author} {\bibfnamefont {P.}~\bibnamefont {Ziherl}}, \ and\ \bibinfo {author} {\bibfnamefont {J.}~\bibnamefont {Prost}},\ }\href@noop {} {\bibfield  {journal} {\bibinfo  {journal} {Physical Review E}\ }\textbf {\bibinfo {volume} {98}},\ \bibinfo {pages} {022409} (\bibinfo {year} {2018})}\BibitemShut {NoStop}%
\bibitem [{\citenamefont {Sussman}(2020)}]{sussman2020interplay}%
  \BibitemOpen
  \bibfield  {author} {\bibinfo {author} {\bibfnamefont {D.~M.}\ \bibnamefont {Sussman}},\ }\href@noop {} {\bibfield  {journal} {\bibinfo  {journal} {Physical Review Research}\ }\textbf {\bibinfo {volume} {2}},\ \bibinfo {pages} {023417} (\bibinfo {year} {2020})}\BibitemShut {NoStop}%
\bibitem [{\citenamefont {Fiore}\ \emph {et~al.}(2020)\citenamefont {Fiore}, \citenamefont {Krajnc}, \citenamefont {Quiroz}, \citenamefont {Levorse}, \citenamefont {Pasolli}, \citenamefont {Shvartsman},\ and\ \citenamefont {Fuchs}}]{fiore2020mechanics}%
  \BibitemOpen
  \bibfield  {author} {\bibinfo {author} {\bibfnamefont {V.~F.}\ \bibnamefont {Fiore}}, \bibinfo {author} {\bibfnamefont {M.}~\bibnamefont {Krajnc}}, \bibinfo {author} {\bibfnamefont {F.~G.}\ \bibnamefont {Quiroz}}, \bibinfo {author} {\bibfnamefont {J.}~\bibnamefont {Levorse}}, \bibinfo {author} {\bibfnamefont {H.~A.}\ \bibnamefont {Pasolli}}, \bibinfo {author} {\bibfnamefont {S.~Y.}\ \bibnamefont {Shvartsman}}, \ and\ \bibinfo {author} {\bibfnamefont {E.}~\bibnamefont {Fuchs}},\ }\href@noop {} {\bibfield  {journal} {\bibinfo  {journal} {Nature}\ }\textbf {\bibinfo {volume} {585}},\ \bibinfo {pages} {433} (\bibinfo {year} {2020})}\BibitemShut {NoStop}%
\bibitem [{\citenamefont {Sahu}\ \emph {et~al.}(2020)\citenamefont {Sahu}, \citenamefont {Sussman}, \citenamefont {R{\"u}bsam}, \citenamefont {Mertz}, \citenamefont {Horsley}, \citenamefont {Dufresne}, \citenamefont {Niessen}, \citenamefont {Marchetti}, \citenamefont {Manning},\ and\ \citenamefont {Schwarz}}]{Sahu2020}%
  \BibitemOpen
  \bibfield  {author} {\bibinfo {author} {\bibfnamefont {P.}~\bibnamefont {Sahu}}, \bibinfo {author} {\bibfnamefont {D.~M.}\ \bibnamefont {Sussman}}, \bibinfo {author} {\bibfnamefont {M.}~\bibnamefont {R{\"u}bsam}}, \bibinfo {author} {\bibfnamefont {A.~F.}\ \bibnamefont {Mertz}}, \bibinfo {author} {\bibfnamefont {V.}~\bibnamefont {Horsley}}, \bibinfo {author} {\bibfnamefont {E.~R.}\ \bibnamefont {Dufresne}}, \bibinfo {author} {\bibfnamefont {C.~M.}\ \bibnamefont {Niessen}}, \bibinfo {author} {\bibfnamefont {M.~C.}\ \bibnamefont {Marchetti}}, \bibinfo {author} {\bibfnamefont {M.~L.}\ \bibnamefont {Manning}}, \ and\ \bibinfo {author} {\bibfnamefont {J.~M.}\ \bibnamefont {Schwarz}},\ }\href@noop {} {\bibfield  {journal} {\bibinfo  {journal} {Soft Matter}\ }\textbf {\bibinfo {volume} {16}},\ \bibinfo {pages} {3325} (\bibinfo {year} {2020})}\BibitemShut {NoStop}%
\bibitem [{\citenamefont {Tong}\ \emph {et~al.}(2022)\citenamefont {Tong}, \citenamefont {Singh}, \citenamefont {Sknepnek},\ and\ \citenamefont {Ko{\v{s}}mrlj}}]{tong2022linear}%
  \BibitemOpen
  \bibfield  {author} {\bibinfo {author} {\bibfnamefont {S.}~\bibnamefont {Tong}}, \bibinfo {author} {\bibfnamefont {N.~K.}\ \bibnamefont {Singh}}, \bibinfo {author} {\bibfnamefont {R.}~\bibnamefont {Sknepnek}}, \ and\ \bibinfo {author} {\bibfnamefont {A.}~\bibnamefont {Ko{\v{s}}mrlj}},\ }\href@noop {} {\bibfield  {journal} {\bibinfo  {journal} {PLoS computational biology}\ }\textbf {\bibinfo {volume} {18}},\ \bibinfo {pages} {e1010135} (\bibinfo {year} {2022})}\BibitemShut {NoStop}%
\bibitem [{\citenamefont {P{\'e}rez-Verdugo}\ and\ \citenamefont {Banerjee}(2023)}]{perez2023tension}%
  \BibitemOpen
  \bibfield  {author} {\bibinfo {author} {\bibfnamefont {F.}~\bibnamefont {P{\'e}rez-Verdugo}}\ and\ \bibinfo {author} {\bibfnamefont {S.}~\bibnamefont {Banerjee}},\ }\href@noop {} {\bibfield  {journal} {\bibinfo  {journal} {PRX Life}\ }\textbf {\bibinfo {volume} {1}},\ \bibinfo {pages} {023006} (\bibinfo {year} {2023})}\BibitemShut {NoStop}%
\bibitem [{\citenamefont {Zhang}\ and\ \citenamefont {Schwarz}(2022)}]{Zhang2023}%
  \BibitemOpen
  \bibfield  {author} {\bibinfo {author} {\bibfnamefont {T.}~\bibnamefont {Zhang}}\ and\ \bibinfo {author} {\bibfnamefont {J.~M.}\ \bibnamefont {Schwarz}},\ }\href {\doibase 10.1103/PhysRevResearch.4.043148} {\bibfield  {journal} {\bibinfo  {journal} {Phys. Rev. Res.}\ }\textbf {\bibinfo {volume} {4}},\ \bibinfo {pages} {043148} (\bibinfo {year} {2022})}\BibitemShut {NoStop}%
\bibitem [{\citenamefont {Staddon}\ \emph {et~al.}(2023)\citenamefont {Staddon}, \citenamefont {Hernandez}, \citenamefont {Bowick}, \citenamefont {Moshe},\ and\ \citenamefont {Marchetti}}]{Staddon2023}%
  \BibitemOpen
  \bibfield  {author} {\bibinfo {author} {\bibfnamefont {M.~F.}\ \bibnamefont {Staddon}}, \bibinfo {author} {\bibfnamefont {A.}~\bibnamefont {Hernandez}}, \bibinfo {author} {\bibfnamefont {M.~J.}\ \bibnamefont {Bowick}}, \bibinfo {author} {\bibfnamefont {M.}~\bibnamefont {Moshe}}, \ and\ \bibinfo {author} {\bibfnamefont {M.~C.}\ \bibnamefont {Marchetti}},\ }\href {\doibase 10.1039/D2SM01580C} {\bibfield  {journal} {\bibinfo  {journal} {Soft Matter}\ } (\bibinfo {year} {2023}),\ 10.1039/D2SM01580C}\BibitemShut {NoStop}%
\bibitem [{\citenamefont {Mongera}\ \emph {et~al.}(2018)\citenamefont {Mongera}, \citenamefont {Rowghanian}, \citenamefont {Gustafson}, \citenamefont {Shelton}, \citenamefont {Kealhofer}, \citenamefont {Carn}, \citenamefont {Serwane}, \citenamefont {Lucio}, \citenamefont {Giammona},\ and\ \citenamefont {Camp\'as}}]{Mongera2018}%
  \BibitemOpen
  \bibfield  {author} {\bibinfo {author} {\bibfnamefont {A.}~\bibnamefont {Mongera}}, \bibinfo {author} {\bibfnamefont {P.}~\bibnamefont {Rowghanian}}, \bibinfo {author} {\bibfnamefont {H.~J.}\ \bibnamefont {Gustafson}}, \bibinfo {author} {\bibfnamefont {E.}~\bibnamefont {Shelton}}, \bibinfo {author} {\bibfnamefont {D.~A.}\ \bibnamefont {Kealhofer}}, \bibinfo {author} {\bibfnamefont {E.~K.}\ \bibnamefont {Carn}}, \bibinfo {author} {\bibfnamefont {F.}~\bibnamefont {Serwane}}, \bibinfo {author} {\bibfnamefont {A.~A.}\ \bibnamefont {Lucio}}, \bibinfo {author} {\bibfnamefont {J.}~\bibnamefont {Giammona}}, \ and\ \bibinfo {author} {\bibfnamefont {O.}~\bibnamefont {Camp\'as}},\ }\href {\doibase 10.1038/s41586-018-0479-2} {\bibfield  {journal} {\bibinfo  {journal} {Nature}\ }\textbf {\bibinfo {volume} {561}},\ \bibinfo {pages} {401} (\bibinfo {year} {2018})}\BibitemShut {NoStop}%
\bibitem [{\citenamefont {Lenne}\ and\ \citenamefont {Trivedi}(2022)}]{Lenne2022}%
  \BibitemOpen
  \bibfield  {author} {\bibinfo {author} {\bibfnamefont {P.-F.}\ \bibnamefont {Lenne}}\ and\ \bibinfo {author} {\bibfnamefont {V.}~\bibnamefont {Trivedi}},\ }\href {\doibase 10.1038/s41467-022-28151-9} {\bibfield  {journal} {\bibinfo  {journal} {Nature Communications}\ }\textbf {\bibinfo {volume} {13}},\ \bibinfo {pages} {2041} (\bibinfo {year} {2022})}\BibitemShut {NoStop}%
\bibitem [{\citenamefont {Angelini}\ \emph {et~al.}(2011)\citenamefont {Angelini}, \citenamefont {Hannezo}, \citenamefont {Trepat}, \citenamefont {Marquez}, \citenamefont {Fredberg},\ and\ \citenamefont {Weitz}}]{Angelini2011}%
  \BibitemOpen
  \bibfield  {author} {\bibinfo {author} {\bibfnamefont {T.~E.}\ \bibnamefont {Angelini}}, \bibinfo {author} {\bibfnamefont {E.}~\bibnamefont {Hannezo}}, \bibinfo {author} {\bibfnamefont {X.}~\bibnamefont {Trepat}}, \bibinfo {author} {\bibfnamefont {M.}~\bibnamefont {Marquez}}, \bibinfo {author} {\bibfnamefont {J.~J.}\ \bibnamefont {Fredberg}}, \ and\ \bibinfo {author} {\bibfnamefont {D.~A.}\ \bibnamefont {Weitz}},\ }\href {\doibase 10.1073/pnas.1010059108} {\bibfield  {journal} {\bibinfo  {journal} {Proceedings of the National Academy of Sciences}\ }\textbf {\bibinfo {volume} {108}},\ \bibinfo {pages} {4714} (\bibinfo {year} {2011})}\BibitemShut {NoStop}%
\bibitem [{\citenamefont {Latorre}\ \emph {et~al.}(2018)\citenamefont {Latorre}, \citenamefont {Kale}, \citenamefont {Casares}, \citenamefont {G{\'o}mez-Gonz{\'a}lez}, \citenamefont {Uroz}, \citenamefont {Valon}, \citenamefont {Nair}, \citenamefont {Garreta}, \citenamefont {Montserrat}, \citenamefont {Del~Campo} \emph {et~al.}}]{latorre2018active}%
  \BibitemOpen
  \bibfield  {author} {\bibinfo {author} {\bibfnamefont {E.}~\bibnamefont {Latorre}}, \bibinfo {author} {\bibfnamefont {S.}~\bibnamefont {Kale}}, \bibinfo {author} {\bibfnamefont {L.}~\bibnamefont {Casares}}, \bibinfo {author} {\bibfnamefont {M.}~\bibnamefont {G{\'o}mez-Gonz{\'a}lez}}, \bibinfo {author} {\bibfnamefont {M.}~\bibnamefont {Uroz}}, \bibinfo {author} {\bibfnamefont {L.}~\bibnamefont {Valon}}, \bibinfo {author} {\bibfnamefont {R.~V.}\ \bibnamefont {Nair}}, \bibinfo {author} {\bibfnamefont {E.}~\bibnamefont {Garreta}}, \bibinfo {author} {\bibfnamefont {N.}~\bibnamefont {Montserrat}}, \bibinfo {author} {\bibfnamefont {A.}~\bibnamefont {Del~Campo}},  \emph {et~al.},\ }\href@noop {} {\bibfield  {journal} {\bibinfo  {journal} {Nature}\ }\textbf {\bibinfo {volume} {563}},\ \bibinfo {pages} {203} (\bibinfo {year} {2018})}\BibitemShut {NoStop}%
\bibitem [{\citenamefont {Sussman}\ \emph {et~al.}(2018)\citenamefont {Sussman}, \citenamefont {Paoluzzi}, \citenamefont {Marchetti},\ and\ \citenamefont {Manning}}]{sussman2018anomalous}%
  \BibitemOpen
  \bibfield  {author} {\bibinfo {author} {\bibfnamefont {D.~M.}\ \bibnamefont {Sussman}}, \bibinfo {author} {\bibfnamefont {M.}~\bibnamefont {Paoluzzi}}, \bibinfo {author} {\bibfnamefont {M.~C.}\ \bibnamefont {Marchetti}}, \ and\ \bibinfo {author} {\bibfnamefont {M.~L.}\ \bibnamefont {Manning}},\ }\href@noop {} {\bibfield  {journal} {\bibinfo  {journal} {Europhysics Letters}\ }\textbf {\bibinfo {volume} {121}},\ \bibinfo {pages} {36001} (\bibinfo {year} {2018})}\BibitemShut {NoStop}%
\bibitem [{\citenamefont {Malinverno}\ \emph {et~al.}(2017)\citenamefont {Malinverno}, \citenamefont {Corallino}, \citenamefont {Giavazzi}, \citenamefont {Bergert}, \citenamefont {Li}, \citenamefont {Leoni}, \citenamefont {Disanza}, \citenamefont {Frittoli}, \citenamefont {Oldani}, \citenamefont {Martini}, \citenamefont {Lendenmann}, \citenamefont {Deflorian}, \citenamefont {Beznoussenko}, \citenamefont {Poulikakos}, \citenamefont {Ong}, \citenamefont {Uroz}, \citenamefont {Trepat}, \citenamefont {Parazzoli}, \citenamefont {Maiuri}, \citenamefont {Yu}, \citenamefont {Ferrari}, \citenamefont {Cerbino},\ and\ \citenamefont {Scita}}]{Malinverno2017}%
  \BibitemOpen
  \bibfield  {author} {\bibinfo {author} {\bibfnamefont {C.}~\bibnamefont {Malinverno}}, \bibinfo {author} {\bibfnamefont {S.}~\bibnamefont {Corallino}}, \bibinfo {author} {\bibfnamefont {F.}~\bibnamefont {Giavazzi}}, \bibinfo {author} {\bibfnamefont {M.}~\bibnamefont {Bergert}}, \bibinfo {author} {\bibfnamefont {Q.}~\bibnamefont {Li}}, \bibinfo {author} {\bibfnamefont {M.}~\bibnamefont {Leoni}}, \bibinfo {author} {\bibfnamefont {A.}~\bibnamefont {Disanza}}, \bibinfo {author} {\bibfnamefont {E.}~\bibnamefont {Frittoli}}, \bibinfo {author} {\bibfnamefont {A.}~\bibnamefont {Oldani}}, \bibinfo {author} {\bibfnamefont {E.}~\bibnamefont {Martini}}, \bibinfo {author} {\bibfnamefont {T.}~\bibnamefont {Lendenmann}}, \bibinfo {author} {\bibfnamefont {G.}~\bibnamefont {Deflorian}}, \bibinfo {author} {\bibfnamefont {G.~V.}\ \bibnamefont {Beznoussenko}}, \bibinfo {author} {\bibfnamefont {D.}~\bibnamefont {Poulikakos}}, \bibinfo {author} {\bibfnamefont {K.~H.}\ \bibnamefont {Ong}}, \bibinfo {author} {\bibfnamefont
  {M.}~\bibnamefont {Uroz}}, \bibinfo {author} {\bibfnamefont {X.}~\bibnamefont {Trepat}}, \bibinfo {author} {\bibfnamefont {D.}~\bibnamefont {Parazzoli}}, \bibinfo {author} {\bibfnamefont {P.}~\bibnamefont {Maiuri}}, \bibinfo {author} {\bibfnamefont {W.}~\bibnamefont {Yu}}, \bibinfo {author} {\bibfnamefont {A.}~\bibnamefont {Ferrari}}, \bibinfo {author} {\bibfnamefont {R.}~\bibnamefont {Cerbino}}, \ and\ \bibinfo {author} {\bibfnamefont {G.}~\bibnamefont {Scita}},\ }\href@noop {} {\bibfield  {journal} {\bibinfo  {journal} {Nature Materials}\ }\textbf {\bibinfo {volume} {16}},\ \bibinfo {pages} {587} (\bibinfo {year} {2017})}\BibitemShut {NoStop}%
\bibitem [{\citenamefont {Belteton}\ \emph {et~al.}(2021)\citenamefont {Belteton}, \citenamefont {Li}, \citenamefont {Yanagisawa}, \citenamefont {Hatam}, \citenamefont {Quinn}, \citenamefont {Szymanski}, \citenamefont {Marley}, \citenamefont {Turner},\ and\ \citenamefont {Szymanski}}]{Belteton2021}%
  \BibitemOpen
  \bibfield  {author} {\bibinfo {author} {\bibfnamefont {S.~A.}\ \bibnamefont {Belteton}}, \bibinfo {author} {\bibfnamefont {W.}~\bibnamefont {Li}}, \bibinfo {author} {\bibfnamefont {M.}~\bibnamefont {Yanagisawa}}, \bibinfo {author} {\bibfnamefont {F.~A.}\ \bibnamefont {Hatam}}, \bibinfo {author} {\bibfnamefont {M.~I.}\ \bibnamefont {Quinn}}, \bibinfo {author} {\bibfnamefont {M.~K.}\ \bibnamefont {Szymanski}}, \bibinfo {author} {\bibfnamefont {M.~W.}\ \bibnamefont {Marley}}, \bibinfo {author} {\bibfnamefont {J.~A.}\ \bibnamefont {Turner}}, \ and\ \bibinfo {author} {\bibfnamefont {D.~B.}\ \bibnamefont {Szymanski}},\ }\href {\doibase 10.1038/s41477-021-00931-z} {\bibfield  {journal} {\bibinfo  {journal} {Nature Plants}\ }\textbf {\bibinfo {volume} {7}},\ \bibinfo {pages} {826} (\bibinfo {year} {2021})}\BibitemShut {NoStop}%
\bibitem [{\citenamefont {Rigato}\ \emph {et~al.}(2022)\citenamefont {Rigato}, \citenamefont {Meng}, \citenamefont {Abouakil},\ and\ \citenamefont {LeGoff}}]{rigato2022}%
  \BibitemOpen
  \bibfield  {author} {\bibinfo {author} {\bibfnamefont {A.}~\bibnamefont {Rigato}}, \bibinfo {author} {\bibfnamefont {H.}~\bibnamefont {Meng}}, \bibinfo {author} {\bibfnamefont {F.}~\bibnamefont {Abouakil}}, \ and\ \bibinfo {author} {\bibfnamefont {L.}~\bibnamefont {LeGoff}},\ }\href@noop {} {\bibfield  {journal} {\bibinfo  {journal} {bioRxiv}\ ,\ \bibinfo {pages} {2022}} (\bibinfo {year} {2022})}\BibitemShut {NoStop}%
\bibitem [{\citenamefont {Spaargaren}\ and\ \citenamefont {Bos}(1999)}]{spaargaren1999rab5}%
  \BibitemOpen
  \bibfield  {author} {\bibinfo {author} {\bibfnamefont {M.}~\bibnamefont {Spaargaren}}\ and\ \bibinfo {author} {\bibfnamefont {J.~L.}\ \bibnamefont {Bos}},\ }\href@noop {} {\bibfield  {journal} {\bibinfo  {journal} {Molecular biology of the cell}\ }\textbf {\bibinfo {volume} {10}},\ \bibinfo {pages} {3239} (\bibinfo {year} {1999})}\BibitemShut {NoStop}%
\bibitem [{\citenamefont {Svitkina}\ and\ \citenamefont {Borisy}(1999)}]{svitkina1999arp2}%
  \BibitemOpen
  \bibfield  {author} {\bibinfo {author} {\bibfnamefont {T.~M.}\ \bibnamefont {Svitkina}}\ and\ \bibinfo {author} {\bibfnamefont {G.~G.}\ \bibnamefont {Borisy}},\ }\href@noop {} {\bibfield  {journal} {\bibinfo  {journal} {The Journal of cell biology}\ }\textbf {\bibinfo {volume} {145}},\ \bibinfo {pages} {1009} (\bibinfo {year} {1999})}\BibitemShut {NoStop}%
\bibitem [{\citenamefont {Gopinathan}\ \emph {et~al.}(2007)\citenamefont {Gopinathan}, \citenamefont {Lee}, \citenamefont {Schwarz},\ and\ \citenamefont {Liu}}]{gopinathan2007branching}%
  \BibitemOpen
  \bibfield  {author} {\bibinfo {author} {\bibfnamefont {A.}~\bibnamefont {Gopinathan}}, \bibinfo {author} {\bibfnamefont {K.-C.}\ \bibnamefont {Lee}}, \bibinfo {author} {\bibfnamefont {J.}~\bibnamefont {Schwarz}}, \ and\ \bibinfo {author} {\bibfnamefont {A.~J.}\ \bibnamefont {Liu}},\ }\href@noop {} {\bibfield  {journal} {\bibinfo  {journal} {Physical review letters}\ }\textbf {\bibinfo {volume} {99}},\ \bibinfo {pages} {058103} (\bibinfo {year} {2007})}\BibitemShut {NoStop}%
\bibitem [{\citenamefont {Honda}\ \emph {et~al.}(2004)\citenamefont {Honda}, \citenamefont {Tanemura},\ and\ \citenamefont {Nagai}}]{honda2004three}%
  \BibitemOpen
  \bibfield  {author} {\bibinfo {author} {\bibfnamefont {H.}~\bibnamefont {Honda}}, \bibinfo {author} {\bibfnamefont {M.}~\bibnamefont {Tanemura}}, \ and\ \bibinfo {author} {\bibfnamefont {T.}~\bibnamefont {Nagai}},\ }\href@noop {} {\bibfield  {journal} {\bibinfo  {journal} {Journal of theoretical biology}\ }\textbf {\bibinfo {volume} {226}},\ \bibinfo {pages} {439} (\bibinfo {year} {2004})}\BibitemShut {NoStop}%
\bibitem [{\citenamefont {Shraiman}(2005)}]{shraiman2005mechanical}%
  \BibitemOpen
  \bibfield  {author} {\bibinfo {author} {\bibfnamefont {B.~I.}\ \bibnamefont {Shraiman}},\ }\href@noop {} {\bibfield  {journal} {\bibinfo  {journal} {Proceedings of the National Academy of Sciences}\ }\textbf {\bibinfo {volume} {102}},\ \bibinfo {pages} {3318} (\bibinfo {year} {2005})}\BibitemShut {NoStop}%
\bibitem [{\citenamefont {Lin}\ \emph {et~al.}(2017)\citenamefont {Lin}, \citenamefont {Li},\ and\ \citenamefont {Feng}}]{lin2017dynamic}%
  \BibitemOpen
  \bibfield  {author} {\bibinfo {author} {\bibfnamefont {S.-Z.}\ \bibnamefont {Lin}}, \bibinfo {author} {\bibfnamefont {B.}~\bibnamefont {Li}}, \ and\ \bibinfo {author} {\bibfnamefont {X.-Q.}\ \bibnamefont {Feng}},\ }\href@noop {} {\bibfield  {journal} {\bibinfo  {journal} {Acta Mechanica Sinica}\ }\textbf {\bibinfo {volume} {33}},\ \bibinfo {pages} {250} (\bibinfo {year} {2017})}\BibitemShut {NoStop}%
\bibitem [{\citenamefont {Ishimoto}\ and\ \citenamefont {Morishita}(2014)}]{Ishimoto2014}%
  \BibitemOpen
  \bibfield  {author} {\bibinfo {author} {\bibfnamefont {Y.}~\bibnamefont {Ishimoto}}\ and\ \bibinfo {author} {\bibfnamefont {Y.}~\bibnamefont {Morishita}},\ }\href {\doibase 10.1103/PhysRevE.90.052711} {\bibfield  {journal} {\bibinfo  {journal} {Phys. Rev. E}\ }\textbf {\bibinfo {volume} {90}},\ \bibinfo {pages} {052711} (\bibinfo {year} {2014})}\BibitemShut {NoStop}%
\bibitem [{\citenamefont {Perrone}\ \emph {et~al.}(2016)\citenamefont {Perrone}, \citenamefont {Veldhuis},\ and\ \citenamefont {Brodland}}]{Perrone2016}%
  \BibitemOpen
  \bibfield  {author} {\bibinfo {author} {\bibfnamefont {M.~C.}\ \bibnamefont {Perrone}}, \bibinfo {author} {\bibfnamefont {J.~H.}\ \bibnamefont {Veldhuis}}, \ and\ \bibinfo {author} {\bibfnamefont {G.~W.}\ \bibnamefont {Brodland}},\ }\href {\doibase 10.1007/s10237-015-0697-6} {\bibfield  {journal} {\bibinfo  {journal} {Biomechanics and Modeling in Mechanobiology}\ }\textbf {\bibinfo {volume} {15}},\ \bibinfo {pages} {405} (\bibinfo {year} {2016})}\BibitemShut {NoStop}%
\bibitem [{\citenamefont {Boromand}\ \emph {et~al.}(2018)\citenamefont {Boromand}, \citenamefont {Signoriello}, \citenamefont {Ye}, \citenamefont {O’Hern},\ and\ \citenamefont {Shattuck}}]{boromand2018jamming}%
  \BibitemOpen
  \bibfield  {author} {\bibinfo {author} {\bibfnamefont {A.}~\bibnamefont {Boromand}}, \bibinfo {author} {\bibfnamefont {A.}~\bibnamefont {Signoriello}}, \bibinfo {author} {\bibfnamefont {F.}~\bibnamefont {Ye}}, \bibinfo {author} {\bibfnamefont {C.~S.}\ \bibnamefont {O’Hern}}, \ and\ \bibinfo {author} {\bibfnamefont {M.~D.}\ \bibnamefont {Shattuck}},\ }\href@noop {} {\bibfield  {journal} {\bibinfo  {journal} {Physical review letters}\ }\textbf {\bibinfo {volume} {121}},\ \bibinfo {pages} {248003} (\bibinfo {year} {2018})}\BibitemShut {NoStop}%
\bibitem [{\citenamefont {Vetter}\ \emph {et~al.}(2023)\citenamefont {Vetter}, \citenamefont {Runser},\ and\ \citenamefont {Iber}}]{vetter2023polyhoop}%
  \BibitemOpen
  \bibfield  {author} {\bibinfo {author} {\bibfnamefont {R.}~\bibnamefont {Vetter}}, \bibinfo {author} {\bibfnamefont {S.~V.}\ \bibnamefont {Runser}}, \ and\ \bibinfo {author} {\bibfnamefont {D.}~\bibnamefont {Iber}},\ }\href@noop {} {\bibfield  {journal} {\bibinfo  {journal} {arXiv preprint arXiv:2307.15006}\ } (\bibinfo {year} {2023})}\BibitemShut {NoStop}%
\bibitem [{\citenamefont {{Karl Az}}(2014)}]{estomatos}%
  \BibitemOpen
  \bibfield  {author} {\bibinfo {author} {\bibnamefont {{Karl Az}}},\ }\href {https://commons.wikimedia.org/wiki/File:Est%C3%B4matos.jpg} {\enquote {\bibinfo {title} {Est{\^o}matos},}\ } (\bibinfo {year} {2014}),\ \bibinfo {note} {a cropped image of the original version. This work is licensed under the Creative Commons Attribution 4.0 International License.}\BibitemShut {Stop}%
\bibitem [{\citenamefont {{Ben Amar}}\ \emph {et~al.}(2019)\citenamefont {{Ben Amar}}, \citenamefont {Nassoy},\ and\ \citenamefont {LeGoff}}]{ben2019physics}%
  \BibitemOpen
  \bibfield  {author} {\bibinfo {author} {\bibfnamefont {M.}~\bibnamefont {{Ben Amar}}}, \bibinfo {author} {\bibfnamefont {P.}~\bibnamefont {Nassoy}}, \ and\ \bibinfo {author} {\bibfnamefont {L.}~\bibnamefont {LeGoff}},\ }\href@noop {} {\bibfield  {journal} {\bibinfo  {journal} {Philosophical Transactions of the Royal Society A}\ }\textbf {\bibinfo {volume} {377}},\ \bibinfo {pages} {20180070} (\bibinfo {year} {2019})}\BibitemShut {NoStop}%
\bibitem [{\citenamefont {Hufnagel}\ \emph {et~al.}(2007)\citenamefont {Hufnagel}, \citenamefont {Teleman}, \citenamefont {Rouault}, \citenamefont {Cohen},\ and\ \citenamefont {Shraiman}}]{hufnagel2007mechanism}%
  \BibitemOpen
  \bibfield  {author} {\bibinfo {author} {\bibfnamefont {L.}~\bibnamefont {Hufnagel}}, \bibinfo {author} {\bibfnamefont {A.~A.}\ \bibnamefont {Teleman}}, \bibinfo {author} {\bibfnamefont {H.}~\bibnamefont {Rouault}}, \bibinfo {author} {\bibfnamefont {S.~M.}\ \bibnamefont {Cohen}}, \ and\ \bibinfo {author} {\bibfnamefont {B.~I.}\ \bibnamefont {Shraiman}},\ }\href@noop {} {\bibfield  {journal} {\bibinfo  {journal} {Proceedings of the National Academy of Sciences}\ }\textbf {\bibinfo {volume} {104}},\ \bibinfo {pages} {3835} (\bibinfo {year} {2007})}\BibitemShut {NoStop}%
\bibitem [{\citenamefont {Zehnder}\ \emph {et~al.}(2015)\citenamefont {Zehnder}, \citenamefont {Suaris}, \citenamefont {Bellaire},\ and\ \citenamefont {Angelini}}]{zehnder2015cell}%
  \BibitemOpen
  \bibfield  {author} {\bibinfo {author} {\bibfnamefont {S.~M.}\ \bibnamefont {Zehnder}}, \bibinfo {author} {\bibfnamefont {M.}~\bibnamefont {Suaris}}, \bibinfo {author} {\bibfnamefont {M.~M.}\ \bibnamefont {Bellaire}}, \ and\ \bibinfo {author} {\bibfnamefont {T.~E.}\ \bibnamefont {Angelini}},\ }\href@noop {} {\bibfield  {journal} {\bibinfo  {journal} {Biophysical journal}\ }\textbf {\bibinfo {volume} {108}},\ \bibinfo {pages} {247} (\bibinfo {year} {2015})}\BibitemShut {NoStop}%
\bibitem [{\citenamefont {Bonnet}\ \emph {et~al.}(2012)\citenamefont {Bonnet}, \citenamefont {Marcq}, \citenamefont {Bosveld}, \citenamefont {Fetler}, \citenamefont {Bella{\"i}che},\ and\ \citenamefont {Graner}}]{bonnet2012mechanical}%
  \BibitemOpen
  \bibfield  {author} {\bibinfo {author} {\bibfnamefont {I.}~\bibnamefont {Bonnet}}, \bibinfo {author} {\bibfnamefont {P.}~\bibnamefont {Marcq}}, \bibinfo {author} {\bibfnamefont {F.}~\bibnamefont {Bosveld}}, \bibinfo {author} {\bibfnamefont {L.}~\bibnamefont {Fetler}}, \bibinfo {author} {\bibfnamefont {Y.}~\bibnamefont {Bella{\"i}che}}, \ and\ \bibinfo {author} {\bibfnamefont {F.}~\bibnamefont {Graner}},\ }\href@noop {} {\bibfield  {journal} {\bibinfo  {journal} {Journal of The Royal Society Interface}\ }\textbf {\bibinfo {volume} {9}},\ \bibinfo {pages} {2614} (\bibinfo {year} {2012})}\BibitemShut {NoStop}%
\bibitem [{\citenamefont {Manning}\ \emph {et~al.}(2010)\citenamefont {Manning}, \citenamefont {Foty}, \citenamefont {Steinberg},\ and\ \citenamefont {Schoetz}}]{manning_2010}%
  \BibitemOpen
  \bibfield  {author} {\bibinfo {author} {\bibfnamefont {M.~L.}\ \bibnamefont {Manning}}, \bibinfo {author} {\bibfnamefont {R.~A.}\ \bibnamefont {Foty}}, \bibinfo {author} {\bibfnamefont {M.~S.}\ \bibnamefont {Steinberg}}, \ and\ \bibinfo {author} {\bibfnamefont {E.-M.}\ \bibnamefont {Schoetz}},\ }\href {\doibase 10.1073/pnas.1003743107} {\bibfield  {journal} {\bibinfo  {journal} {Proceedings of the National Academy of Sciences}\ }\textbf {\bibinfo {volume} {107}},\ \bibinfo {pages} {12517} (\bibinfo {year} {2010})}\BibitemShut {NoStop}%
\bibitem [{\citenamefont {Bouchet}\ and\ \citenamefont {Akhmanova}(2017)}]{bouchet2017microtubules}%
  \BibitemOpen
  \bibfield  {author} {\bibinfo {author} {\bibfnamefont {B.~P.}\ \bibnamefont {Bouchet}}\ and\ \bibinfo {author} {\bibfnamefont {A.}~\bibnamefont {Akhmanova}},\ }\href@noop {} {\bibfield  {journal} {\bibinfo  {journal} {Journal of cell science}\ }\textbf {\bibinfo {volume} {130}},\ \bibinfo {pages} {39} (\bibinfo {year} {2017})}\BibitemShut {NoStop}%
\bibitem [{\citenamefont {Kopf}\ \emph {et~al.}(2020)\citenamefont {Kopf}, \citenamefont {Renkawitz}, \citenamefont {Hauschild}, \citenamefont {Girkontaite}, \citenamefont {Tedford}, \citenamefont {Merrin}, \citenamefont {Thorn-Seshold}, \citenamefont {Trauner}, \citenamefont {H{\"a}cker}, \citenamefont {Fischer} \emph {et~al.}}]{kopf2020microtubules}%
  \BibitemOpen
  \bibfield  {author} {\bibinfo {author} {\bibfnamefont {A.}~\bibnamefont {Kopf}}, \bibinfo {author} {\bibfnamefont {J.}~\bibnamefont {Renkawitz}}, \bibinfo {author} {\bibfnamefont {R.}~\bibnamefont {Hauschild}}, \bibinfo {author} {\bibfnamefont {I.}~\bibnamefont {Girkontaite}}, \bibinfo {author} {\bibfnamefont {K.}~\bibnamefont {Tedford}}, \bibinfo {author} {\bibfnamefont {J.}~\bibnamefont {Merrin}}, \bibinfo {author} {\bibfnamefont {O.}~\bibnamefont {Thorn-Seshold}}, \bibinfo {author} {\bibfnamefont {D.}~\bibnamefont {Trauner}}, \bibinfo {author} {\bibfnamefont {H.}~\bibnamefont {H{\"a}cker}}, \bibinfo {author} {\bibfnamefont {K.-D.}\ \bibnamefont {Fischer}},  \emph {et~al.},\ }\href@noop {} {\bibfield  {journal} {\bibinfo  {journal} {Journal of Cell Biology}\ }\textbf {\bibinfo {volume} {219}},\ \bibinfo {pages} {e201907154} (\bibinfo {year} {2020})}\BibitemShut {NoStop}%
\bibitem [{\citenamefont {Ackermann}\ \emph {et~al.}(2022)\citenamefont {Ackermann}, \citenamefont {Qu}, \citenamefont {LeGoff},\ and\ \citenamefont {{Ben Amar}}}]{ackermann2022modeling}%
  \BibitemOpen
  \bibfield  {author} {\bibinfo {author} {\bibfnamefont {J.}~\bibnamefont {Ackermann}}, \bibinfo {author} {\bibfnamefont {P.-Q.}\ \bibnamefont {Qu}}, \bibinfo {author} {\bibfnamefont {L.}~\bibnamefont {LeGoff}}, \ and\ \bibinfo {author} {\bibfnamefont {M.}~\bibnamefont {{Ben Amar}}},\ }\href@noop {} {\bibfield  {journal} {\bibinfo  {journal} {The European Physical Journal Plus}\ }\textbf {\bibinfo {volume} {137}},\ \bibinfo {pages} {1} (\bibinfo {year} {2022})}\BibitemShut {NoStop}%
\bibitem [{\citenamefont {Winkler}(1867)}]{winkler1867lehre}%
  \BibitemOpen
  \bibfield  {author} {\bibinfo {author} {\bibfnamefont {E.}~\bibnamefont {Winkler}},\ }\href@noop {} {\emph {\bibinfo {title} {Die Lehre von der Elasticitaet und Festigkeit: mit besonderer R{\"u}cksicht auf ihre Anwendung in der Technik, f{\"u}r polytechnische Schulen, Bauakademien, Ingenieure, Maschinenbauer, Architecten, etc}}}\ (\bibinfo  {publisher} {H. Dominicus},\ \bibinfo {year} {1867})\BibitemShut {NoStop}%
\bibitem [{\citenamefont {Kim}\ \emph {et~al.}(2023)\citenamefont {Kim}, \citenamefont {Zhang},\ and\ \citenamefont {Schwarz}}]{kim2023mean}%
  \BibitemOpen
  \bibfield  {author} {\bibinfo {author} {\bibfnamefont {K.}~\bibnamefont {Kim}}, \bibinfo {author} {\bibfnamefont {T.}~\bibnamefont {Zhang}}, \ and\ \bibinfo {author} {\bibfnamefont {J.}~\bibnamefont {Schwarz}},\ }\href@noop {} {\bibfield  {journal} {\bibinfo  {journal} {arXiv preprint arXiv:2308.12892}\ } (\bibinfo {year} {2023})}\BibitemShut {NoStop}%
\bibitem [{\citenamefont {Sato}\ and\ \citenamefont {Umetsu}(2021)}]{Sato2021}%
  \BibitemOpen
  \bibfield  {author} {\bibinfo {author} {\bibfnamefont {K.}~\bibnamefont {Sato}}\ and\ \bibinfo {author} {\bibfnamefont {D.}~\bibnamefont {Umetsu}},\ }\href {\doibase 10.3389/fphy.2021.704878} {\bibfield  {journal} {\bibinfo  {journal} {Frontiers in Physics}\ }\textbf {\bibinfo {volume} {9}} (\bibinfo {year} {2021}),\ 10.3389/fphy.2021.704878}\BibitemShut {NoStop}%
\bibitem [{\citenamefont {Lin}\ \emph {et~al.}(2022)\citenamefont {Lin}, \citenamefont {Merkel},\ and\ \citenamefont {Rupprecht}}]{Lin2022}%
  \BibitemOpen
  \bibfield  {author} {\bibinfo {author} {\bibfnamefont {S.-Z.}\ \bibnamefont {Lin}}, \bibinfo {author} {\bibfnamefont {M.}~\bibnamefont {Merkel}}, \ and\ \bibinfo {author} {\bibfnamefont {J.-F.}\ \bibnamefont {Rupprecht}},\ }\href {\doibase 10.1140/epje/s10189-021-00154-2} {\bibfield  {journal} {\bibinfo  {journal} {The European Physical Journal E}\ }\textbf {\bibinfo {volume} {45}} (\bibinfo {year} {2022}),\ 10.1140/epje/s10189-021-00154-2}\BibitemShut {NoStop}%
\bibitem [{\citenamefont {{Ben Amar}}\ and\ \citenamefont {Bianca}(2016)}]{ben2016onset}%
  \BibitemOpen
  \bibfield  {author} {\bibinfo {author} {\bibfnamefont {M.}~\bibnamefont {{Ben Amar}}}\ and\ \bibinfo {author} {\bibfnamefont {C.}~\bibnamefont {Bianca}},\ }\href@noop {} {\bibfield  {journal} {\bibinfo  {journal} {Scientific reports}\ }\textbf {\bibinfo {volume} {6}},\ \bibinfo {pages} {33849} (\bibinfo {year} {2016})}\BibitemShut {NoStop}%
\bibitem [{\citenamefont {Inc.}(2022)}]{Mathematica}%
  \BibitemOpen
  \bibfield  {author} {\bibinfo {author} {\bibfnamefont {W.~R.}\ \bibnamefont {Inc.}},\ }\href {https://www.wolfram.com/mathematica} {\enquote {\bibinfo {title} {Mathematica, {V}ersion 13.2},}\ } (\bibinfo {year} {2022}),\ \bibinfo {note} {champaign, IL, 2022}\BibitemShut {NoStop}%
\bibitem [{\citenamefont {Lin}\ \emph {et~al.}(2023)\citenamefont {Lin}, \citenamefont {Merkel},\ and\ \citenamefont {Rupprecht}}]{Lin2023}%
  \BibitemOpen
  \bibfield  {author} {\bibinfo {author} {\bibfnamefont {S.-Z.}\ \bibnamefont {Lin}}, \bibinfo {author} {\bibfnamefont {M.}~\bibnamefont {Merkel}}, \ and\ \bibinfo {author} {\bibfnamefont {J.-F. m.~c.}\ \bibnamefont {Rupprecht}},\ }\href {\doibase 10.1103/PhysRevLett.130.058202} {\bibfield  {journal} {\bibinfo  {journal} {Phys. Rev. Lett.}\ }\textbf {\bibinfo {volume} {130}},\ \bibinfo {pages} {058202} (\bibinfo {year} {2023})}\BibitemShut {NoStop}%
\bibitem [{\citenamefont {Yan}\ and\ \citenamefont {Bi}(2019)}]{Yan2019}%
  \BibitemOpen
  \bibfield  {author} {\bibinfo {author} {\bibfnamefont {L.}~\bibnamefont {Yan}}\ and\ \bibinfo {author} {\bibfnamefont {D.}~\bibnamefont {Bi}},\ }\href {\doibase 10.1103/PhysRevX.9.011029} {\bibfield  {journal} {\bibinfo  {journal} {Phys. Rev. X}\ }\textbf {\bibinfo {volume} {9}},\ \bibinfo {pages} {011029} (\bibinfo {year} {2019})}\BibitemShut {NoStop}%
\bibitem [{\citenamefont {Huang}\ \emph {et~al.}(2022)\citenamefont {Huang}, \citenamefont {Cochran}, \citenamefont {Fielding}, \citenamefont {Marchetti},\ and\ \citenamefont {Bi}}]{Huang2022}%
  \BibitemOpen
  \bibfield  {author} {\bibinfo {author} {\bibfnamefont {J.}~\bibnamefont {Huang}}, \bibinfo {author} {\bibfnamefont {J.~O.}\ \bibnamefont {Cochran}}, \bibinfo {author} {\bibfnamefont {S.~M.}\ \bibnamefont {Fielding}}, \bibinfo {author} {\bibfnamefont {M.~C.}\ \bibnamefont {Marchetti}}, \ and\ \bibinfo {author} {\bibfnamefont {D.}~\bibnamefont {Bi}},\ }\href@noop {} {\bibfield  {journal} {\bibinfo  {journal} {Physical Review Letters}\ }\textbf {\bibinfo {volume} {128}},\ \bibinfo {pages} {178001} (\bibinfo {year} {2022})}\BibitemShut {NoStop}%
\bibitem [{\citenamefont {Gandikota}\ \emph {et~al.}(2022)\citenamefont {Gandikota}, \citenamefont {Parker},\ and\ \citenamefont {Schwarz}}]{Gandikota2022}%
  \BibitemOpen
  \bibfield  {author} {\bibinfo {author} {\bibfnamefont {M.}~\bibnamefont {Gandikota}}, \bibinfo {author} {\bibfnamefont {A.}~\bibnamefont {Parker}}, \ and\ \bibinfo {author} {\bibfnamefont {J.}~\bibnamefont {Schwarz}},\ }\href@noop {} {\bibfield  {journal} {\bibinfo  {journal} {Physical Review E}\ }\textbf {\bibinfo {volume} {106}},\ \bibinfo {pages} {055003} (\bibinfo {year} {2022})}\BibitemShut {NoStop}%
\bibitem [{\citenamefont {Bidhendi}\ \emph {et~al.}(2019)\citenamefont {Bidhendi}, \citenamefont {Altartouri}, \citenamefont {Gosselin},\ and\ \citenamefont {Geitmann}}]{Bidhendi563403}%
  \BibitemOpen
  \bibfield  {author} {\bibinfo {author} {\bibfnamefont {A.~J.}\ \bibnamefont {Bidhendi}}, \bibinfo {author} {\bibfnamefont {B.}~\bibnamefont {Altartouri}}, \bibinfo {author} {\bibfnamefont {F.~P.}\ \bibnamefont {Gosselin}}, \ and\ \bibinfo {author} {\bibfnamefont {A.}~\bibnamefont {Geitmann}},\ }\href {\doibase 10.1101/563403} {\bibfield  {journal} {\bibinfo  {journal} {bioRxiv}\ } (\bibinfo {year} {2019}),\ 10.1101/563403}\BibitemShut {NoStop}%
\bibitem [{\citenamefont {Zuch}\ \emph {et~al.}(2021)\citenamefont {Zuch}, \citenamefont {Doyle}, \citenamefont {Majda}, \citenamefont {Smith}, \citenamefont {Robert},\ and\ \citenamefont {Torii}}]{koab250}%
  \BibitemOpen
  \bibfield  {author} {\bibinfo {author} {\bibfnamefont {D.~T.}\ \bibnamefont {Zuch}}, \bibinfo {author} {\bibfnamefont {S.~M.}\ \bibnamefont {Doyle}}, \bibinfo {author} {\bibfnamefont {M.}~\bibnamefont {Majda}}, \bibinfo {author} {\bibfnamefont {R.~S.}\ \bibnamefont {Smith}}, \bibinfo {author} {\bibfnamefont {S.}~\bibnamefont {Robert}}, \ and\ \bibinfo {author} {\bibfnamefont {K.~U.}\ \bibnamefont {Torii}},\ }\href {\doibase 10.1093/plcell/koab250} {\bibfield  {journal} {\bibinfo  {journal} {The Plant Cell}\ }\textbf {\bibinfo {volume} {34}},\ \bibinfo {pages} {209} (\bibinfo {year} {2021})}\BibitemShut {NoStop}%
\bibitem [{\citenamefont {Kim}(2022)}]{kimthesis}%
  \BibitemOpen
  \bibfield  {author} {\bibinfo {author} {\bibfnamefont {K.}~\bibnamefont {Kim}},\ }\href {https://surface.syr.edu/etd/1656} {\enquote {\bibinfo {title} {Geometry of discrete and continuous bounded surfaces},}\ } (\bibinfo {year} {2022}),\ \bibinfo {note} {dissertations - ALL. 1656.}\BibitemShut {Stop}%
\bibitem [{\citenamefont {Brander}\ \emph {et~al.}(2018)\citenamefont {Brander}, \citenamefont {B{\ae}rentzen}, \citenamefont {Fisker},\ and\ \citenamefont {Gravesen}}]{brander2018bezier}%
  \BibitemOpen
  \bibfield  {author} {\bibinfo {author} {\bibfnamefont {D.}~\bibnamefont {Brander}}, \bibinfo {author} {\bibfnamefont {J.~A.}\ \bibnamefont {B{\ae}rentzen}}, \bibinfo {author} {\bibfnamefont {A.-S.}\ \bibnamefont {Fisker}}, \ and\ \bibinfo {author} {\bibfnamefont {J.}~\bibnamefont {Gravesen}},\ }\href@noop {} {\bibfield  {journal} {\bibinfo  {journal} {Computer-Aided Design}\ }\textbf {\bibinfo {volume} {104}},\ \bibinfo {pages} {36} (\bibinfo {year} {2018})}\BibitemShut {NoStop}%
\bibitem [{\citenamefont {Mimura}\ and\ \citenamefont {Inoue}(2023)}]{Mimura2023}%
  \BibitemOpen
  \bibfield  {author} {\bibinfo {author} {\bibfnamefont {T.}~\bibnamefont {Mimura}}\ and\ \bibinfo {author} {\bibfnamefont {Y.}~\bibnamefont {Inoue}},\ }\href {\doibase https://doi.org/10.1016/j.jtbi.2023.111560} {\bibfield  {journal} {\bibinfo  {journal} {Journal of Theoretical Biology}\ }\textbf {\bibinfo {volume} {571}},\ \bibinfo {pages} {111560} (\bibinfo {year} {2023})}\BibitemShut {NoStop}%
\bibitem [{\citenamefont {Murisic}\ \emph {et~al.}(2015)\citenamefont {Murisic}, \citenamefont {Hakim}, \citenamefont {Kevrekidis}, \citenamefont {Shvartsman},\ and\ \citenamefont {Audoly}}]{Murisic2015}%
  \BibitemOpen
  \bibfield  {author} {\bibinfo {author} {\bibfnamefont {N.}~\bibnamefont {Murisic}}, \bibinfo {author} {\bibfnamefont {V.}~\bibnamefont {Hakim}}, \bibinfo {author} {\bibfnamefont {I.}~\bibnamefont {Kevrekidis}}, \bibinfo {author} {\bibfnamefont {S.~Y.}\ \bibnamefont {Shvartsman}}, \ and\ \bibinfo {author} {\bibfnamefont {B.}~\bibnamefont {Audoly}},\ }\href {\doibase 10.1016/j.bpj.2015.05.019} {\bibfield  {journal} {\bibinfo  {journal} {Biophysical Journal}\ }\textbf {\bibinfo {volume} {109}},\ \bibinfo {pages} {154} (\bibinfo {year} {2015})}\BibitemShut {NoStop}%
\end{thebibliography}%

\end{document}